\documentclass{SCIS2023}

\usepackage{multirow}
\usepackage{subfloat}

\begin{document}
\ArticleType{REVIEW}
\Year{2024}
\Month{}
\Vol{}
\No{}
\DOI{}
\ArtNo{}
\ReceiveDate{}
\ReviseDate{}
\AcceptDate{}
\OnlineDate{}

\title{Perceptual Video Quality Assessment: A Survey}{Perceptual Video Quality Assessment: A Survey}

\author[]{Xiongkuo Min}{}
\author[]{Huiyu Duan}{}
\author[]{Wei Sun}{}
\author[]{Yucheng Zhu}{}
\author[]{Guangtao Zhai}{zhaiguangtao@sjtu.edu.cn}

\AuthorMark{Min X K, \textit{et al.}}

\AuthorCitation{Min X K, Duan H Y, Sun W, Zhu Y C, Zhai G T, et al}


\address[]{Department of Electronic Engineering, Shanghai Jiao Tong University, Shanghai 200240, China}

\abstract{Perceptual video quality assessment plays a vital role in the field of video processing due to the existence of quality degradations introduced in various stages of video signal acquisition, compression, transmission and display.
With the advancement of internet communication and cloud service technology, video content and traffic are growing exponentially, which further emphasizes the requirement for accurate and rapid assessment of video quality.
Therefore, numerous subjective and objective video quality assessment studies have been conducted over the past two decades for both generic videos and specific videos such as streaming, user-generated content (UGC), 3D, virtual and augmented reality (VR and AR), high frame rate (HFR), audio-visual, \textit{etc}.
This survey provides an up-to-date and comprehensive review of these video quality assessment studies.
Specifically, we first review the subjective video quality assessment methodologies and databases, which are necessary for validating the performance of video quality metrics.
Second, the objective video quality assessment algorithms for general purposes are surveyed and concluded according to the methodologies utilized in the quality measures.
Third, we overview the objective video quality assessment measures for specific applications and emerging topics.
Finally, the performances of the state-of-the-art video quality assessment measures are compared and analyzed.
This survey provides a systematic overview of both classical works and recent progresses in the realm of video quality assessment, which can help other researchers quickly access the field and conduct relevant research.
}


\maketitle

\section{Introduction}
Video is an electronic medium that involves the recording, copying, playback, broadcasting, and display of moving visual information, which is one of the most important forms of media.
It is estimated that video traffic contributed 65\% of the internet traffic \cite{prnewswire,saha3perceptual} due to evolution of internet communication, cloud service, and the popularization of video-sharing platforms.
Recommending and delivering high-quality video content is important for retaining the interest of users.
However, considering the uneven video shooting quality, huge user upload volume and congested network status, the video quality on the user side may not be satisfactory, which may cause negative quality of experience (QoE) and reduce the engagement.
Therefore, video quality assessment (VQA) is crucial in video communication systems to ensure and improve the quality of video contents delivered to the end-users.

Video quality assessment can be performed subjectively or objectively \cite{zhai2020perceptual}.
Subjective video quality assessment is usually considered as the most reliable and accurate evaluation method.
However, performing subjective evaluation is time-consuming and expensive, which makes it hard to be used in visual communication systems.
Thus, subjective video quality assessment is generally used as the evaluation method for objective video quality assessment, which aims to objectively assess the perceived quality of videos of the human visual system (HVS).
Video quality can be affected by many factors such as spatial and temporal resolution, frame rate, blur, noise, compression artifacts, \textit{etc.}, which brings challenges for designing objective video quality evaluation algorithms.
Moreover, video categories are diverse, which include user-generated content (UGC), virtual and augmented reality (VR and AR), high frame rate (HFR), audio-visual, gaming, \textit{etc.}.
HVS has different perceptual characteristics for different types of videos, thus the influence of degradation on the perceived quality of these videos is also different, which further increases the difficulties of devising objective VQA measures.
Therefore, numerous works have conducted studies on subjective and objective video quality assessment considering different conditions and factors \cite{duan2017ivqad,chen2012study,nasiri2015perceptual,min2020study,zhu2023perceptual,shang2022subjective,yu2023subjective}.

\begin{figure}[t]
    \centering
    \includegraphics[width=1\textwidth]{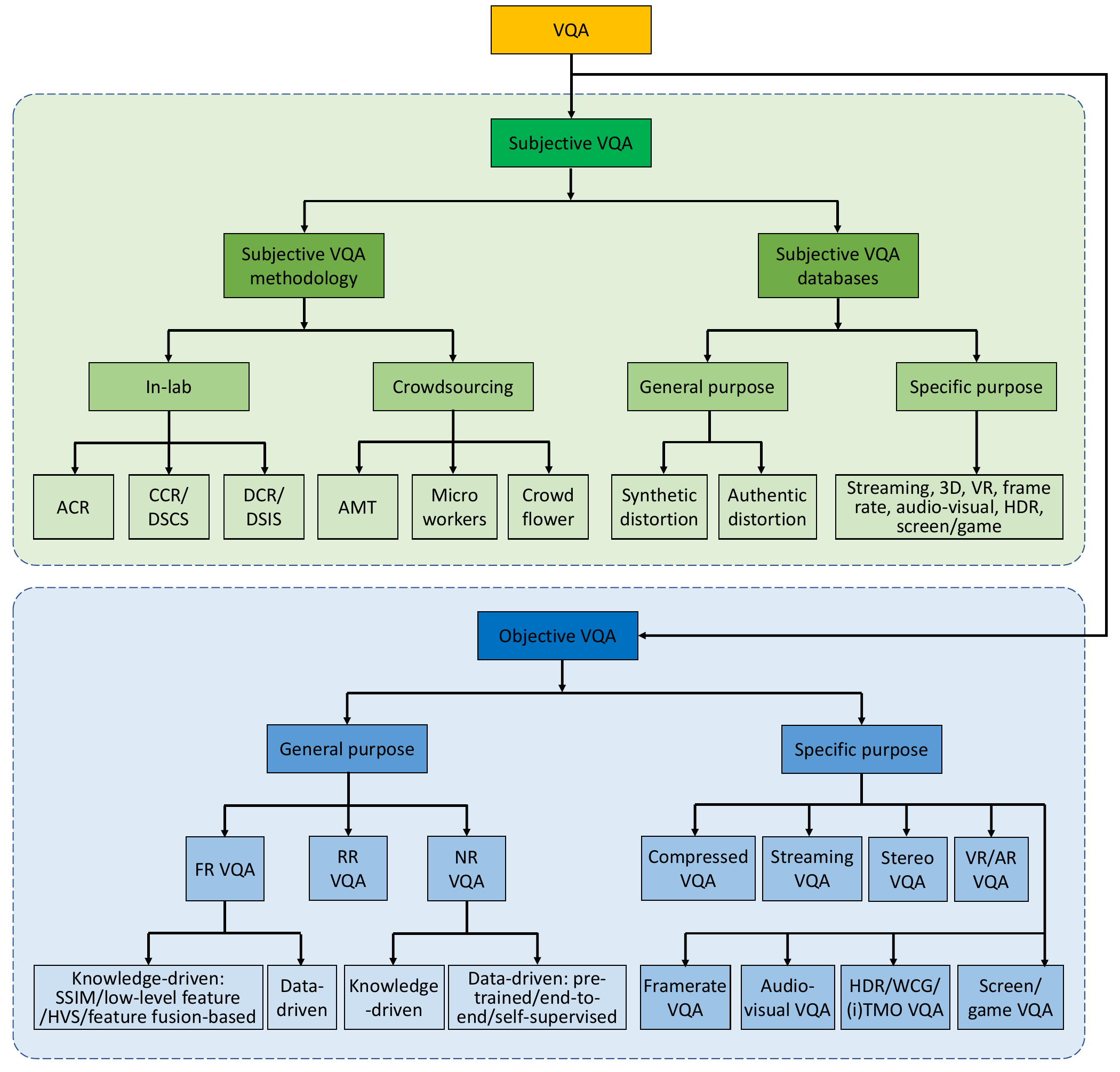}
    \caption{Scope of this survey.}
    \label{fig:scope}
\end{figure}

\subsection{Related Surveys}
Since a large number of IQA and VQA studies have been carried out, some papers have also surveyed these works.
Some papers have reviewed related research on image quality assessment.
Wang and Bovik \cite{wang2009mean} provided an initial analysis of full-reference (FR) image fidelity assessment from the aspect of signal fidelity.
They further gave a comprehensive introduction to reduced-reference (RR) and no-reference (NR) image quality assessment (IQA) \cite{wang2011reduced} in 2011.
Lin and Kuo \cite{lin2011perceptual} discussed several influence factors of perceptual visual quality measure including signal decomposition \cite{mallat1989theory,hyvarinen2000independent,duan2022develop}, just-noticeable distortions \cite{wu2013just,wu2017enhanced}, visual attention \cite{liu2011visual,duan2022confusing}, feature and artifact extraction \cite{zhu2014no}, viewing conditions \cite{gu2015quality,zhu2016closing,duan2022confusing,duan2022augmented}, \textit{etc.} 
Moorthy and Bovik \cite{moorthy2011visual} presented their perspectives of the future trends of the visual quality assessment field.
Chandler \cite{chandler2013seven} summarized seven challenges in image quality assessment.
Zhai \textit{et al.} \cite{zhai2020perceptual} gave a comprehensive survey of classical algorithms and recent progress in the realm of perceptual image quality assessment.

Some researchers have also surveyed the studies on perceptual video quality assessment.
Chikkerur \textit{et al.} \cite{chikkerur2011objective} classified, reviewed the objective video quality assessment methods, and compared the performance of these metrics.
Shahid \textit{et al.} \cite{shahid2014no} discusses classical and well-know NR video quality assessment algorithms.
Chen \textit{et al.} \cite{chen2014qos} wrote a tutorial for video quality assessment discussing the relationship between quality of service (QoS) and quality of experience (QoE).
Fan \textit{et al.} \cite{fan2019metrics} briefly reviewed the metrics and methods of video quality assessment.
Li \textit{et al.} \cite{li2019recent} summarized recent advances and challenges in video quality assessment.
Zhou \textit{et al.} \cite{zhou2022brief} introduced a brief survey on adaptive video streaming quality assessment.
Saha \textit{et al.} \cite{saha3perceptual} surveyed the recent progress of video quality assessment and discussed future research trends.

Most of these existing surveys for VQA research discuss classical VQA techniques, or only provide an overview of specific VQA topics.
With the advancement of deep learning, many state-of-the-art VQA models have adopted deep neural networks (DNN) to predict perceptual quality, which is rarely addressed in most previous reviews.
Moreover, with the recent progress in multimedia, many works have also conducted VQA research for specific applications, such as VR/AR, HFR, UGC, audio-visual, \textit{etc.}, which failed to be reviewed in most surveys.
Therefore, a systematic, comprehensive, and up-to-date survey is still needed.

\subsection{Scope and Organization of This Survey}
This survey provides an up-to-date and comprehensive review of VQA studies.
Since VQA has varied application-specific use cases, we only thoroughly overview these studies, but refrain from benchmarking them.
The scope of this survey is shown in Figure \ref{fig:scope}. 
The organization of this survey is introduced as follows.
Section \ref{sec:subjective} summarizes subjective VQA methodologies, and reviews subjective VQA databases for general purpose and specific applications.
In Section \ref{sec:objective_traditional}, we review objective VQA measures for traditional topics (\textit{i.e.}, general purposes), including FR, RR, and NR metrics.
Section \ref{sec:objective_specific} surveys objective VQA measures for emerging topics (\textit{i.e.,} specific applications), including compression VQA, streaming VQA, stereoscopic VQA, virtual reality VQA, high frame rate VQA, audio-visual VQA, high dynamic range (HDR) VQA, screen and game VQA.
In Section \ref{sec:evaluation}, evaluation process of VQA measures is discussed, with a comparison of their performances.
Section \ref{sec:future} outlooks future trends in VQA and Section \ref{sec:summary} summarizes the whole paper.

\section{Subjective Video Quality Assessment}
\label{sec:subjective}
Subjective quality assessment is the most reliable way for evaluating the perceptual quality of images or videos, since human eyes are usually the ultimate receiver of these contents \cite{zhai2020perceptual,quan2023evaluation,quan2023binocular}.
Different application systems may require different subjective assessment methods \cite{chen2021muiqa,chen2022mriqa,wu2021accurate,duan2023masked,li2023boosting,hu2018combination,hu2017dual}.
In this section, we first review the general methodology of subjective quality assessment suggested by ITU-R BT.500 \cite{series2012methodology} including subjective viewing environment setup, subject recruitment, subject grading, and subjective result processing, \textit{etc.}
Then, 20 subjective VQA databases for general contents are reviewed.

\subsection{Subjective VQA Methodology}

Subjective quality assessment usually needs a large number of subjects to rate the quality of the target objects according to certain standards, and take the mean opinion score (MOS) or difference mean opinion score (DMOS) as the result of the subjective quality.
The MOS means the average score from all subjects for specific stimuli, while the DMOS refers to the average value of the difference between the scores of the reference stimuli and the scores of the corresponding distorted stimuli.
For DMOS, the influence of the video content can be effectively reduced by subtracting the scores of the reference material.
The measurement of the perceived quality of videos requires the use of subjective scaling methods, and several test methods are usually adopted, including absolute category rating (ACR), comparison category rating (CCR, also known as double stimulus comparison scale (DSCS)), degradation category rating (DCR, also known as double stimulus impairment scale (DSIS)), \textit{etc.}, whereas ACR is better suited to obtain a general, unbiased judgment of the overall quality, DCR and CCR might be better suited for smaller, subtle differences.
The subjective quality assessment process generally includes five steps according to the recommendations given by the ITU \cite{series2012methodology}:
(1) Build evaluation environment. Experimental instructors need to set up and calibrate the test environment and test equipment to achieve the corresponding viewing requirements.
(2) Prepare test stimuli. Generally, experimental instructors prepare test stimuli according to the problem to be evaluated, such as raw videos and distorted videos.
(3) Invite or recruit subjects to give opinion scores. The subjects can be experts or non-experts according to the requirements of the experiment, but in any case the subjects should not know the purpose of the experiment. All subjects should have normal or corrected to normal vision. Generally, the number of subjects should not be less than 15.
(4) Conduct subjective experiments. The subjects should give their subjective quality ratings according to predetermined test methods and evaluation scales. There are many types of test methods, such as single-stimulus method and double-stimulus method, which are generally selected according to the needs of the experiment. The evaluation scale generally adopts a five grade quality and impairment scale, and the detailed scale can be continuous or discrete according to the requirements of the experiment.
(5) Process subjective data. First, the subjective data needs to be screened to remove abnormal subjects and abnormal scores, and the screening criteria can refer to the recommendations provided by the ITU. Then the MOS values or DMOS values can be calculated as the final subjective quality results.

Since it is hard to recruit numerous subjects to join a lab experiment, many recent studies have adopted to use crowdsourcing methods to conduct subjective video quality assessment experiments.
Crowdsourcing offers fast, low cost, and scalable approaches by outsourcing tasks to a large number of participants. 
In addition, crowdsourcing also provides a large diverse source of participants, and a practical environment for quality assessment of multimedia services and applications. 
Nevertheless, crowdsourcing subjective quality assessment methods may not obtain data with equal quality compared to laboratory testing methods, due to factors they inherit from the nature of crowdsourcing.
Therefore, crowdsourcing experiments should be designed differently and carefully.
ITU has given some recommendations for crowdsourcing subjective quality assessment \cite{ITU2018}.
The subjective crowdsourcing video quality assessment method generally includes the following steps:
(1) Choose crowdsourcing platforms and crowdworkers. There are many crowdsourcing platforms that can be adopted to conduct subjective VQA experiments, such as Amazon Mechanical Turk (AMT) \cite{AMT}, Microworkers \cite{Microworkers}, CrowdFlower \cite{CrowdFlower}, Crowdee \cite{Crowdee}, \textit{etc}. Moreover, most platforms provide the function of choosing target subjects, including the conditions of household size, educational level, yearly income, \textit{etc.}
(2) Prepare subjective experiment. Considering the crowdsourcing condition, some aspects should be considered when preparing the experimental stimuli, including overall experiment enjoyability, task duration and complexity, user interface (UI) logistics, compensation relative to the test duration and complexity, test clarity and the user's ability to understand the task.
(3) Conduct subjective experiment. The experiment procedure includes a qualification step, a training step, and a rating step. It should be noted that experimenters should incorporate different validity check methods in these steps for ensuring the worker's full attentiveness to the task, and the adequateness of the environment and system used by the worker.
(4) Screen subjective data. It is important to screen the data obtained from the crowdsourcing methods, including examining entire votes for each video stimulus to remove score outliers, and calculate the correlation coefficient between the MOS from one work and the global MOS to remove subject outliers.

\begin{table*}[t]
    \centering
    \caption{An overview of popular public video quality assessment databases for general contents, including lab-introduced video datasets with \textbf{synthetic} distortions and large-scale crowdsourced user-generated content (UGC) video datasets with \textbf{authentic} distortions.}
    \label{tab:tab2.1}
    \setlength{\tabcolsep}{0.35em}
    \scalebox{0.6}{
    \renewcommand{\arraystretch}{1.5}
    \begin{tabular}{l l l l l l l l l l l l l l}
    \toprule
        Database & Type & Year & \#Cont. & \#Total & Resolution & FR & Dur. & Format & Distortions & \#Subj. & \#Ratings & Data & Env. \\
        \hline
        LIVE-VQA~\cite{seshadrinathan2010study} & Syn. & 2008 & 10 & 160 & 768$\times$432 & 25/50 & 10 & YUV$+$264 & Compression, transmission & 38 & 29 & DMOS$+$$\sigma$ & In-lab \\
        EPFL-PoliMI~\cite{de2010h} & Syn. & 2009 & 12 & 156 & CIF/4CIF & 25/30 & 10 & YUV$+$264 & Compression, transmission & 40 & 34 & MOS & In-lab \\
        VQEG-HDTV~\cite{VQEGHDTV} & Syn. & 2010 & 49 & 740 & 1080i/p & 25/30 & 10 & AVI & Compression, transmission & 120 & 24 & RAW & In-lab \\
        IVP~\cite{IVP} & Syn. & 2011 & 10 & 138 & 1080p & 25 & 10 & YUV & Compression, transmission & 42 & 35 & DMOS$+$$\sigma$ & In-lab \\
        TUM 1080p50~\cite{keimel2012tum} & Syn. & 2012 & 5 & 25 & 1080p & 50 & 10 & YUV & Compression & 21 & 21 & MOS & In-lab \\
        LIVE Mobile~\cite{moorthy2012video} & Syn. & 2012 & 10 & 200 & 720p & 30 & 15 & YUV & Compression, transmission & 30$+$17 & 30$+$17 & DMOS$+$$\sigma$ & In-lab \\
        CSIQ~\cite{vu2014vis} & Syn. & 2014 & 12 & 228 & 832$\times$480 & 24-60 & 10 & YUV & Compression, transmission & 35 & N/A & DMOS$+$$\sigma$ & In-lab \\
        MCL-V~\cite{lin2015mcl} & Syn. & 2015 & 12 & 108 & 1080p & 24-30 & 6 & YUV & Compression, scaling & 45 & 32 & MOS & In-lab \\
        MCL-JCV~\cite{wang2016mcl} & Syn. & 2016 & 30 & 1560 & 1080p & 24-30 & 5 & MP4 & Compression & 150 & 50 & RAW-JND & In-lab \\
        \hline
        CVD2014~\cite{nuutinen2016cvd2014} & Aut. & 2014 & 5 & 234 & 720p, 480p & 9-30 & 10-25 & AVI & In-capture & 210 & 30 & MOS & In-lab \\
        LIVE-Qualcomm~\cite{ghadiyaram2017capture} & Aut. & 2016 & 54 & 208 & 1080p & 30 & 15 & YUV & In-capture & 39 & 39 & MOS & In-lab \\
        KoNViD-1k~\cite{hosu2017konstanz} & Aut. & 2017 & 1200 & 1200 & 540p & 24-30 & 8 & MP4 & In-the-wild & 642 & 114 & MOS$+$$\sigma$ & Crowd \\
        LIVE-VQC~\cite{sinno2018large} & Aut. & 2018 & 585 & 585 & 1080p-240p & 19-30 & 10 & MP4 & In-the-wild & 4776 & 240 & MOS & Crowd \\
        YouTube-UGC~\cite{wang2019youtube} & Aut. & 2019 & 1380 & 1380 & 4k-360p & 15-60 & 20 & MKV & In-the-wild & $>$8k & 123 & MOS$+$$\sigma$ & Crowd \\
        LSVQ~\cite{ying2021patch} & Aut. & 2021 &  39075 & 39075 & Diverse & Diverse & 5-12 & MP4 & In-the-wild & 6284 & 35 & MOS & Crowd \\
        \hline
        UGC-VIDEO~\cite{li2020ugc} & Syn.$+$Aut. & 2019 & 50 & 550 & 720p & 30 & 10 & N/A & UGC$+$compression & 30 & 30 & DMOS & In-lab \\
        LIVE-WC~\cite{yu2021predicting} & Syn.$+$Aut. & 2020 & 55 & 275 & 1080p & 30 & 10 & MP4 & UGC$+$compression & 40 & 40 & MOS & In-lab \\
        YT-UGC$+$(Subset)~\cite{wang2021rich} & Syn.$+$Aut. & 2021 & 189 & 567 & 1080p, 720p & Diverse & 20 & RAW$+$264 & UGC$+$compression & N/A & 30 & DMOS & In-lab \\
        ICME2021~\cite{icme21} & Syn.$+$Aut. & 2021 & 1000 & 8000 & N/A & N/A & N/A & N/A & UGC$+$compression & N/A & N/A & MOS & In-lab \\
        TaoLive~\cite{zhang2023md} & Syn.$+$Aut. & 2022 & 418 & 3762 & 1080p, 720p & 20 & 8 & MP4 & UGC$+$compression & 44 & 44 & MOS & In-lab \\
    \bottomrule
    \end{tabular}
    }
    \raggedright
    \scalebox{0.59}{
    \begin{tabular}{l}
        Note: \#Cont.: The number of unique video contents. \quad \#Total: Total number of test video sequences. \quad FR: Framerate (in fps). \quad Dur.: Video duration/length (in seconds). \\
         \quad  \quad  \quad \#Subj.: Total number of subjects in the study. \quad \#Ratings: Average number of subjective ratings per video. \quad Env.: Subjective experiment environment. \\
         \quad  \quad  \quad In-lab: Experiment was conducted in a laboratory. \quad Crowd: Experiment was conducted by crowdsourcing. \quad Syn.: Synthetic. \quad Aut.: Authentic.
    \end{tabular}
    }
\end{table*}

\subsection{Subjective VQA Databases for General Purpose}

Table~\ref{tab:tab2.1}~\cite{seshadrinathan2010study,de2010h,VQEGHDTV,IVP,keimel2012tum,moorthy2012video,vu2014vis,lin2015mcl,wang2016mcl,nuutinen2016cvd2014,ghadiyaram2017capture,hosu2017konstanz,sinno2018large,wang2019youtube,ying2021patch,li2020ugc,yu2021predicting,wang2021rich,icme21,zhang2023md} gives an overview of 20 databases that are widely used in the research of visual quality assessment for general video contents. 
Information including the type of databases, years, the number of unique video contents, total numbers of test video sequences, video resolutions, video frame rates, video durations, video formats, distortion types, subject numbers, average numbers of subjective ratings per video, subjective score types and subjective experiment environments is summarized.

\subsubsection{General VQA Databases with Synthetic Distortions}

Many early VQA studies have only considered synthetic distortions in their databases.
We first review 9 popular VQA databases with synthetic distortions as follows \cite{seshadrinathan2010study,de2010h,VQEGHDTV,IVP,keimel2012tum,moorthy2012video,vu2014vis,lin2015mcl,wang2016mcl}.

\begin{itemize}
    \item LIVE video quality assessment database (LIVE-VQA) \cite{seshadrinathan2010study}.
    LIVE-VQA is a synthetic database, which includes 10 pristine videos and 160 distorted videos corrupted by compression and transmission distortions.
    All videos have a resolution of 768$\times$432, a frame rate of 25 or 50 fps, and a duration of 10 seconds. The video formats are YUV and h.264.
    The subjective experiment was conducted in a lab environment, and the subjective data contains DMOS and $\sigma$.

    \item EPFL-PoliMI database \cite{de2010h}.
    EPFL-PoliMI is a synthetic database, which includes 12 pristine videos and 156 distorted videos corrupted by compression and transmission distortions.
    All videos have a resolution of 360$\times$240 or 704$\times$480, a frame rate of 25 or 30, and a duration of 10 seconds. The video formats are YUV and h.264.
    The subjective experiment was conducted in a lab environment, and the subjective data is MOS.

    \item VQEG-HDTV database \cite{VQEGHDTV}.
    VQEG-HDTV is a synthetic database, which includes 49 pristine videos and 740 distorted videos degraded by compression and transmission distortions.
    All videos have a resolution of 1080i or 1080p, a frame rate of 25 or 30, and a duration of 10 seconds. The video format is AVI.
    The subjective experiment was conducted in a lab environment, and the raw subjective data is available.

    \item IVP subjective quality video database \cite{IVP}.
    IVP is a synthetic database, which includes 10 pristine videos and 138 distorted videos generated by compression and transmission distortions.
    All videos have a resolution of 1080p, a frame rate of 25, and a duration of 10 seconds. The video format is YUV.
    The subjective experiment was conducted in a lab environment, and the subjective data contains DMOS and $\sigma$.

    \item TUM high definition video datasets (TUM 1080p50)~\cite{keimel2012tum}.
    It is a synthetic database, which includes 5 pristine videos and 25 distorted videos degraded by compression.
    All videos have a resolution of 1080p, a frame rate of 50, and a duration of 10 seconds. The video format is YUV.
    The subjective experiment was conducted in a lab environment, and the subjective data available is MOS.

    \item LIVE mobile video quality assessment database (LIVE Mobile)~\cite{moorthy2012video}.
    LIVE mobile is a synthetic database, which includes 10 pristine videos and 200 distorted videos corrupted by compression and transmission distortions.
    All videos have a resolution of 720p, a frame rate of 30, and a duration of 15 seconds. The video format is YUV.
    The subjective experiment was conducted in a lab environment, and the subjective data contains DMOS and $\sigma$.

    \item CSIQ video database \cite{vu2014vis}.
    CSIQ is a synthetic database, which includes 12 pristine videos and 228 distorted videos corrupted by compression and transmission distortions.
    All videos have a resolution of 832$\times$480, and a duration of 10 seconds. 
    The frame rate ranges from 24 to 60.
    The video format is YUV.
    The subjective experiment was conducted in a lab environment, and the subjective data available is MOS.

    \item MCL-V: a streaming video quality assessment database~\cite{lin2015mcl}.
    MCL-V is a synthetic database, which includes 12 pristine videos and 108 distorted videos corrupted by compression and scaling distortions.
    All videos have a resolution of 1080p and a duration of 6 seconds. 
    The frame rate ranges from 24 to 30.
    The video format is YUV.
    The subjective experiment was conducted in a lab environment, and the subjective data available is MOS.
    
    \item MCL-JCV: a jnd-based h.264/avc video quality assessment datase~\cite{wang2016mcl}.
    MCL-JCV is a synthetic database, which includes 30 pristine videos and 1560 distorted videos degraded by compression.
    All videos have a resolution of 1080p and a duration of 5 seconds. 
    The frame rate ranges from 24 to 30.
    The video format is YUV.
    The subjective experiment was conducted in a lab environment, and the subjective data format is RAW-JND.
\end{itemize}

\subsubsection{General VQA Databases with Authentic Distortions}

With the popularity of UGC, many recent VQA studies have focused on authentic distortions \cite{nuutinen2016cvd2014,ghadiyaram2017capture,hosu2017konstanz,sinno2018large,wang2019youtube,ying2021patch}, \textit{i.e.,} in-capture or in-the-wild distortions, and some studies have also considered both synthetic and authentic distortions \cite{li2020ugc,yu2021predicting,wang2021rich,icme21,zhang2023md}.
These databases are summarized as follows.

\begin{itemize}
    \item CVD2014~\cite{nuutinen2016cvd2014}.
    It is an authentic database, which includes 5 scenes and 234 test video sequences with camera in-capture distortions.
    The resolution of the videos in the CVD2014 is 720p or 480p.
    The frame rate ranges from 9-30 fps.
    The video length ranges from 10 seconds to 25 seconds.
    The video format is AVI.
    The subjective experiment was conducted in a lab environment, and the subjective data format is MOS.
    
    \item LIVE-Qualcomm~\cite{ghadiyaram2017capture}.
    LIVE-Qualcomm is an authentic database, which includes 54 scenes and 208 test video sequences with in-capture distortions.
    All videos have a resolution of 1080p, a frame rate of 30 fps, and a duration of 15 seconds. The video format is YUV.
    The subjective experiment was conducted in a lab environment, and the subjective data format is MOS.
    
    \item The konstanz natural video database (KoNViD-1k)~\cite{hosu2017konstanz}.
    KoNViD-1k is an authentic database, which includes 1200 unique test video sequences with diverse authentic distortions.
    All videos were sampled from YFCC100m (Flickr) via a feature space of blur, colorfulness, contrast, SI, TI, and NIQE, and content was clipped from the original and resized to 540p with the landscape layout.
    The frame rates of these videos are 24, 25, and 30 fps, and the duration is 8 seconds.
    All videos are in MP4 format.
    The subjective experiment was conducted by crowdsourcing using CrowdFlower.
    A total of 642 subjects were included in the experiment, and 136800 subjective quality ratings were collected with about 114 votes per video.
    The MOS and $\sigma$ values are available in the database.
    
    \item LIVE-VQC~\cite{sinno2018large}.
    LIVE-VQC is an authentic database, which includes 585 unique test video sequences with diverse authentic distortions.
    All videos were manually captured by certain people, which includes many camera motion distortions and some night scene distortions.
    The resolution is not uniformly distributed and ranges from 240p to 1080p with the landscape or portrait layouts.
    The frame rates of these videos are 20, 24, 25, and 30 fps, and the duration is 10 seconds.
    All videos are in MP4 format.
    The subjective experiment was conducted by crowdsourcing using AMT.
    A total of 4776 subjects were included in the experiment, and 205000 subjective quality ratings were collected with about 240 votes per video.
    The MOS values are available in the database.
    
    \item Youtube UGC dataset for video compression research (YouTube-UGC)~\cite{wang2019youtube}.
    YouTube-UGC is an authentic database, which includes 1380 unique test video sequences with diverse authentic distortions.
    All videos were sampled from YouTube via a feature space of spatial, color, temporal, and chunk variation with diverse video contents including HDR, screen content, animation, gaming videos, \textit{etc}.
    All videos are in resolutions of 4k, 1080p, 720p, 480p, and 360p with the landscape and portrait layouts.
    The frame rates of these videos are 15, 20, 24, 25, 30, 50, and 60 fps, and the duration is 20 seconds.
    All videos are in MKV format.
    The subjective experiment was conducted by crowdsourcing using AMT.
    More than 8000 subjects participated in the experiment, and 170159 subjective quality ratings were collected with about 123 votes per video.
    The MOS and $\sigma$ values are available in the database.
    
    \item LSVQ~\cite{ying2021patch}.
    LSVQ is a large-scale authentic video quality assessment database, which includes 39075 unique video sequences with diverse authentic distortions.
    The videos in LSVQ have diverse resolutions and frame rates.
    The duration ranges from 5 seconds to 12 seconds. 
    All videos are in MP4 format.
    The subjective experiment was conducted by crowdsourcing using AMT.
    A total of 6284 subjects were included in the experiment, and each video was evaluated by 35 subjects.
    The subjective data format is MOS.
    
    \item UGC-VIDEO~\cite{li2020ugc}.
    UGC-VIDEO is a video quality assessment database with both synthetic and authentic distortions.
    It contains 50 UGC videos and 550 distorted videos corrupted by compression.
    All videos have a resolution of 720p, a frame rate of 30 fps, and a duration of 10 seconds.
    The subjective experiment was conducted in a lab environment, and each video was assessed by 30 subjects.
    The subjective data format is DMOS.
    
    \item LIVE-WC~\cite{yu2021predicting}.
    LIVE-WC is a video quality assessment database with both synthetic and authentic distortions.
    It contains 55 UGC videos and 275 distorted videos corrupted by compression.
    All videos have a resolution of 1080p, a frame rate of 30 fps, and a duration of 10 seconds.
    All videos are in MP4 format.
    The subjective experiment was conducted in a lab environment, and each video was assessed by 40 subjects.
    The subjective data format is MOS.
    
    \item YT-UGC+(Subset)~\cite{wang2021rich}.
    YT-UGC+(Subset) is a video quality assessment database with both synthetic and authentic distortions.
    It contains 189 UGC videos and 567 distorted videos corrupted by compression.
    All videos are in the resolutions of 1080p or 720p.
    The videos have diverse frame rates and a fixed duration of 20 seconds.
    The subjective experiment was conducted in a lab environment, and each video was assessed by 30 subjects.
    The subjective data format is DMOS.
    
    \item ICME2021~\cite{icme21}.
    It is a video quality assessment database with both synthetic and authentic distortions.
    It contains 1000 UGC videos and 8000 distorted videos corrupted by compression.
    The subjective experiment was conducted in a lab environment, and the data format is MOS.
    
    \item TaoLive~\cite{zhang2023md}.
    TaoLive is a video quality assessment database with both synthetic and authentic distortions, which contains 418 UGC videos and 3762 distorted videos corrupted by compression.
    All videos are in the resolutions of 1080p or 720p.
    The frame rate is 20 fps and the video length is 8 seconds.
    The subjective experiment was conducted in a lab environment, and each video was assessed by 44 subjects.
    The subjective data format is MOS.

\end{itemize}


\begin{table*}[!t]
    \centering
    \caption{An overview of popular public video quality assessment databases for specific applications.}
    \label{tab:tab2.2}
    \setlength{\tabcolsep}{0.3em}
    \scalebox{0.57}{
    \renewcommand{\arraystretch}{1.5}
    \begin{tabular}{l l l l l l l l l l}
    \toprule
        Category & Database & Year & \#Ref. & \#Total & Resolution & Dur. & \#Dist. Type & \#Subj. & Data \\
        \hline
        \multirow{9}{*}{Streaming} & LIVE Mobile \cite{moorthy2012video} & 2012 & 10 & 200 & 720p & 10 & H.264 compression, switching, stalling & 47 & DMOS \\
        & LIVE-TVSQ \cite{chen2014modeling} & 2014 & 3 & 15 & 720p & 300 & H.264 compression, switching & 25 & RDMOS \\
        & LIVE-AMV \cite{ghadiyaram2014study} & 2014 & 24 & 180 & 720p, 360p & 29-134 & stalling & 27 & DMOS \\
        & LIVE-NFLX-I \cite{bampis2017study} & 2017 & 14 & 112 & 1080p & $>$60 & H.264, initial buffering, stalling, switching & 55 & MOS \\
        & LIVE-NFLX-II \cite{bampis2021towards} & 2018 & 15 & 420 & 1080p & diverse & Video encoding, network simulation, \textit{etc.} & 65 & MOS \\
        & WaterlooSQoE-I \cite{duanmu2016quality} & 2016 & 20 & 180 & 1080p & 10 & H.264, initial buffering, stalling & 25 & MOS \\
        & WaterlooSQoE-II \cite{duanmu2017quality} & 2017 & 12 & 588 & 1080p & diverse & H.264, switching & 35 & MOS \\
        & WaterlooSQoE-III \cite{duanmu2018quality} & 2018 & 20 & 450 & diverse & diverse & H.264, initial buffering, stalling, switching & 34 & MOS \\
        & WaterlooSQoE-IV \cite{WaterlooSQoEIV} & 2019 & 5 & 1450 & N/A & N/A & video encoders, network traces, ABR algorithms, \textit{etc.} & N/A & MOS \\

        \hline
        \multirow{8}{*}{3D} & LIVE 3D \cite{chen2012study} & 2012 & 6 & 54 & 480p & 10, 15 & H.264 & 27 & DMOS \\
        & StSD 3D \cite{de2013toward} & 2013 & 14 & 116 & 1080p & 8 & H.264, HEVC & 16 & DMOS \\
        & Tampere 3D \cite{jumisko2011subjective} & 2011 & 4 & 60 & 1080p & 10 & H.264, Depth level & 30 & N/A \\
        & MMSPG 3D \cite{goldmann2010comprehensive} & 2010 & 6 & 30 & 1080p & 10 & Camer distance & 17 & MOS \\
        & NAMA3DS1-COSPAD1 \cite{urvoy2012nama3ds1} & 2012 & 10 & 110 & 1080p & 13, 16 & H.264, JPEG2000 & 29 & MOS \\
        & UBC DML 3D \cite{banitalebi2014effect} & 2014 & 5 & 64 & 1080p & 10 & HEVC, frame rates & 16 & MOS \\
        & 3DVCL$@$FER \cite{banitalebi2015effect,dumic20173d} & 2015 & 6 & 184 & 1080p & 10 & H.264, JPEG2000, geometric distortion, \textit{etc.} & 35 & MOS \\
        & WATERLOO-IVC 3D \cite{wang2017asymmetrically} & 2017 & 10 & 704 & 1080p, 1024$\times$768 & 6, 10 & HEVC, Gaussian low-pass filtering & 54 & MOS \\

        \hline
        \multirow{8}{*}{VR} & IVQAD2017 \cite{duan2017ivqad} & 2017 & 10 & 160 & 4096$\times$2048 & 15 & MPEG-4, difference resolutions, different frame rates & 13 & MOS \\
        & Zhang \textit{et al.} \cite{zhang2018subjective} & 2018 & 10 & 60 & 3600$\times$1800 & 10 & H.265 with different QPs & 30 & DMOS, HM \\
        & Zhang \textit{et al.} \cite{zhang2017subjective} & 2017 & 16 & 400 & 4096$\times$2048 & N/A & VP9, H.264, H.265, Gaussian noise, box blur & 23 & MOS, DMOS \\
        & Lopes \textit{et al.} \cite{lopes2018subjective} & 2018 & 6 & 85 & 8192$\times$4096 & 10 & H.265, different resolutions, different frame rates & 37 & MOS, DMOS \\
        & Singla \textit{et al.} \cite{singla2017comparison} & 2017 & 6 & 66 & 1080p, 4k & 10 & H.265 with different bitrates & 30 & MOS, HM \\
        & VR-VQA48 \cite{xu2018assessing} & 2018 & 12 & 48 & 4096$\times$2048 & 12 & H.265 with different QPs & 48 & MOS, DMOS \\
        & Tran \textit{et al.} \cite{tran2018study} & 2018 & 6 & 126 & N/A & 30 & H.265 with different QPs, different resolutions & 37 & MOS \\
        & VQA-ODV \cite{li2018bridge} & 2018 & 60 & 600 & 7680$\times$3840-3840$\times$1920 & 10-23 & H.265 with different QPs; different projections & 221 & MOS, DMOS \\
        & LIVE-FBT-FCVR 2D \& 3D \cite{2021Subjective} & 2021 & 20 & 360 & 7680$\times$3840, 5376$\times$5376 & 10 & 3 foveated samples, 4 radii, 5 VP9 QPs & N/A & MOS \\

        \hline
        \multirow{6}{*}{\shortstack[l]{Frame rate,\\frame inter-\\polation}} & Waterloo-IVC-HFR \cite{nasiri2015perceptual} & 2015 & 7 & 336 & 1080p, 480p & 10 & Frame rate, QP, resolution & 25 & MOS \\

        & BVI-HFR \cite{mackin2018study} & 2018 & 22 & 88 & 1080p & 10s & Frame rate & 51 & MOS \\

        & LIVE-YouTube-HFR \cite{madhusudana2021subjective} & 2021 & 16 & 480 & UHD-1, HD & 6-8, 10 & Frame rate, VP9 & 85 & MOS \\

        & ETRI-LIVE STSVQ \cite{lee2021subjective} & 2021 & 15 & 437 & 3840$\times$2160 & 4.5-7 & Spatial subsampling, temporal subsampling, HEVC & 34 & DMOS \\


        & KosMo-1k \cite{men2020visual} & 2020 & 30 & 1350 & 480$\times$540 & 8 & Slow motion & N/A & MOS \\

        & BVI-VFI \cite{danier2022bvi} & 2022 & 108 & 540 & UHD-1, HD, 960$\times$540 & N/A & Frame repeating, averaging, interpolation & 189 & DMOS \\

        \hline
        \multirow{7}{*}{Audio-Visual} & Winkler \textit{et al.} \cite{winkler2006perceived} & 2006 & 6 & 48 & QCIF (176$\times$144)  & $\sim$8 & H.264, MPEG-4 & 24 & MOS \\
        & VQEGMM2 \cite{pinson2012influence,pinson2013subjective} & 2012 & 10 & 60 & 640$\times$480 & 10 & H.264 (AVC), AAC & 10 & MOS \\ 
        & Demirbilek \textit{et al.} \cite{demirbilek2016towards} & 2016 & 6 & 144 & 1080p, 720p & N/A & Resolution, bit rate, bandwidth, \textit{etc.} & 24 & MOS \\ 
        & Martinez \textit{et al.} \cite{martinez2014full,martinez2018combining} & 2014 & 6 & 132 & 720p & N/A & H.264, MPEG-1 layer-3 & 17 & MOS \\ 
        & LIVE-SJTU A/V-QA \cite{min2020study} & 2020 & 14 & 336 & 1080p & 8 & HEVC, scaling, AAC & 35 & MOS \\ 
        & Fela \textit{et al.} \cite{fela2022perceptual} & 2022 & 12 & 576 & 6144$\times$3072 & 20 & H.265/HEVC, AAC-LC, resolution & 20 & MOS \\ 
        & OAVQAD \cite{zhu2023perceptual} & 2023 & 15 & 375 & 7680$\times$3840 & 6 & HEVC, AAC, resolution, noise, blur, stalling & 22 & MOS \\ 

        \hline
        \multirow{7}{*}{\shortstack[l]{HDR/\\WCG/\\iTMO/\\TMO}} & DML-HDR \cite{banitalebi2014compression} & 2014 & 4 & 32 & 1080p & 10, 17 & HEVC, H.264 with different QPs  & 17 & MOS \\ 
        & Narwaria \textit{et al.} \cite{narwaria2015study} & 2015 & 9 & 153 & 1080p & N/A & TMO, iTMO, compression and decompression & 25 & MOS \\ 
        & Mukherjee \textit{et al.} \cite{mukherjee2016objective} & 2016 & 39 & 429 & 1080p & 5 & H.264 & 64 & Rank \\ 
        & Yeganeh \textit{et al.} \cite{yeganeh2016objective} & 2016 & 10 & 40 & N/A & N/A & TMO & 30 & MOS \\ 
        & DML-HDR 2 \cite{azimi2018evaluating} & 2018 & 5 & N/A & 2048$\times$1080 & 10 & AWGN, mean intensity shift, low Pass filter, \textit{etc.} & 18 & MOS \\ 
        & Waterloo UHD-HDR-WCG \cite{athar2019perceptual} & 2019 & 14 & 140 & 3840$\times$2160 & 10 & H.264, HEVC & 51 & MOS \\ 
        & LIVE-HDR \cite{shang2022subjective} & 2022 & 31 & 310 & 4k, 1080p, 720p, 540p & 1-10 & HEVC with different bitrates, resolution & 66 & DMOS \\ 

        \hline
        \multirow{7}{*}{\shortstack[l]{Screen/game}} & SCVD \cite{cheng2020screen} & 2020 & 16 & 800 & 1080p & 10 & GN, GB, MB, CC, CSC, CQD, H.264, \textit{etc.} & 32 & MOS \\ 
        & CSCVQ \cite{li2020subjective} & 2020 & 11 & 165 & 720p & 10 & H.264, HEVC, HEVC-SCC & 20 & MOS \\ 
        & GamingVideoSET \cite{barman2018gamingvideoset} & 2018 & 24 & 600 & 480p, 720p, 1080p & 30 & H.264 compression & 25 & MOS \\ 
        & KUGVD \cite{barman2019no} & 2019 & 6 & 150 & 480p, 720p, 1080p & 30 & H.264 compression & 17 & MOS \\ 
        & CGVDS \cite{zadtootaghaj2020quality} & 2020 & 15 & 255 & 480p, 720p, 1080p & 30 & H.264 compression & $>$100 & MOS \\ 
        & TGV \cite{wen2022subjective} & 2022 & 150 & 1293 & 480p, 720p, 1080p & 5 &  H264, H265, Tencent codec & 19 & N/A \\ 
        & LIVE-YOUTUBE GVQA \cite{yu2023subjective} & 2023 & 600 & 600 & 360p, 480p, 720p, 1080p & 8-9 & PGC, UGC & 61 & MOS \\

    \bottomrule
    \end{tabular}
    }
\end{table*}

\subsection{Subjective VQA Databases for Specific Applications}
With the advancement of multimedia video services, video categories have gradually enriched, therefore, many studies have studied VQA for specific applications.
In this section, we mainly discuss VQA databases for specific applications as demonstrated in Table \ref{tab:tab2.2}.

\subsubsection{Streaming VQA Databases}
Many databases have considered the temporal degradations of the videos during streaming services.
\begin{itemize}
    \item LIVE mobile video quality assessment database (LIVE Mobile)~\cite{moorthy2012video}.
    LIVE mobile consists of 10 reference videos and 200 distorted videos with 5 distortion types including H.264 compression, stalling, frame drop, rate adaptation, and wireless channel packet-loss.
    The resolution of the videos is 720p, and the duration is 15 seconds.
    A total of 47 subjects were included in the subjective experiment.

    \item LIVE time-varying subjective quality database (LIVE-TVSQ) \cite{chen2014modeling}.
    LIVE-TVSQ consists of 3 reference videos constructed by concatenating 8 high-quality high-definition video clips of different content, and 15 distorted videos corrupted by adjusting the encoding bitrate of H.264 video encoder with 5 bitrate-varying levels.
    The resolution is 720p.
    Each video is 5 minutes long and is viewed by 25 subjects.
    The subjective data format is Reversed DMOS (RDMOS).

    \item LIVE-Avvasi Mobile Video database (LIVE-AMV) \cite{ghadiyaram2014study}.
    LIVE-AMV consists of 24 reference videos and 180 distorted videos generated with 26 hand-crafted stalling events.
    17 videos have a resolution of 720p, and 7 videos have a resolution of 360p.
    The video lengths range between 29-134 seconds.
    The single stimulus continuous quality evaluation procedure was adopted, where the reference videos were also evaluated to obtain a DMOS for each distorted video sequence. 

    \item LIVE-Netflix Video QoE Database I (LIVE-NFLX-I) \cite{bampis2017study}.
    LIVE-NFLX-I consists of 112 distorted videos derived from 14 source content with 8 handcrafted playout patterns including dynamically changing H.264 compression rates, rebuffering events and mixtures of both.
    The resolution of the videos is 1080p.
    The video sequences were displayed on a small mobile screen at low bitrates, and were viewed for at least one minute by 55 subjects.
    MOS values were obtained for the videos.
    
    \item LIVE-Netflix Video QoE Database II (LIVE-NFLX-II) \cite{bampis2021towards}.
    LIVE-NFLX-II consists of 420 streaming videos derived from 15 source content with various streaming degradations including content-adaptive encoding profiles, bitrate adaptation algorithms, and various network conditions.
    The videos have a resolution of 1080p and diverse video lengths.
    A total of 65 subjects were included in the experiment and MOS values were collected.
    
    \item Waterloo Streaming QoE Database I (WaterlooSQoE-I) \cite{duanmu2016quality}.
    WaterlooSQoE-I contains 20 pristine videos and 180 distorted videos including 60 compressed videos, 60 initial buffering videos, and 60 mid-stalling videos.
    The resolution of the videos is 1080p, and the duration is 10 seconds.
    Each video was evaluated by 25 subjects, and MOS values were obtained.
    
    \item Waterloo Streaming QoE Database II (WaterlooSQoE-II) \cite{duanmu2017quality}.
    WaterlooSQoE-II contains 12 pristine videos and 588 distorted videos corrupted by various compression levels, spatial resolutions, and frame rates.
    The resolution is 1080p.
    The videos have diverse lengths.
    The subjective data format is MOS.
    
    \item Waterloo Streaming QoE Database III (WaterlooSQoE-III) \cite{duanmu2018quality}.
    WaterlooSQoE-III contains 20 source videos and 450 streaming videos corrupted by various encoding configurations, bandwidth shaping, and ABR algorithms.
    The videos have diverse resolutions and lengths.
    The subjective data format is MOS.

    \item Waterloo Streaming QoE Database IV (WaterlooSQoE-IV) \cite{WaterlooSQoEIV}.
    WaterlooSQoE-IV dataset contains 1350 highly realistic streaming videos generated from 5 pristine videos with the combinations of 2 video encoders, 9 real-world network traces, 5 ABR algorithms, and 3 viewing devices. 
    The 5 ABR algorithms include RB, BB, FastMPC, Pensieve, and RDOS.
    
\end{itemize}

\subsubsection{3D VQA Databases}
Traditional videos are plane videos without stereoscopic depth cues.
With the advancement of display techniques, many 3D videos have emerged, and many 3D VQA databases have been established.
\begin{itemize}
    \item LIVE 3D Video Database (LIVE 3D) \cite{chen2012study}.
    LIVE 3D contains 6 pristine videos and 54 distorted videos corrupted by 9 different quantization parameter (QP) levels.
    The resolution of the videos is 480p.
    The length of two source videos is 15 seconds, while the length of the remaining four source videos is 10 seconds.
    A total of 27 subjects were recruited and divided into two groups.
    In group A, 13 subjects were asked to evaluate the spatial video quality (SVQ), depth quality (DQ), and visual comfort (VC) 
    thirteen subjects, while in group B, 14 subjects were asked to give their ratings of the overall 3D video quality (3DVQ).

    \item StSD 3D Video Database (StSD 3D) \cite{de2013toward}.
    StSD 3D contains 14 pristine videos and 116 distorted videos corrupted by H.264 and HEVC compressions.
    The resolution of the videos is 1080p.
    The duration is 8 seconds.
    A total of 16 subjects were included in the experiment and DMOS values were collected.

    \item Tampere 3D Video Database (Tampere 3D) \cite{jumisko2011subjective}.
    Tampere 3D contains 4 pristine videos and 60 distorted videos corrupted by the H.264 compression and various depth levels.
    All videos have a resolution of 1080p and a duration of 10 seconds.
    A total of 30 subjects were included in the experiment.

    \item MMSPG 3D Video Quality Assessment Database (MMSPG 3D) \cite{goldmann2010comprehensive}.
    MMSPG 3D contains 6 pristine scenes, and 5 different stimuli were generated for each scene with different camera distances including 10, 20, 30, 40, 50 cm.
    All videos have a resolution of 1080p and a duration of 10 seconds.
    MOS values were calculated by subjective quality ratings collected from 17 qualified subjects.

    \item NAMA3DS1-COSPAD1 \cite{urvoy2012nama3ds1}.
    NAMA3DS1-COSPAD1 contains 10 pristine scenes and 110 distorted videos corrupted by H.264 and JPEG 2000 compressions.
    The resolution of the videos is 1080p.
    The video lengths are 13 seconds or 16 seconds. 
    A total of 29 subjects were included in the experiment and MOS values were collected.

    \item UBC Digital Multimedia Lab 3D Video Database (UBC DML 3D) \cite{banitalebi2014effect}.
    UBC DML 3D contains 5 pristine videos and 64 distorted videos corrupted by HEVC compression and different frame rates.
    All videos have a resolution of 1080p and a duration of 10 seconds.
    A total of 16 subjects were included in the experiment and MOS values were collected.

    \item 3DVCL$@$FER Video Database (3DVCL$@$FER) \cite{banitalebi2015effect,dumic20173d}.
    3DVCL$@$FER contains 6 pristine videos and 184 distorted videos corrupted by H.264 compression, JPEG2000 compression, Geometric distortion, packet losses, different frame rates and frame freeze.
    All videos have a resolution of 1080p and a duration of 10 seconds.
    A total of 35 subjects were included in the experiment and MOS values were collected.

    \item WATERLOO-IVC 3D Video Quality Database (WATERLOO-IVC 3D) \cite{wang2017asymmetrically}.
    WATERLOO-IVC 3D contains two sub-databases.
    Waterloo-IVC 3D Video Database Phase I contains 4 pristine multi-view 3D videos and 176 distorted videos corrupted by symmetric and asymmetric transform-domain quantization coding followed by different levels of low-pass filtering.
    Waterloo-IVC 3D Video Database Phase II includes 6 pristine 3D videos and various distorted stereoscopic 3D videos obtained from mixed-resolution coding, asymmetric transform-domain quantization coding, their combinations, and different levels of low-pass filtering.
    The videos in database Phase I have a resolution of 1024$\times$768, and the videos in database Phase II have a resolution of 1080p.
    22 subjects were recruited in the Phase I experiment, while 32 subjects were recruited in the Phase II experiment.
    MOS values were obtained in the experiments.
    
\end{itemize}

\subsubsection{VR VQA Databases}

Virtual Reality (VR) allows users to perceive 360$^\circ$ digital content immersively via head-mounted displays (HMDs), which is a gradually popular display media.
Omnidirectional videos are important digital contents in VR, thus many omnidirectional VQA databases have also been established \cite{duan2024quick}.

\begin{itemize}
    \item Immersive Video Quality Assessment Database 2017 (IVQAD2017) \cite{duan2017ivqad}.
    IVQAD2017 is a large-scale immersive video quality assessment database, which contains 10 pristine videos and 160 distorted videos corrupted by MPEG-4 compression, different resolutions, and different frame rates.
    All videos in IVQAD2017 have a resolution of 4096$\times$2048 and a duration of 15 seconds.
    The VR device used in the subjective experiment was HTC VIVE.
    A total of 13 subjects participated in the experiment and MOS values were obtained.

    \item Zhang \textit{et al.} \cite{zhang2018subjective}.
    Zhang \textit{et al.} \cite{zhang2018subjective} established a VR VQA database, which contains 10 pristine videos and 60 distorted videos corrupted by H.265 compression with different QPs.
    All videos in the database have a resolution of 3600$\times$1800 and a duration of 10 seconds.
    The VR device used in the subjective experiment was HTC VIVE.
    A total of 30 subjects participated in the experiment, and DMOS values and head movement data were obtained.
    
    \item Zhang \textit{et al.} \cite{zhang2017subjective}.
    Zhang \textit{et al.} \cite{zhang2017subjective} established a VR VQA database, which contains 16 pristine videos and 400 distorted videos corrupted by VP9, H.264, H.265 compressions with different bitrates, and different levels of Gaussian noise and box blur.
    All videos in the database have a resolution of 4096$\times$2048.
    The VR device used in the subjective experiment was HTC VIVE.
    A total of 23 subjects participated in the experiment, and MOS as well as DMOS values were obtained.
    
    \item Lopes \textit{et al.} \cite{lopes2018subjective}.
    Lopes \textit{et al.} \cite{lopes2018subjective} established a VR VQA database, which contains 6 pristine videos and 85 distorted videos corrupted by H.265 compression with different QPs, different resolutions, and different frame rates.
    All videos in the database have a resolution of 8192$\times$4096 and a duration of 10 seconds.
    The VR device used in the subjective experiment was Oculus Rift.
    A total of 37 subjects participated in the experiment, and MOS as well as DMOS values were obtained.
    
    \item Singla \textit{et al.} \cite{singla2017comparison}.
    Singla \textit{et al.} \cite{singla2017comparison} established a VR VQA database, which contains 6 pristine videos and 66 distorted videos corrupted by H.265 compression with different bitrates.
    All videos in the database have the resolutions of 4096$\times$2048 and 2048$\times$1024, and a duration of 10 seconds.
    The VR device used in the subjective experiment was Oculus Rift CV1.
    A total of 30 subjects participated in the experiment, and MOS values and head movement data were obtained.
    
    \item Virtual Reality Video Quality Assessment Database 48 (VR-VQA48) \cite{xu2018assessing}.
    VR-VQA48 contains 12 pristine videos and 48 distorted videos corrupted by H.265 compression with different QPs.
    All videos in the database have a resolution of 4096$\times$2048, and a duration of 12 seconds.
    The VR device used in the subjective experiment was HTC Vive.
    A total of 48 subjects participated in the experiment, and MOS values, DMOS values and head movement data were obtained.
    
    \item Tran \textit{et al.} \cite{tran2018study}.
    Tran \textit{et al.} \cite{tran2018study} established a VR VQA database, which contains 6 pristine videos and 126 distorted videos corrupted by H.265 with different QPs and different resolutions.
    All videos in the database have a duration of 30 seconds.
    The VR devices used in the subjective experiment were Samsung Gear VR and Samsung Galaxy S6.
    A total of 37 subjects participated in the experiment, and MOS values were obtained.
    
    \item Video Quality Assessment - Omnidirectional Videos (VQA-ODV) \cite{li2018bridge}.
    VQA-ODV contains 60 pristine videos and 600 distorted videos corrupted by H.265 compression with different QPs and different projections.
    All videos in the database have the resolutions of 7680$\times$3840 and 3840$\times$1920.
    The video lengths range from 10 seconds to 23 seconds.
    The VR device used in the subjective experiment was HTC Vive.
    A total of 221 subjects participated in the experiment, and MOS values, DMOS values, head movement data and eye movement data were obtained.

    \item LIVE-FBT-FCVR 2D \& 3D \cite{2021Subjective}. It contains 10 pristine 2D omnidirectional videos, 10 pristine 3D omnidirectional videos, and generated 180 distorted 2D omnidirectional videos as well as 180 distorted 3D omnidirectional videos.
    The resolution of 2D omnidirectional videos is 7680$\times$ 3840, and the resolution of 3D omnidirectional videos is 5376$\times$ 5376.
    The duration is 10 seconds.
    The used display device is HTC VIVE.
\end{itemize}

\subsubsection{High Frame Rate $\&$ Frame interpolation VQA Databases}

Users are pursuing higher frame-rate videos.
With the improvement of video communication technologies, high frame rate (HFR) videos can be displayed at 50 fps or more, rather than traditional videos which are typically displayed at 30 fps or 24 fps.
Therefore, some HFR VQA databases have also been constructed.

\begin{itemize}
    \item Waterloo-IVC High Frame Rate Video Quality Database (Waterloo-IVC-HFR) \cite{nasiri2015perceptual}.
    Waterloo-IVC-HFR contains 7 pristine 60fps source videos and their generated 336 test video sequences corrupted by the combination of 6 frame rate levels, 4 QP levels, and 2 resolution levels.
    The videos in the database have two different resolutions including 1080p and 480p.
    The video length is 10 seconds.
    A total of 25 subjects participated in the experiment, and MOS values were obtained.
    
    \item Bristol Vision Institute High Frame Rate Video Database (BVI-HFR) \cite{mackin2018study}.
    BVI-HFR contains 22 120 fps source sequences and 88 distorted videos with 4 different frame rates varying from 15 fps to 120 fps obtained by subsampling the source videos via frame averaging.
    All videos in the database have a resolution of 1080p, and a duration of 10 seconds.
    A total of 51 subjects participated in the experiment, and MOS values were obtained.

    \item LIVE-YouTube-HFR Database (LIVE-YouTube-HFR) \cite{madhusudana2021subjective}.
    LIVE-YouTube-HFR contains 16 source videos and 480 test sequences with 6 levels of frame rate and 5 levels (lossless+4 CRF) of VP9 compression.
    11 sequences were borrowed from the BVI-HFR video database \cite{mackin2018study}, which have a resolution of 1920$\times$1080 (HD), and a duration of 10 seconds.
    5 other sequences were high-motion sports content captured by the Fox Media Group, which have a resolution of 3840$\times$2160 (UHD-1), and video lengths of 6-8 seconds.
    A total of 85 subjects participated in the experiment, and MOS values were obtained.

    \item ETRI-LIVE STSVQ \cite{lee2021subjective}.
    ETRI-LIVE STSVQ contains 15 high-quality 4K 10-bit source contents, which have a resolution of 3840$\times$2160, and a chroma format of YUV420p.
    The video lengths range from 4.5 seconds to 7 seconds.
    437 distorted videos were generated by spatial subsampling, temporal subsampling, and HEVC compression with different QPs.
    A total of 34 subjects participated in the experiment, and DMOS values were obtained.


    \item KosMo-1k \cite{men2020visual}.
    KosMo-1k contains 30 source videos and 1350 distorted videos corrupted by slow motion.
    All videos in KosMo-1k have a resolution of 480$\times$540, and a duration of 8 seconds.
    MOS values were calculated as the subjective data.
    
    \item BVI-VFI \cite{danier2022bvi}.
    BVI-VFI contains 108 source videos and 540 distorted videos corrupted by dropping every second frame, then reconstructing the dropped frames using five VFI algorithms: frame repeating, frame averaging (where the middle frame is generated by averaging every two frames), DVF \cite{liu2017video}, QVI \cite{xu2019quadratic} and ST-MFNet \cite{danier2022st}.
    The resolutions of the videos include UHD-1, HD, and 960$\times$540.
    A total of 189 subjects participated in the experiment, and DMOS values were obtained.
    
\end{itemize}

\subsubsection{Audio-Visual VQA Databases}

Videos are generally accompanied by audios, and the degradation of audio can also affect the overall QoE.
Thus, some works have also explored the audio-visual VQA.

\begin{itemize}
    \item Winkler \textit{et al.} \cite{winkler2006perceived}.
    Winkler \textit{et al.} \cite{winkler2006perceived} established an audio-visual VQA database, which contains 6 pristine videos and 8 distorted videos.
    The video track was corrupted by H.264 compression with different bitrates, and the audio track was corrupted by MPEG-4 AAC-LC with different sampling rates and bitrates.
    All videos in the database have a resolution of 176$\times$144 (QCIF), and a duration of about 8 seconds.
    A total of 24 subjects participated in the experiment, and MOS values were obtained.

    \item Video Quality Experts Group Multimedia Phase II (VQEGMM2) \cite{pinson2012influence,pinson2013subjective}.
    Pinson \textit{et al.} \cite{pinson2012influence} have established an audio-visual VQA database, which contains 10 pristine videos and 60 test videos (10 pristine + 50 distorted).
    The video track was corrupted by H.264 compression (advanced video coding (AVC)), and the audio track was corrupted by advanced audio coding (AAC).
    All videos in the database have a resolution of 640$\times$480, and a duration of 10 seconds.
    A total of 10 subjects participated in the experiment, and MOS values were obtained.

    \item Demirbilek \textit{et al.} \cite{demirbilek2016towards}.
    Demirbilek \textit{et al.} \cite{pinson2012influence} established an audio-visual VQA database, which contains 6 pristine videos and 144 test videos corrupted by different resolutions, bitrates, bandwidths, packet loss rates, jitter cases.
    The resolutions of the videos include 1080p and 720p.
    A total of 24 subjects participated in the experiment, and MOS values were obtained.

    \item Martinez \textit{et al.} \cite{martinez2014full,martinez2018combining}.
    Martinez \textit{et al.} \cite{martinez2014full} established an audio-visual VQA database, which contains 6 pristine videos and 132 test videos.
    All videos in the database have a resolution of 720p.
    The experiment includes three sessions.
    For session I, each of the original video test sequences (no audio) was compressed using the H.264 codec with four different bitrate values including 30, 2, 1, and 0.8 Mbps, resulting in 30 test conditions (6 pristine $\times$ 5 levels (4 bitrate levels + 1 pristine)), and a total of 16 subjects participated in this session.
    For session II, only the audio components of the videos were compressed using the MPEG-1 layer-3 coding standard with three different bitrate values including 128, 96, and 48 kbps, resulting in 24 test conditions (6 pristine $\times$ 4 levels (3 bitrate levels + 1 pristine)), and a total of 16 subjects participated in this session.
    For session III,  both audio and video components of the test sequences were compressed, where the video components were compressed using H.264 and the audio components were compressed using the MPEG-1 layer-3 coding standard, resulting in 78 test conditions (6 pristine $\times$ 13 levels (3 audio bitrates $\times$ 4 video bitrates + 1 pristine)), and a total of 16 subjects participated in this session.
    The MOS values were obtained.

    \item LIVE-SJTU A/V-QA \cite{min2020study}.
    LIVE-SJTU A/V-QA contains 14 pristine videos and 336 distorted videos corrupted by HEVC compression with 4 different constant rate factor (CRF) levels, video compression plus scaling, and AAC compression with 3 different constant bit rate (CBR) levels.
    The videos have a resolution of 1080p, a duration of 8 seconds, and are provided in the raw YUV 4:2:0 format.
    A total of 35 subjects participated in the experiment, and MOS values were obtained.

    \item Fela \textit{et al.} \cite{fela2022perceptual}.
    Fela \textit{et al.} \cite{fela2022perceptual} established an audio-visual VQA database for 360 videos, which contains 12 pristine videos and 576 test videos corrupted by 3 different resolutions, 4 QP levels, and four AAC-LC levels.
    The resolution of the pristine videos is 6144$\times$3072.
    The duration is 20 seconds.
    A total of 20 subjects participated in the experiment, and MOS values were obtained.
    
    \item Omnidirectional Audio-visual Quality Assessment Database (OAVQAD) \cite{zhu2023perceptual}.
    OAVQAD contains 15 pristine videos and 375 distorted videos corrupted by HEVC compression, AAC compression, different resolutions, noise, blur, and stalling.
    The resolution of the pristine videos is 7680$\times$3840.
    The duration is 6 seconds.
    A total of 22 subjects participated in the experiment, and MOS values were obtained.
    
\end{itemize}

\subsubsection{HDR, WCG, iTMO, and TMO VQA Databases}

With the increasing requirements for video experience, high dynamic range (HDR) and wide color gamut (WCG) video technologies have been gradually developed.
Many studies have investigated the VQA problem for HDR and WCG videos, as well as tone mapping operation and inverse tone mapping operation.

\begin{itemize}
    \item Digital Multimedia Lab HDR (DML-HDR) \cite{banitalebi2014compression}.
    DML-HDR contains 4 pristine videos and 32 distorted videos corrupted by HEVC, H.264 compressions with different QPs.
    The resolution of the pristine videos is 1080p.
    The video lengths include 10 and 17 seconds.
    A total of 17 subjects participated in the experiment, and MOS values were obtained.
    
    \item Narwaria \textit{et al.} \cite{narwaria2015study}.
    Narwaria \textit{et al.} \cite{narwaria2015study} established an HDR VQA database, which contains 9 pristine videos and 153 test videos corrupted by different TMO, iTMO, compression and decompression methods.
    All videos have a resolution of 1080p.
    A total of 25 subjects participated in the experiment, and MOS values were obtained.

    \item Mukherjee \textit{et al.} \cite{mukherjee2016objective}.
    Mukherjee \textit{et al.} \cite{mukherjee2016objective} established an HDR VQA database, which contains 39 pristine videos and 429 test videos corrupted by different H.264 compressions.
    All videos have a resolution of 1080p.
    A total of 64 subjects participated in the experiment, and rank values were obtained.

    \item Yeganeh \textit{et al.} \cite{yeganeh2016objective}.
    Yeganeh \textit{et al.} \cite{yeganeh2016objective} established an HDR VQA database, which contains 10 pristine videos and 40 test videos corrupted by different TMO methods.
    A total of 30 subjects participated in the experiment, and MOS values were obtained.

    \item Digital Multimedia Lab HDR 2 (DML-HDR 2) \cite{azimi2018evaluating}.
    DML-HDR 2 contains 5 pristine videos and various distorted videos corrupted by AWGN, mean intensity shift, salt and pepper noise, low Pass filter, and compression.
    All videos have a resolution of 2048$\times$1080 and a duration of 10 seconds.
    A total of 18 subjects participated in the experiment, and MOS values were obtained.

    \item Waterloo UHD-HDR-WCG \cite{athar2019perceptual}.
    Waterloo UHD-HDR-WCG contains 14 pristine videos and 140 distorted videos corrupted by H.264 compression and HEVC compression.
    All videos have a resolution of 3840$\times$2160 and a duration of 10 seconds.
    A total of 51 subjects participated in the experiment, and MOS values were obtained.

    \item LIVE-HDR \cite{shang2022subjective}.
    LIVE-HDR contains 310 test video sequences including 31 pristine videos and 279 distorted videos corrupted by different resolutions and the HEVC compression with different bitrates.
    The resolutions include 4k, 1080p, 720p, 540p.
    The video lengths range from 1 to 10 seconds.
    A total of 66 subjects participated in the experiment, and DMOS values were obtained.

\end{itemize}

\subsubsection{Screen and Game VQA Databases}

Screen graphics and cloud gaming are other popular video applications, and pursue high-quality video experience.
Therefore, many screen content or game content VQA databases have been constructed.

\begin{itemize}
    \item Screen Content Video Database (SCVD) \cite{cheng2020screen}.
    SCVD contains 16 pristine videos and 800 distorted videos corrupted by 10 different distortions including Gaussian noise (GN), Gaussian blur (GB), motion blur (MB), contrast change (CC), color saturation change (CSC), color quantization with dithering (CQD), H.264, high efficiency video coding (HEVC), screen content coding (SCC), and packet loss (PL).
    All videos have a resolution of 1080p and a duration of 10 seconds.
    A total of 32 subjects participated in the experiment, and MOS values were obtained.
    
    \item Compressed Screen Content Video Quality (CSCVQ) Database \cite{li2020subjective}.
    CSCVQ contains 11 pristine videos and 165 distorted videos corrupted by H.264 compression, HEVC compression, and HEVC Screen Content Coding (HEVC-SCC).
    All videos have a resolution of 720p and a duration of 10 seconds.
    A total of 20 subjects participated in the experiment, and MOS values were obtained.

    \item GamingVideoSET \cite{barman2018gamingvideoset}.
    GamingVideoSET contains 24 pristine videos and 600 distorted videos corrupted by H.264 compression.
    The resolutions of the videos include 480p, 720p, and 1080p.
    All videos have a duration of 30 seconds.
    A total of 25 subjects participated in the experiment, and MOS values were obtained.
    
    \item Kingston University Gaming Video Dataset (KUGVD) \cite{barman2019no}.
    KUGVD contains 6 pristine videos and 150 distorted videos corrupted by H.264 compression.
    The resolutions of the videos include 480p, 720p, and 1080p.
    All videos have a duration of 30 seconds.
    A total of 17 subjects participated in the experiment, and MOS values were obtained.
    
    \item CGVDS \cite{zadtootaghaj2020quality}.
    CGVDS contains 15 pristine videos and 255 distorted videos corrupted by H.264 compression.
    The resolutions of the videos include 480p, 720p, and 1080p.
    All videos have a duration of 30 seconds.
    Over 100 subjects participated in the experiment, and MOS values were obtained.
    
    \item Tencent Gaming Video (TGV) \cite{wen2022subjective}.
    TGV contains 150 pristine videos and 1293 distorted videos corrupted by H.264 compression, H.265 compression, and Tencent codec.
    The resolutions of the videos include 480p, 720p, and 1080p.
    All videos have a duration of 5 seconds.
    A total of 19 subjects participated in the experiment.

    \item LIVE-YOUTUBE Gaming Video Quality Database (LIVE-YOUTUBE GVQA) \cite{yu2023subjective}.
    LIVE-YOUTUBE GVQA contains 600 Professionally-Generated-Content (PGC) or User-Generated-Content (UGC) gaming videos.
    The resolutions of the videos include 360p, 480p, 720p, and 1080p.
    The videos were clipped into 8-9 seconds.
    A total of 61 subjects participated in the experiment, and MOS values were obtained.

\end{itemize}

\begin{figure}[t]
    \centering
    \includegraphics[width=0.55\textwidth]{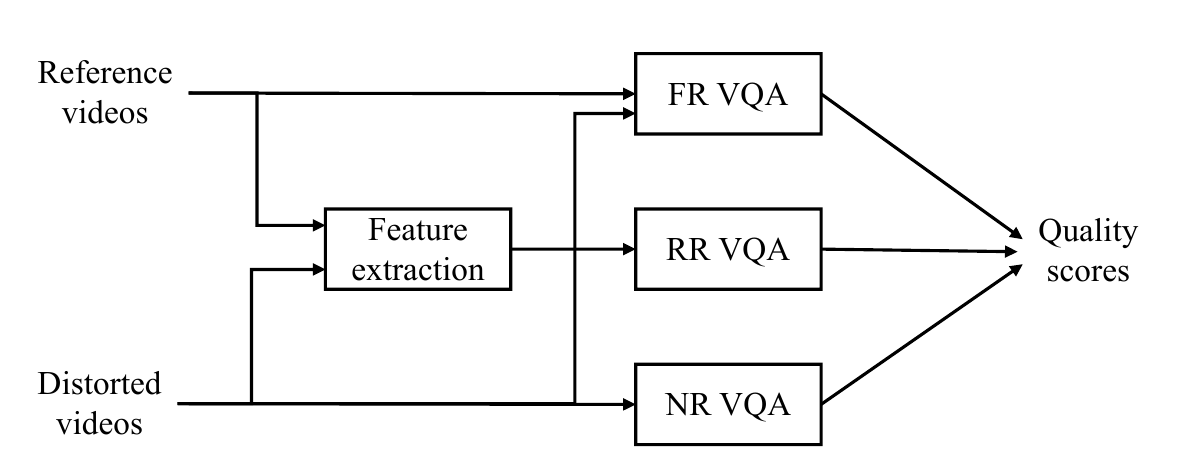}
    \caption{The categories of objective video quality assessment models.}
    \label{categories_VQA}
\end{figure}

\section{Objective Video Quality Assessment: General-purpose Models} \label{sec:objective_traditional}
In this section, we review the general-purpose objective VQA models that are designed to handle various video distortions. As illustrated in Figure \ref{categories_VQA}, depending on whether reference video information is required or not, we category the objective VQA models into three types, full-reference (FR) VQA, reduced-reference (RR) VQA, and no-reference (NR) VQA.
\subsection{Full-reference Video Quality Assessment}
\label{FR_VQA}
\begin{figure}[t]
    \centering
    \includegraphics[width=0.92\textwidth]{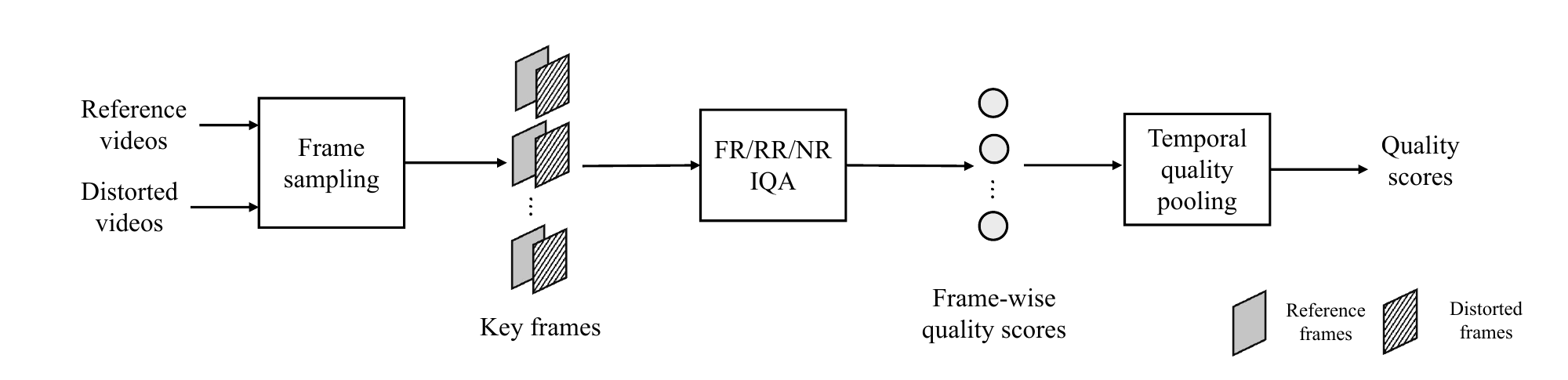}
    \caption{The framework of IQA-based objective VQA methods.}
    \label{IQA_based_VQA_methods}
\end{figure}

Full reference video quality assessment models aim to evaluate the quality of a video signal by comparing it to its reference (original or pristine) video, which are commonly employed in various domains such as video broadcasting, video streaming, video compression, video enhancement, and quality control in video production. Generally, FR VQA models measure the fidelity between distorted and reference videos. As shown in Figure \ref{IQA_based_VQA_methods}, one prevalent approach involves applying FR image quality assessment methods to individual or sampled video frames and subsequently aggregating the frame-level quality scores into the video-level quality score. Well-known FR IQA methods include PSNR, SSIM \cite{wang2004image}, MS-SSIM \cite{wang2003multiscale}, VIF \cite{sheikh2006image}, LPIPS \cite{zhang2018unreasonable}, \textit{etc.}, and a comprehensive survey on FR IQA methods can refer to \cite{zhai2020perceptual}. However, video quality is intricately related to the temporal distortions like jitter, flicker, etc., which are not effectively captured by these IQA-based methods. Therefore, to address temporal distortions in videos and achieve a better evaluation ability, lots of FR VQA models have been proposed in literature. These models can be roughly classified them into knowledge-driven and data-driven methods based on their types of feature extraction. For knowledge-driven FR VQA methods, quality-aware features are extracted based on the characteristics of the human visual system, whereas data-driven FR VQA methods employ the machine learning techniques to directly acquire quality-aware features from video data.

%
%
%
%
%

\subsubsection{Knowledge-driven FR VQA}
\textbf{(1) SSIM-based FR VQA:} Structure Similarity (SSIM) \cite{wang2004image} has been the most popular FR IQA methods over the last two decades. It calculates luminance, contrast and structure similarities between distorted and reference images. Due to its simplicity and effectiveness, numerous efforts have been made to extend SSIM to the video domain. 
Wang \textit{et al.} \cite{wang2004video} investigated two pooling strategies to apply SSIM to video quality assessment. Specifically, they calculate frame-level SSIM values and subsequently aggregate them into the video-level score based on the luminance intensity and motion degree of the frames. 
Wang and Li \cite{wang2007video} introduced a motion-based pooling strategy for well-known FR IQA methods (\textit{e.g.} PSNR and SSIM). They incorporate a human visual speed perception model into an information framework and estimate motion information and perceptual uncertainty as the weighting factors. 
Moorthy and Bovik proposed a motion-compensated SSIM (MC-SSIM) \cite{moorthy2010efficient,moorthy2009motion} to assess both spatial and temporal quality scores, where the spatial quality scores are calculated using SSIM and the temporal quality scores are evaluated by evaluating structural retention between motion-compensated regions. 
Seshadrinathan \textit{et al.} \cite{seshadrinathan2011temporal} observed a hysteresis effect in human user study of subjective VQA and propose a hysteresis temporal pooling strategy to aggregate frame-level quality scores into the video-level quality score. 
Park \textit{et al.} \cite{park2012video} introduced a content-adaptive spatial and temporal pooling strategy named Video Quality Pooling (VQPooling), which emphasizes the influence of the ``worst" quality scores along both the spatial and temporal dimensions of a video sequence on the overall video quality. 
Manasa and Channappayya \cite{manasa2016optical} employed MS-SSIM \cite{wang2003multiscale} to characterize spatial quality estimation and utilize local flow statistics defined by the mean, the standard deviation, the coefficient of variation, and the minimum eigenvalue of the local flow patches to represent temporal distortions. 
Instead of calculating the quality scores frame-by-frame and then merging them into the video-level quality scores, Zeng \textit{et al.} \cite{zeng20123d} treated the video as the 3D volume data and directly calculated SSIM values of 3D volume data. Different with 2D SSIM, they utilized local information content and local distortion based weighting methods to pool the quality map into the quality score. 

\begin{figure}[t]
    \centering
    \includegraphics[width=0.65\textwidth]{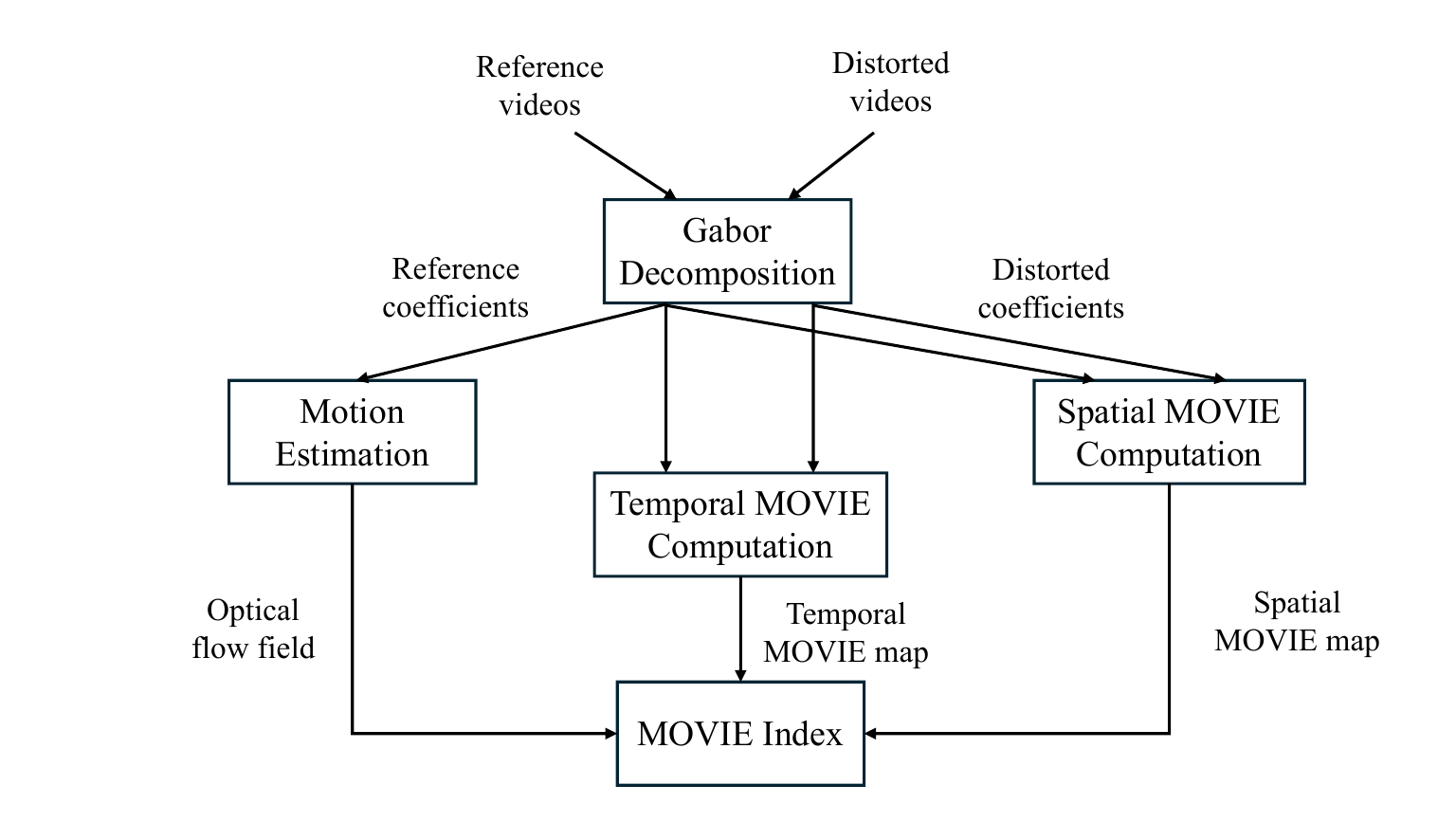}
    \caption{The framework of MOVIE~\cite{seshadrinathan2009motion}.}
    \label{movie_framework}
\end{figure}

\textbf{(2) Low-level feature-based FR VQA:} Some FR VQA models attempt to leverage abundant low-level features like optical flow, gradient, etc. to represent video quality. 
For example, Seshadrinathan and Bovik introduce a motion-based video integrity evaluation (MOVIE) index \cite{seshadrinathan2009motion,seshadrinathan2009motionconference}, which uses Gabor filters to decompose the video and calculate corresponding spatial, temporal, and motion features. In particular, motion estimation is computed in the optical flow field. The framework of MOVIE is illustrated in Figure~\ref{movie_framework}.
To handle local flicker distortions, Choi \textit{et al.} \cite{choi2018video,choi2016flicker} developed flicker sensitive MOVIE (FS-MOVIE) by integrating a perceptual flicker masking index into MOVIE Index, where the flicker masking mechanism is derived from the responses of neurons in primary visual cortex to video flicker. 
Wang \textit{et al.} \cite{wang2012novel} proposed a VQA model by leveraging structural features in localized spacetime regions to jointly represent spatial edge features and temporal motion characteristics, thus having a relatively low computational complexity. 
Vu \textit{et al.} \cite{vu2011spatiotemporal} developed spatial-temporal most apparent distortion (ST-MAD) by applying MAD \cite{larson2010most} to each frame to obtain spatial MAD and utilizing an optical-flow-derived weighting scheme to emphasize the appearance component of spatial MAD in fast-moving regions to derive temporal MAD. 
Yan and Mou \cite{yan2018video} partitioned the spatiotemporal slice images into regions with simple motion and complex motion. They then utilized gradient magnitude standard deviation (GMSD) \cite{xue2013gradient} index to evaluate distortions within these distinct segments. 

\textbf{(3) HVS-based FR VQA:} The human visual system plays a crucial role in guiding the design of FR VQA models. Zhang and Bull \cite{zhang2015perception} introduce a perception-based FR VQA model, which utilizes an enhanced nonlinear model to combine noticeable distortion and blurring artifacts, simulating simulate the HVS perception process. 
The visual attention mechanism \cite{duan2018learning,fang2020identifying} reflects how human allocate their attention to regions of the video, and several works utilize visual attention or saliency mechanism to develop FR VQA models. 
Since the human visual system is sensitive to motion objects, Wu \textit{et al.} \cite{wu2019quality} propose a full-reference assessor along salient trajectories (FAST) model, which computes the motion object trajectories in the optical flow domain, employ the motion velocity to represent temporal quality, and apply the 3D filters to motion content to represent sptio-temporal quality. Additionally, spatial quality is represented by calculating GMSD \cite{xue2013gradient} for each frame. Finally, they combine three three quality metrics to obtain an overall video quality score.
For example, You \textit{et al.} \cite{you2013attention} proposed an attention-driven foveated FR VQA models by integrating the attention-driven contrast sensitivity function into a wavelet-based distortion visibility measure. 
Peng \textit{et al.} \cite{peng2017efficient} developed an attention-guided and motion-tuned temporal distortion metric based on \textit{spacetime texture}, which serves as a uniform and distributive descriptor of a wide set of spacetime structures. 
Zhang and Liu \cite{zhang2017study} conducted a video saliency experiment to gather reliable eye-tracking data for distorted videos and integrate the eye-tracking data into FR VQA models to improve their performance.

\textbf{(4) Features fusion based FR VQA:} Recently, some studies attempt to extract various types of features and subsequently employ a learning-based regressor to map these features to video quality scores, thereby capitalizing on the strengths of different extracted feature types. 
Freitas \textit{et al.} \cite{freitas2018using} extracted a set of features including multiscale salient local binary patterns, MS-SSIM, GMSD, Riesz pyramids similarity deviation, spatial activity and temporal distortion measures and then employ a random forest regression algorithm to derive the video quality score. 
Video Multi-method Assessment Fusion (VMAF) \cite{li2016toward} extracts two kinds of FR IQA features including Visual Information Fidelity (VIF) \cite{sheikh2006image} and Detail Loss Metric (DLM) \cite{li2011image} along with motion features quantified by temporal difference between consecutive frames. Then it learns a Support Vector Regressor (SVR) to map these features into the video quality score.
Bampis \textit{et al.} \cite{bampis2018spatiotemporal,bampis2018simple} further made enhancements to VMAF from two aspects, known as spatiotemporal VMAF (ST-VMAF) and ensemble VMAF (E-VMAF), by incorporating space–time features at multiple scales.
In order to reduce the computational complexity of VMAF, Venkataramanan \textit{et al.} \cite{venkataramanan2022funque} proposed a VQA model named fusion of unified quality evaluators (FUNQUE), which calculates the features in VMAF including VIF, DLM, motion features, and SSIM on a common transform domain that accounts for the human visual system. 
Liu \textit{et al.} \cite{liu2021video} introduced a serial dependence modeling framework for FR VQA, which first extracts static appearance features and two kinds of motion information features (represented by an explicit content-based 3D structure and an implicit feature-based 2D structure) and subsequently utilizes the LSTM and attention-based quality pooling strategy to obtain the video quality score.

\begin{figure}[t]
    \centering
    \includegraphics[width=0.76\textwidth]{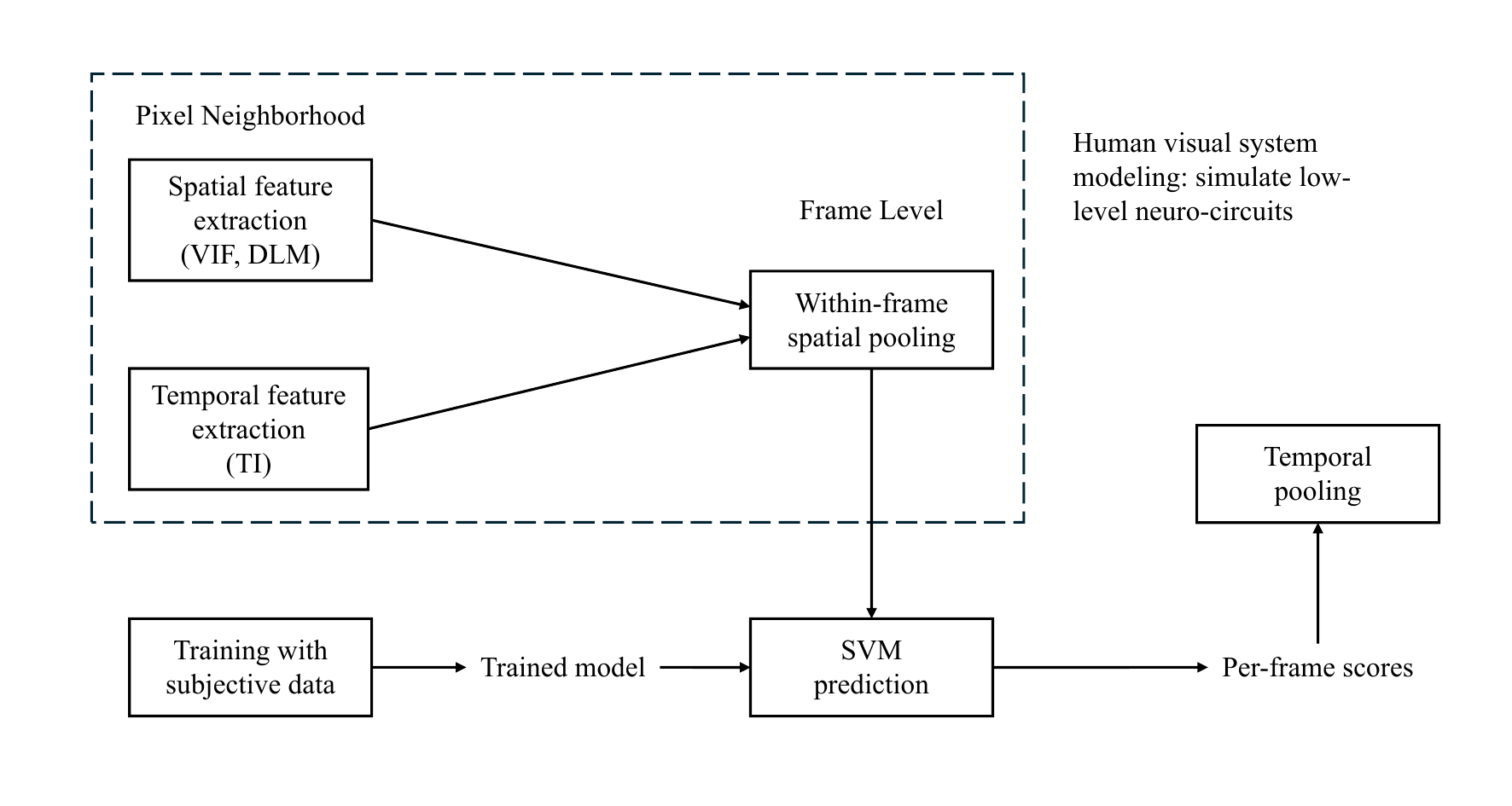}
    \caption{The framework of VMAF~\cite{li2016toward}. Image credit: NETFLX$^\circledR$.}
    \label{vmaf_framework}
\end{figure}

\subsubsection{Data-driven FR VQA}
Data-driven FR VQA models rely on large-scale video datasets to automatically learn quality-aware features for video quality evaluation. 
In recent years, with popularity of deep neural network, convolutional neural network (CNN) and Vision Transformer (ViT) have become the two dominant approaches for data-driven FR VQA models. 
For instance, Kim \textit{et al.} \cite{kim2018deep} proposed a deep video quality assessor named DeepVQA, which employs CNNs to generate spatio-temporal sensitivity maps for tackling temporal motion artifacts and introduces a convolutional neural aggregation network to capture temporal memory effects for quality judgment. 
Xu \textit{et al.} \cite{xu2020c3dvqa} introduced C3DVQA, which leverages CNNs with the 3D kernels for FR VQA. In particular, C3DVQA utilizes 2D CNNs to extract spatial features from both distorted frames and residual frames (\textit{i.e.} the difference between distorted and reference frames), and it utilizes 3D CNN to learn spatio-temporal features from extracted spatial features for video quality evaluation. 
Zhang \textit{et al.} \cite{zhang2019objective} presented a transfer learning framework for FR VQA to address challenges posed by imbalanced and limited samples of VQA datasets. They utilized distorted images as the related domain to enrich the distorted samples and train a six-layer CNN to extract high-level spatiotemporal features from distorted image blocks and video blocks annotated by classic FR IQA metrics.  
Zhang \textit{et al.} \cite{zhang2021video} proposed a FR VQA model that integrates DenseNet with the spatial pyramid pooling strategy and RankNet, where the former is used to extract high-level distortion representations and the latter acts as a temporal pooling method to characterize the high-level relevance among frames. 
Wu \textit{et al.} \cite{wo2022video} developed a quality aggregation network for FR VQA. It employs a 3D CNN to extract spatiotemporal features and utilizes a LSTM-based temporal quality pooling network to capture the nonlinearities and temporal dependencies inherent in the video quality evaluation process. 

\begin{table}[t]
	\footnotesize
	\centering
	\renewcommand{\arraystretch}{1.25}
	\caption{Overview of the FR and RR Video Quality Assessment Models.}
	\label{performance_synthetic_distortion_SRCC}
	\resizebox{1\textwidth}{!}{
	\begin{tabular}{ccccc}
		
		\toprule[.15em]
		Type & Algorithm & Methodology & Extracted quality features & Quality fusion   \\
		\hline
\multirow{29}{*}{FR} &  Wang \textit{et al.} \cite{wang2004video}	&Structural similarity&	Structure similarity, motion vector	&Weighted sum\\
&Wang and Li \cite{wang2007video}&	Structural similarity&	Structure similarity, motion vector&	Weighted sum\\
&MC-SSIM \cite{moorthy2010efficient,moorthy2009motion}&	Structural similarity&	Structure similarity, motion vector	&Weighted sum\\
&Seshadrinathan \textit{et al.} \cite{seshadrinathan2011temporal}&	Structural similarity	&Structure similarity or MOVIE	&Hysteresis temporal pooling\\
&Park \textit{et al.} \cite{park2012video}	&Structural similarity	&Structure similarity or MOVIE&	VQPooling\\
&Manasa and Channappayya \cite{manasa2016optical}&	Structural similarity	&Multi-scale Structure similarity, optical flow	&Weighted sum\\
&Zeng \textit{et al.} \cite{zeng20123d}&	Structural similarity&	3D structural similarity&	Weighted sum\\
&MOVIE \cite{seshadrinathan2009motion,seshadrinathan2009motionconference}	&Low feature extraction	&Gabor filter, optical flow	&Weighted sum\\
&FS-MOVIE \cite{choi2018video,choi2016flicker}	&Low feature extraction	&Gabor filter, optical flow	&Weighted sum\\
&Wang \textit{et al.} \cite{wang2012novel}	&Low feature extraction	&Sobel gradient features, eigenvalue of 3D structure tensor&	Averaged sum\\
&ST-MAD \cite{vu2011spatiotemporal}&	Low feature extraction&	MAD, optical flow	&Weighted sum\\
&Yan and Mou \cite{yan2018video}	&Low feature extraction	&GMAD, motion features	&Weighted sum\\
&Zhang and Bull \cite{zhang2015perception}	&HVS perception behaviour&	DT-CWT features, motion vector&	Weighted sum\\
&Wu \textit{et al.} \cite{wu2019quality}	&HVS perception behaviour&	GMSD, saliency trajectory features, optical flow	&Weighted sum\\
& You \textit{et al.} \cite{you2013attention}&	HVS perception behaviour	&Visual saliency, CSF, DWT&	Weighted sum\\
&Peng \textit{et al.} \cite{peng2017efficient}&	HVS perception behaviour	&G3 filer, spacetime texture, visual attention&	Weighted sum\\
&Zhang and Liu \cite{zhang2017study}	&HVS perception behaviour&	Visual saliency 	&Weighted sum\\
&Freitas \textit{et al.} \cite{freitas2018using}	&Features fusion&	MSLBP, MSSIM, GMSD, RPSD, SA, TDM	&Weighted sum\\
&VMAF \cite{li2016toward}	&Features fusion&	VIF, DLM, TI	&SVR\\
&ST-VMAF \cite{bampis2018spatiotemporal,bampis2018simple}	&Features fusion&	VIF, DLM, T-SpEED	&SVR\\
&FUNQUE \cite{venkataramanan2022funque}	&Features fusion	&DLM, SSIM,VIF, TI	&SVR\\
&Liu \textit{et al.} \cite{liu2021video}&	Features fusion	&3D Prewitt operators, optical flow, &2D-CNN	LSTM\\
&DeepVQA \cite{kim2018deep}	&Deep learning	&2D-CNN&	CNAN\\
&C3DVQA \cite{xu2020c3dvqa}&	Deep learning	&2D-CNN, 3D-CNN	&MLP\\
&Zhang \textit{et al.} \cite{zhang2019objective}	&Deep learning&	2D-CNN, 3D-CNN	&MLP\\
&Zhang \textit{et al.} \cite{zhang2021video}	&Deep learning&	DenseNet with SPP	&RankNet\\
&Wu \textit{et al.} \cite{wo2022video}&	Deep learning&	3D-CNN&	LSTM\\
&Li \textit{et al.} \cite{li2021full}	&Deep learning	&2D-CNN	&Transformer encoder\\
&Sun \textit{et al.} \cite{sun2021deep}	&Deep learning	&2D-CNN	&MLP\\
&Li \textit{et al.} \cite{li2021user}	&Deep learning	&2D-CNN	&MLP\\
\hline
		
\multirow{10}{*}{RR}&VQM \cite{pinson2004new}&	Low feature extraction	&SI, TI, edge features, chroma features	&Weighted sum\\
&Masry \textit{et al.} \cite{masry2006scalable}	&Low feature extraction	&Color, DWT, contrast, visual masking	&Weighted sum\\
&Callet \textit{et al.} \cite{le2006convolutional}	&Low feature extraction	&GHV, GHVP, TI, blockness, 	&CNN, MLP\\
&Gunawan and Ghanbari \cite{gunawan2008reduced}	&Low feature extraction	&Edge, blockness, motion vector	&Weighted sum\\
&Zeng and Wang \cite{zeng2010temporal}	&Low feature extraction	&Complex wavelet transform, circular variance	&Weighted sum\\
&Ma \textit{et al.} \cite{ma2012reduced}	&NSS	&Energy variation descriptor, GGD, City-block distance	&Averaged sum\\
&Zhu \textit{et al.} \cite{zhu2014optimizing}	&Low feature extraction	&Energy, entropy, kurtosis, Jensen–Shannon divergence, SSIM, smoothness	&Weighted sum\\
&STRRED \cite{soundararajan2012video}	&NSS	&GSM, entropies, wavelet coefficients	&Weighted sum\\
&SpEED-QA \cite{bampis2017speed}	&NSS	&GSM, entropies	&Weighted sum\\
&Wang \textit{et al.} \cite{wang2015very}	&Structural similarity	&Structure similarity, CSF	&Weighted sum\\
		
		\bottomrule[.15em]
		
	\end{tabular}
	}
\end{table}
With the popularity of user generated content videos in recent years, the focus of FR VQA research has shifted to UGC videos. For example, Li \textit{et al.} \cite{li2021full} utilized a Siamese CNN to extract features of distorted and reference videos and subsequently employ a Transformer encoder to map these features into the video quality score. 
Sun \textit{et al.} \cite{sun2021deep} extracted the structure and texture similarities of feature maps extracted from all intermediate layers of a CNN model for the quality-aware feature representation and then used a fully connected layer to map the quality-aware features into the video quality score. 
Li \textit{et al.} \cite{li2021user} first used a learned neural network to estimate the quality maps of the pristine and distorted UGC videos. They then assessed the quality of distorted UGC videos based on the estimated quality maps, considering the influence of the pristine and the distorted video quality on the overall quality assessment.

\subsection{Reduced-reference Video Quality Assessment}
Reduced-reference VQA is a special type of FR VQA, which necessitates only partial reference video information for evaluating the quality of distorted videos. 
So, it provides the potential to significantly save transmission bandwidths in the situations of assessing the quality of transmitted videos compared with FR VQA models. 
The National Telecommunications and Information Administration (NTIA) General Model \cite{pinson2004new}, also known as Video Quality Model (VQM), is a RR VQA method that initially calibrates the reference and distorted video and subsequently extracts low-bandwidth spatial and temporal features for video quality evaluation. The General Model necessitates an ancillary data channel bandwidth that accounts for 9.3\% of the uncompressed video sequence, while the associated calibration techniques demand an extra 4.7\%. 
Masry \textit{et al.} \cite{masry2006scalable} utilized wavelet transforms and separable filters to decompose the video into multiple channels, adjusting the bit rate of the reference video decomposition using a coefficient selection strategy. By setting a reference bit rate of 10 kbit/s, the proposed RR VQA model shows impressive performance while maintaining real-time processing capability.
Zeng and Wang \cite{zeng2010temporal} first utilized the complex wavelet transform to decompose the reference and distorted videos and subsequently calculate the conditional histogram and circular variance (CV) curve. They employed a fourth-order polynomial to model the CV curve of the reference video and set the 5 parameters of the fitted polynomial as the RR features. This method measures temporal motion smoothness and achieves good performance on video distortions like frame jittering, dropping, etc. 
Gunawan and Ghanbari \cite{gunawan2008reduced} developed a RR VQA method to measure the quality of encoded videos using harmonic analysis of spatial gradients. It involves extracting local harmonic strength features from images as reduced-reference data, and through discriminative analysis, generating harmonics gain and loss to represent blocking/tiling and blurring/smearing distortions respectively.

Some works employ learning-based methods, such as linear regression, neural network, \textit{etc.}, for the reduced-reference feature fusion. Le Callet \textit{et al.} \cite{le2006convolutional} first extracted three types of features of both reference and distorted videos, including frequency content measures \cite{melcher1995objective}, temporal content measures \cite{webster1993objective, tetsuji2000objective}, and blocking measures \cite{wang2000blind}, and subsequently employ a time-delay neural network which consisting several CNN and multi-layer perception networks to regress the features into video quality scores. 
Zhu \textit{et al.} \cite{zhu2014optimizing} initially employed a NR VQA model \cite{zhu2013no} to extract three intra-subband features, including energy, entropy, and kurtosis, as well as three inter-subband features, encompassing the Jensen-Shannon divergence, the structural similarity index between two subbands, and the smoothness, for both distorted and reference videos. Subsequently, they introduced a feature pooling approach consisting of three components: a global linear model for aggregating the extracted features, a simple linear model for achieving local alignment wherein the local factors are influenced by the source videos, and a non-linear model for quality calibration.

With the popularity of natural scene statistics (NSS) in the IQA filed, many works try to incorporate NSS as part of the features for RR VQA. For example, Ma \textit{et al.} \cite{ma2012reduced} extracted the spatial information loss and the temporal statistics characteristics from the interframe histogram as the reduced features. Specifically, they introduced an energy variation descriptor to assess the energy difference of each individual encoded frame for spatial information loss and employed the generalized Gaussian density (GGD) function to capture the natural statistics of the interframe histogram distribution. 
Soundararajan and Bovik \cite{soundararajan2012video} proposed the spatiotemporal RR entropic differences (STRRED) metric that calculates the wavelet coefficients of frame differences modeled by the Gaussian scale mixture (GSM) distribution to capture temporal information and leverages their previous developed RR IQA method (SRRED) \cite{soundararajan2011rred} to model spatial information.
To mitigate the computational complexity of STRRED, Bampis \textit{et al.} \cite{bampis2017speed} introduced the spatial efficient entropic differencing for quality assessment (SpEED-QA) model, which computes local entropic differences between reference and distorted videos in the spatial domain, resulting in efficient feature calculation.
Wang \textit{et al.} \cite{wang2015very} proposed a RR VQA metric that leverages the contrast and motion sensitivity characteristics of the human visual system to select the reference data. Specifically, the proposed metric first maps the reference video into different frequency using the Discrete Wavelet Transform (DWT) and then selects the image blocks of each frame according to the energy of the wavelet coefficients in the subbands of interest and spatio-temporal information of frame difference. Finally, it calculates the SSIM values between the selected reference image blocks and their corresponding distorted ones as the quality score.

\begin{figure}[t]
    \centering
    \includegraphics[width=0.71\textwidth]{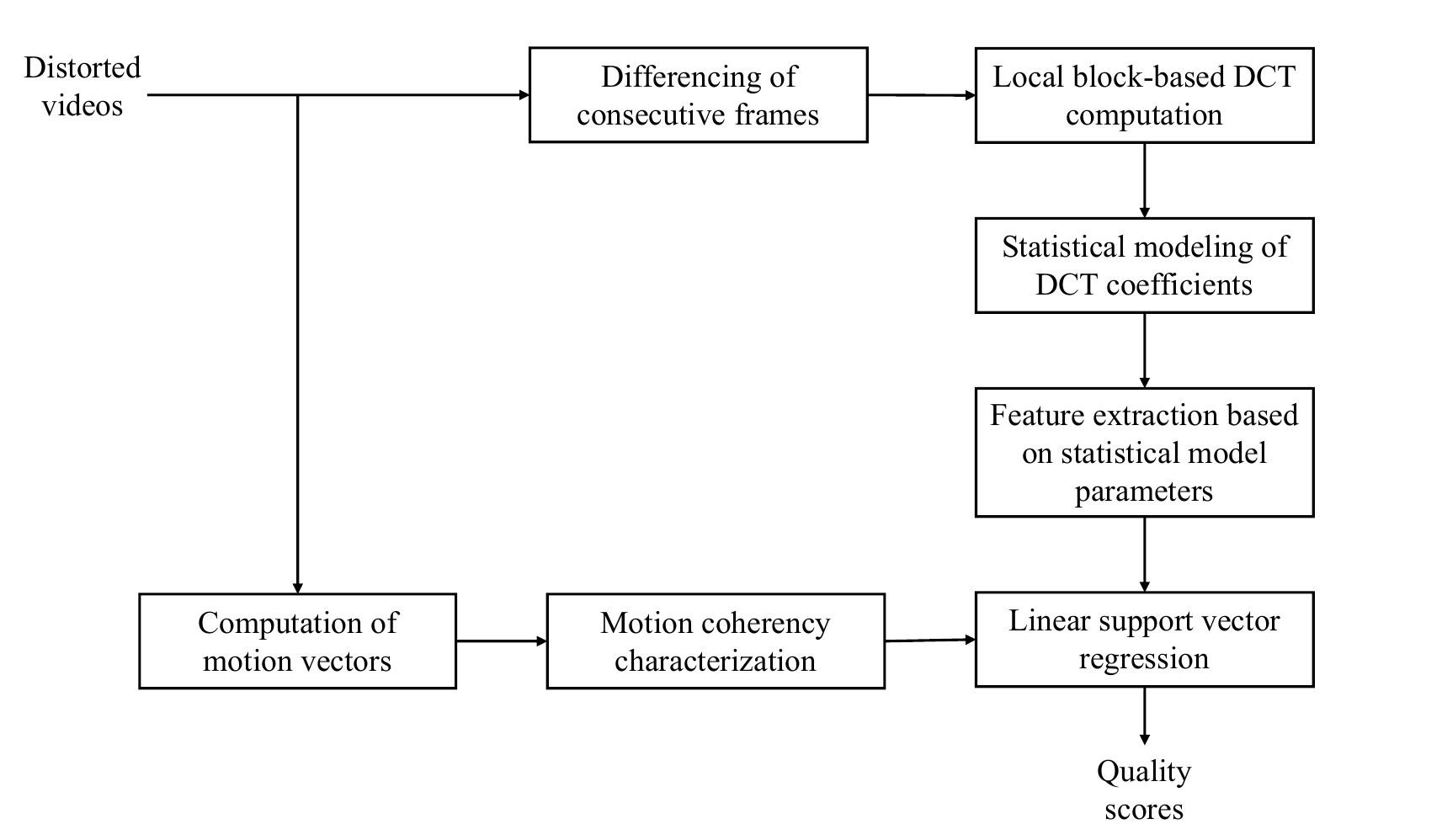}
    \caption{The framework of Video BLIINDS~\cite{saad2014blind}.}
    \label{video_bliinds_framework}
\end{figure}

\begin{table}[t]
	\footnotesize
	\centering
	\renewcommand{\arraystretch}{1.25}
        \setlength{\tabcolsep}{0.4em}
	\caption{Overview of the NR Video Quality Assessment Models.}
	\label{NR_VQA}
	\resizebox{1\textwidth}{!}{
	\begin{tabular}{ccccc}
		
		\toprule[.15em]
		Type & Algorithm & Methodology & Extracted quality features & Quality fusion   \\
		\hline
\multirow{48}{*}{NR} \hspace{10pt}
&  Wang \textit{et al.} \cite{wang2004video} & Hand-crafted feature &	Structure similarity, motion vector	&Weighted sum\\
& V-CORNIA \cite{xu2014no} & Hand-crafted feature & CORNIA, temporal pooling & Weighted sum \\
& Video BLIINDS \cite{saad2014blind} & Hand-crafted feature & Spatial-temporal natural video statistics (NVS), motion vector & SVR \\
& VIIDEO \cite{mittal2015completely} & Hand-crafted feature & Natural video statistics, inter sub-band statistics & Weighted sum \\
& TLVQM \cite{korhonen2019two} & Hand-crafted feature & 45 low complexity \& 30 high complexity features & SVR \\
& VIDEVAL \cite{tu2020ugc} & Hand-crafted feature & BRISQUE, GM-LOG, HIGRADE-GRAD, FRIQUEE, TLVQM & SVR \\
& Chip-QA \cite{ebenezer2021chipqa} & Hand-crafted feature & space-time chip, luma, color, gradient & SVR \\
\cline{2-5}
&VSFA \cite{li2019quality}	&Pre-trained DNN model&	ResNet-50	&GRU\\
&MDVSFA \cite{li2021unified}	&Pre-trained DNN model	&ResNet-50	&GRU\\
&Tang \textit{et al.} \cite{tang2020deep}	&Pre-trained DNN model&	VGG-16&	MLP, temporal memory-based pooling\\
&RIRNet \cite{chen2020rirnet}	&Pre-trained DNN model	&ResNet-50 with SPP	&GRU\\
&Chen \textit{et al.} \cite{chen2021learning}&	Pre-trained DNN model&	VGG-16 with attention module&	GRU\\
&You \cite{you2021long}	&Pre-trained DNN model	&ResNet-50 with FPN and attention module	&Transformer encoder\\
&You and Lin \cite{you2022efficient}	&Pre-trained DNN model	&ResNet-50 with FPN	&Transformer encoder\\
&Wu \textit{et al.} \cite{wu2023discovqa}	&Pre-trained DNN model	&Swin-T	&STDE, TCT\\
&STDAM \cite{xu2021perceptual}	&Pre-trained DNN model	&ResNet-18 with graph convolution module and attention module	&Bi-directional LSTM\\
&PatchVQ \cite{ying2021patch}	&Pre-trained DNN model	&PaQ-2-PiQ, 3D ResNet-18	&InceptionTime\\
&Ying \textit{et al.} \cite{ying2022telepresence}	&Pre-trained DNN model	&MobileNetV3, 3D ResNet-18, MobileNetV1	&GRU-FCN\\
&Li \textit{et al.} \cite{li2022learning,li2022blindly}	&Pre-trained DNN model	&UNIQUE,  SlowFast	&GRU\\
&Liu \textit{et al.} \cite{liu2022quality}	&Pre-trained DNN model	&KonCept512, SlowFast	&Progressively Residual Aggregation\\
&UVQ \cite{wang2021rich}	&Pre-trained DNN model	&EfficientNet-B0, D3D	&MLP\\
&UVQ-lite \cite{wang2022revisiting}	&Pre-trained DNN model	&MobileNet, MoViNet	&MLP\\
&Telili \textit{et al.} \cite{telili20222bivqa}	&Pre-trained DNN model	&ResNet-50	&Bi-LSTM\\
&Lu \textit{et al.} \cite{lu2023bh}	&Pre-trained DNN model	&ResNet-50	&GRU\\
&MD-VQA \cite{zhang2023md}	&Pre-trained DNN model	&EfficientNetV2, ResNet3D-18, blur, noise, block effect, exposure, colorfulness	&MLP\\
&Zhu \textit{et al.} \cite{zhu2022learning}	&Pre-trained DNN model	&ResNeXt-101	&Transformer encoder\\
&Zhang \textit{et al.} \cite{zhang2022hvs}	&Pre-trained DNN model	&ConvNeXt, SAMNet, SlowFast	&GRU\\
&Chen \textit{et al.} \cite{chen2022dynamic}	&Pre-trained DNN model	&ResNet-50, C3D, RIRNet, PVQ, LSCT-PHIQNet	&MLP\\
&Kwong \textit{et al.} \cite{kwong2023quality}	&Pre-trained DNN model	&Multi-channel CNN	&GRU\\
&Wu \textit{et al.} \cite{wu2023towards,wu2023exploring}	&Pre-trained DNN model	&NIQE, TPQI, CLIP	&Weighted sum\\
&Wu \textit{et al.} \cite{wu2023towards}	&Pre-trained DNN model	&FAST-VQA, CLIP-visual, CLIP	&MLP,  cosine similarity\\
&Liu \textit{et al.} \cite{liu2023ada}	&Pre-trained DNN model	&EfficientNet-b7, ir-CSN-152, CLIP, Swin-B, TimeSformer, Video Swin-B, SlowFast	&MLP\\
\cline{2-5}
&Liu \textit{et al.} \cite{liu2018end}	&End-to-end training	&3D-CNN	&MLP\\
&You and Korhonen \cite{you2019deep}	&End-to-end training	&3D-CNN	&LSTM\\
&Yi \textit{et al.} \cite{yi2021attention}	&End-to-end training	&VGG-16 with non-local module	&MLP\\
&Wen and Wang \cite{wen2021strong}	&End-to-end training	&ResNet-18	&MLP\\
&SimpleVQA \cite{sun2022deep}	&End-to-end training	&ResNet-50	&MLP\\
&Minimalistic VQA \cite{sun2023analysis}	&End-to-end training	&ResNet-50 or Swin-B	&MLP\\
&StarVQA \cite{xing2022starvqa}	&End-to-end training	&Transformer	&MLP\\
&Lin \textit{et al.} \cite{lin2022deep}	&End-to-end training	&HED, I3D	&Transformer encoder, MLP\\
&Shen \textit{et al.} \cite{shen2022end}	&End-to-end training	&2D-CNN, 3D-CNN	&MLP\\
&Xian \textit{et al.} \cite{xian2022spatiotemporal}	&End-to-end training	&DeblurGAN-v2, 3D-CNN	&MLP\\
&Guan \textit{et al.} \cite{guan2022end}	&End-to-end training	&ResNet-50	&ConvLSTM, MLP\\
&Lu \textit{et al.} \cite{lu2022deep}	&End-to-end training	&ResNet-18	&MLP\\
&FAST-VQA \cite{wu2022fast}	&End-to-end training	&Video Swin-T	&MLP\\
&DOVER \cite{wu2022disentangling}	&End-to-end training	&Video Swin-T, ConvNeXt-T	&MLP\\
&Kou \textit{et al.} \cite{kou2023stablevqa}	&End-to-end training	&Swin-T, 3D ResNet, blur encoder	&MLP\\
&Yuan \textit{et al.} \cite{yuan2023capturing}	&End-to-end training	&Visual quality transformer	&MLP\\
&Ke \textit{et al.} \cite{ke2023mret}	&End-to-end training	&Spatial and temporal Transformer encoder	&MLP\\
\cline{2-5}
&Liu \textit{et al.} \cite{wu2021no,liu2021spatiotemporal}	&Self-supervised learning	&R(2+1)D	&MLP\\
&Chen \textit{et al.} \cite{chen2021unsupervised}	&Self-supervised learning	&C3D	&MLP\\
&Chen \textit{et al.} \cite{chen2021contrastive}	&Self-supervised learning	&VSFA or RIRNet	&GRU, MLP\\
&Madhusudana \textit{et al.} \cite{madhusudana2022conviqt}	&Self-supervised learning	&ResNet-50	&GRU, regularized linear regressor\\
&Mitra and Soundararajan \cite{mitra2022multiview}	&Self-supervised learning	&ResNet-50	&Weighted sum\\
&Jiang \textit{et al.} \cite{jiang2022self}	&Self-supervised learning	&2D-CNN, 3D-CNN	&Transformer encoder, MLP\\

		\bottomrule[.15em]
		
	\end{tabular}
	}
\end{table}

\subsection{No-reference Video Quality Assessment}

In practical video-enabled applications, reference videos are often unavailable, thereby only NR VQA models are qualified to assess the video quality. Similar to FR VQA, we categorize NR VQA models into two groups: knowledge-driven NR VQA models and data-driven NR VQA models based on their feature extraction modules.

\subsubsection{Knowledge-driven NR VQA}
Classical VQA methods generally adopt the knowledge-driven approach and manually extract hand-crafted features to perform evaluation.
Some early works extended the no-reference image quality assessment (NR-IQA) methods, such as NIQE \cite{mittal2012no} and BRISQUE \cite{mittal2012making} to perform video quality assessment \cite{tu2020comparative}.
To better predict video quality, some classical VQA methods have been proposed by leveraging temporal information in videos.
Xu \textit{et al.} \cite{xu2014no} proposed a V-CORNIA metric, which extracts CORNIA features as spatial features and utilizes the hysteresis temporal pooling method to predict video quality.
As illustrated in Figure~\ref{video_bliinds_framework}, Saad \textit{et al.} \cite{saad2014blind} presented a Video BLIINDS model, which combines spatial-temporal natural video statistic (NVS) features and motion-related features to perform VQA.
Mittal \textit{et al.} \cite{mittal2015completely} introduced a VIIDEO algorithm for VQA, which incorporates natural video statistics and inter sub-band statistics via weighted sum.
Korhonen \textit{et al.} \cite{korhonen2019two} developed a TLVQM VQA measure, which extracts 45 low-complexity features and 30 high-complexity features, and utilizes SVR to integrate them.
Tu \textit{et al.} \cite{tu2020ugc} devised a VIDEVAL VQA method, which extracts BRISQUE, GM-LOG, HIGRADE-GRAD, FRIQUEE, TLVQM features and uses SVR to combine these features.
Ebenezer \textit{et al.} \cite{ebenezer2021chipqa} proposed a Chip-QA metric, which extracts luma, color features as spatial features, and exploits space-time chip to capture temporal motion features, to conduct VQA.
This method achieves better performance compared to existing knowledge-driven NR-VQA models while still keeping low computational complexity.

\begin{figure}[t]
    \centering
    \includegraphics[width=0.70\textwidth]{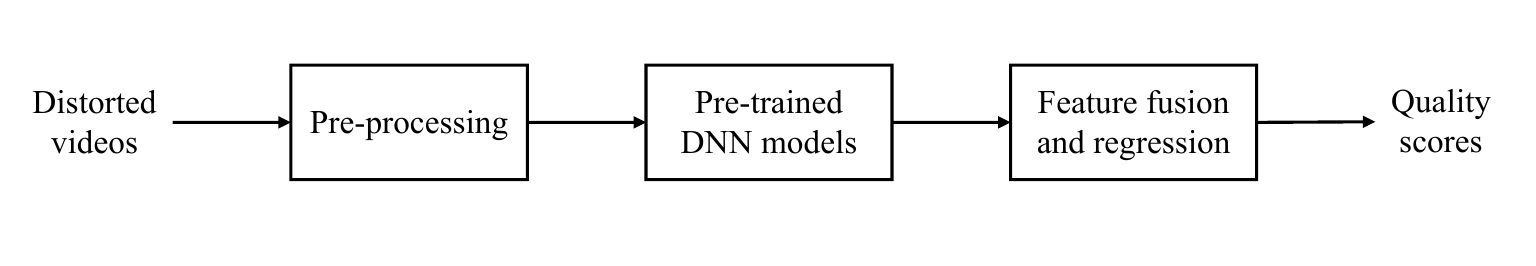}
    \caption{The framework of pre-trained model based NR VQA models.}
    \label{pre_trained_NR_VQA_models}
\end{figure}

\subsubsection{Data-driven NR VQA}
In comparison to knowledge-driven NR VQA methods, data-driven NR VQA models can automatically extract quality-aware features of distorted videos by designed neural networks, which is simper but more powerful. Based on the training methods of the feature extraction network, we can divide data-driven BVQA methods into three categories: pre-trained model based methods, end-to-end training based methods, and unsupervised learning based methods.

\begin{figure}[t]
    \centering
    \includegraphics[width=0.9\textwidth]{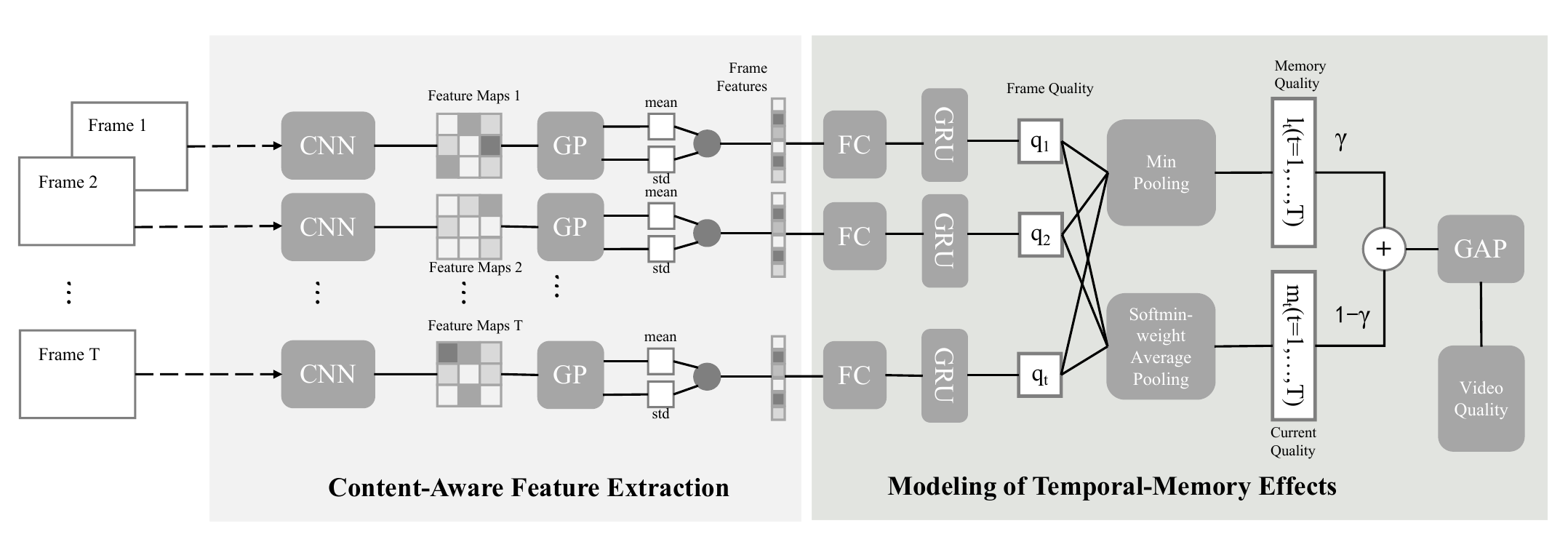}
    \caption{The framework of VSFA~\cite{li2019quality}.}
    \label{VSFA_framework}
\end{figure}

\textbf{(1) Pre-trained model based methods:} 
These methods employ pre-trained quality-aware or semantic-related models as the feature extraction module to extract features, with only the quality regressor requiring training to map the extracted features into video quality scores. 
VSFA \cite{li2019quality} extracts semantic features using a pre-trained ResNet-50 model on ImageNet, subsequently utilizing a gated recurrent unit (GRU) network as the regressor to capture the temporal relationship. It also introduce a differentiable subjectively-inspired temporal pooling strategy to address the temporal hysteresis effect of the human vision system. We present the framework of VSFA in Figure~\ref{VSFA_framework}.
The authors of VSFA further propose MDVSFA \cite{li2021unified}, which enhances the performance and generalization of VSFA by training it on four VQA databases including CVD2014 \cite{nuutinen2016cvd2014},  KoNViD-1k \cite{hosu2017konstanz}, LIVE-Qualcomm \cite{ghadiyaram2017capture}, and LIVE-VQC \cite{sinno2018large}.  
Tang \textit{et al.} \cite{tang2020deep} utilized VGG-16 \cite{simonyan2014very} to extract the content features from frame patches and then employed a patch quality regression network and a patch weight estimation network to derive frame-level quality scores. Finally, they introduced a temporal memory-based pooling method to aggregate the frame-level quality scores into the video quality scores. 
Chen \textit{et al.} \cite{chen2020rirnet} presented a NR VQA framework called Recurrent-In-Recurrent Network (RIRNet), which employs a ResNet-50 followed by a spatial pyramid pooling (SPP) layer to capture the content features of each video frame. Then, these extracted features are divided into multiple groups based on different temporal resolutions and RIRNet utilizes multiple GRU to aggregate the extracted features with different temporal resolutions into video quality score by the deep supervision manner. 
Chen \textit{et al.} \cite{chen2021learning} proposed a generalized spatial-temporal deep feature representation through imposing the Gaussian distribution constraints and a pyramid temporal aggregation module on the spatial-temporal features extracted by the multi-stage layers of VGG-16 and enhanced by a GRU network.

Besides GRU, Transformer has been exploited as a superior feature aggregation model for NR VQA.
You \cite{you2021long} performed a basic Transformer encoder for NR VQA. He first utilized a perceptual hierarchical network with an integrated attention module to extract quality-aware features of each frame and then employed a time-distributed 1D CNN consisting of Conv1D, MaxPool, and Dropout layers to reduce the dimensions of extracted features. Finally, he used a standard Transformer encoder with a mask strategy to drive the video quality scores. 
You and Lin \cite{you2022efficient} further replaced the time distributed 1D CNN module with a shared multi-head attention module. 
Wu \textit{et al.} \cite{wu2023discovqa} first utilized video Swin-T \cite{liu2022video} to extract the spatial-temporal features of the video and then introduced a temporal distorted-content Transformer for aggregating the content features and obtaining the video quality score. To be more specific, the temporal distorted-content Transformer consists of a transformer-based spatial-temporal distortion extraction (STDE) module for discerning various kinds of temporal variations and extracting the temporal distortion features, and encoder-decoder-like temporal content transformer (TCT) for addressing temporal quality attention issues.

Since the image classification model cannot capture the quality-aware and motion-aware features, some studies attempt to leverage NR IQA models to represent quality-aware spatial features and the action recognition model or the optical flow model to extract motion-aware features. 
Xu \textit{et al.} \cite{xu2021perceptual} developed a spatiotemporal distortion-aware model (STDAM) for NR VQA. Specifically, they employed ResNet-18 to extract content features from video frames at two kinds of spatial resolutions and then utilized a graph convolution module and an attention module to aggregate these content features into the frame-level features. Note that the ResNet-18 along with the graph convolution module and the attention module has been pre-trained on KonIQ-10k \cite{hosu2020koniq}, a large-scale IQA dataset. Besides, they computed the optical flow maps of videos and used ResNet-18 to extract the motion features from these optical flow maps. Finally, they utilized a bi-directional LSTM network to map the frame-level features and motion features into the video quality score. 
Ying \textit{et al.} \cite{ying2021patch} introduced PatchVQ, which extracts spatial features using the PaQ-2-PiQ \cite{ying2020patches} backbone pre-trained on the LIVE-FB dataset \cite{ying2020patches} and extracts spatio-temporal features using a 3D ResNet-18 backbone \cite{hara2017learning} pre-trained on the Kinetics dataset \cite{carreira2017quo}. Furthermore, PatchVQ utilizes a region-of-interest pooling (RoIPool) layer and segment-of-interest pooling (SoIPoll) to capture the local interested region of spatial and temporal. Finally, InceptionTime is employed to regress the pooled features into the video quality score. 
Ying \textit{et al.} \cite{ying2022telepresence} further proposed a multi-modal NR VQA model designed for live streaming telepresence content. It consists of three branches, each corresponding to the feature extraction network of the audio, image, and video modalities, respectively. Specifically, the frame-level and patch-level features are extracted by MobileNetV3 \cite{howard2019searching} pre-trained on LIVE-FB dataset \cite{ying2020patches}, the video-level features are extracted by the R(2+1)D model \cite{tran2018closer}, and the audio-level features are extracted by the MobileNetV1 pre-trained on the Google AudioSet dataset \cite{gemmeke2017audio}. 

Li \textit{et al.} \cite{li2022learning,li2022blindly} employed UNIQUE \cite{zhang2021uncertainty}, an IQA model pre-trained on four IQA datasets including BID \cite{ciancio2011no}, LIVE Challenge \cite{ghadiyaram2016massive}, SPAQ \cite{fang2020perceptual}, and KonIQ-10k \cite{hosu2020koniq}, to extract quality-aware spatial features and utilize SlowFast \cite{feichtenhofer2019slowfast} to extract temporal feature. Subsequently, a GRU network is used to model spatial and temporal features and regress them into the video quality scores.  
Similar to the methods of Li \textit{et al.} \cite{li2022learning,li2022blindly}, Liu \textit{et al.} \cite{liu2022quality} also utilized an IQA model named KonCept512 \cite{hosu2020koniq} to capture the static appearance degradation and an action recognition model SlowFast \cite{feichtenhofer2019slowfast} to represent dynamic motion degradation. They then introduced a progressively residual aggregation module to hierarchical merge these two kinds of features to derive the video quality scores.
Zhu \textit{et al.} \cite{zhu2022learning} proposed a spatiotemporal interaction strategy for assessing the quality of user-generated videos. Specifically, they extracted feature maps of video frames using a ResNeXt-101 \cite{xie2017aggregated} pre-trained on KonIQ-10k \cite{hosu2020koniq} and then calculated the mean and standard deviation values of these extracted feature maps as the spatial features and computed the difference between the spatial features of two consecutive frames as to derive motion features. Finally, a Transformer model was utilized to aggregate the spatial and motion features into the video quality scores.
Kwong \textit{et al.} \cite{kwong2023quality} first used the self-supervised learning method to train a multi-channel CNN network on the IQA task without using the quality-rated labels and subsequently fine-tuned the multi-channel CNN model with motion-aware features followed by a GRU network for the NR VQA task.
Lu \textit{et al.} \cite{lu2023bh} calculated deep structural similarities between the feature maps of continuous frames extracted by a quality-aware pre-trained DNN model to capture temporal distortions arising from frame rate variations, object movement, and camera motion.

It is known that a video may exhibit various kinds of distortions. Therefore, employing a broader range of feature descriptors can effectively address complex video distortions. For example, Wang \textit{et al.} \cite{wang2021rich} proposed a feature-rich NR VQA model named UVQ, which incorporates features extracted from three pre-trained models, including compression level classification, action recognition, and distortion type classification.
In \cite{wang2022revisiting}, Wang \textit{et al.} further replaced the backbones from EfficientNet-b0 \cite{tan2019efficientnet} and D3D \cite{stroud2020d3d} to MobileNet \cite{chu2021discovering} and MoViNet \cite{kondratyuk2021movinets} to achieve a light-weight UVQ. 
Telili \textit{et al.} \cite{telili20222bivqa} introduced a double Bi-LSTM network for video quality assessment, where the first Bi-LSTM is employed to spatially pool the features extracted by a ResNet-50 pre-trained on KonIQ-10k \cite{hosu2020koniq} and the second Bi-LSTM is used to temporally pool the spatial features into the video quality scores. 
Zhang \textit{et al.} \cite{zhang2023md} developed a multi-dimensional VQA (MD-VQA) model, which leverages the EfficientNetV2 \cite{tan2021efficientnetv2} to extract the semantic features, utilizes five distortion descriptors including blur \cite{zhan2017no}, noise \cite{chen2015efficient}, block effect \cite{wang2002no}, exposure \cite{korhonen2019two}, and colorfulness \cite{panetta2013no} to measure the distortion level, and employs ResNet3D-18 \cite{hara2018can} to capture the motion information.
Zhang \textit{et al.} \cite{zhang2022hvs} considered five characteristics of HVS for video quality assessment: visual saliency, edge masking, content dependency, motion perception, and temporal hysteresis. For content dependency and edge masking, they used ConvNeXt \cite{liu2022convnet} to extract the content features and edge feature maps from original RGB frames and the corresponding Canny edge maps, respectively. For visual saliency, they performed a saliency detection model SAMNet \cite{liu2021samnet} to extract saliency maps and then used the saliency maps to weight the content and edge feature maps. For motion perception, they utilized SlowFast to extract the motion features. Finally, for temporal hysteresis, they combined the saliency-weighted content and edge features and motion features and employed the GRU module with the subjectively-inspired temporal pooling model in \cite{li2019quality} to regress combined features into video quality scores. 
Chen \textit{et al.} \cite{chen2022dynamic} developed a dynamic expert-knowledge ensemble strategy for generalizable video quality assessment, which relies on one image classification model, ResNet-50 \cite{he2016deep}, one action recognition model, C3D \cite{tran2015learning}, and three trained NR VQA model, RIRNet \cite{chen2020rirnet}, PVQ \cite{ying2021patch}, and LSCT-PHIQNet \cite{you2021long} as the experts. Then, they trained an ensemble model to make full use of complementary information from these experts using the contrast learning method. 

Recently, the visual-language pre-training methods are exploited for NR VQA. 
For example, Wu \textit{et al.} \cite{wu2023towards,wu2023exploring} proposed to combine the spatial naturalness index NIQE \cite{mittal2012making}, the temporal naturalness index TPQI \cite{liao2022exploring}, and the contrastive language-image pre-training (CLIP) model \cite{radford2021learning} with a quality-guided text prompt to achieve zero-shot NR VQA. 
Wu \textit{et al.} \cite{wu2023towards} further introduced a multi-dimensional language-prompted NR VQA model that employs FAST-VQA \cite{wu2022fast} to capture low-level-aware features, CLIP-visual to extract local CLIP features, and CLIP-textual to extract dimensional-oriented quality-guided text features. They calculated the cosine similarity of the visual features fused by low-level-aware features and local CLIP features and text features as the video quality scores.
Liu \textit{et al.} \cite{liu2023ada} extracted a range of quality-aware features from the image modality, video modality, and text-to-image modality. Seven pre-trained models were employed to diverse features from these three modalities and a quality-aware acquisition module was designed to adaptively capture the diversity and complementary information among them. They further utilized a knowledge distillation method to transfer the knowledge from these modalities to a lightweight VQA model.

\begin{figure}[t]
    \centering
    \includegraphics[width=0.7\textwidth]{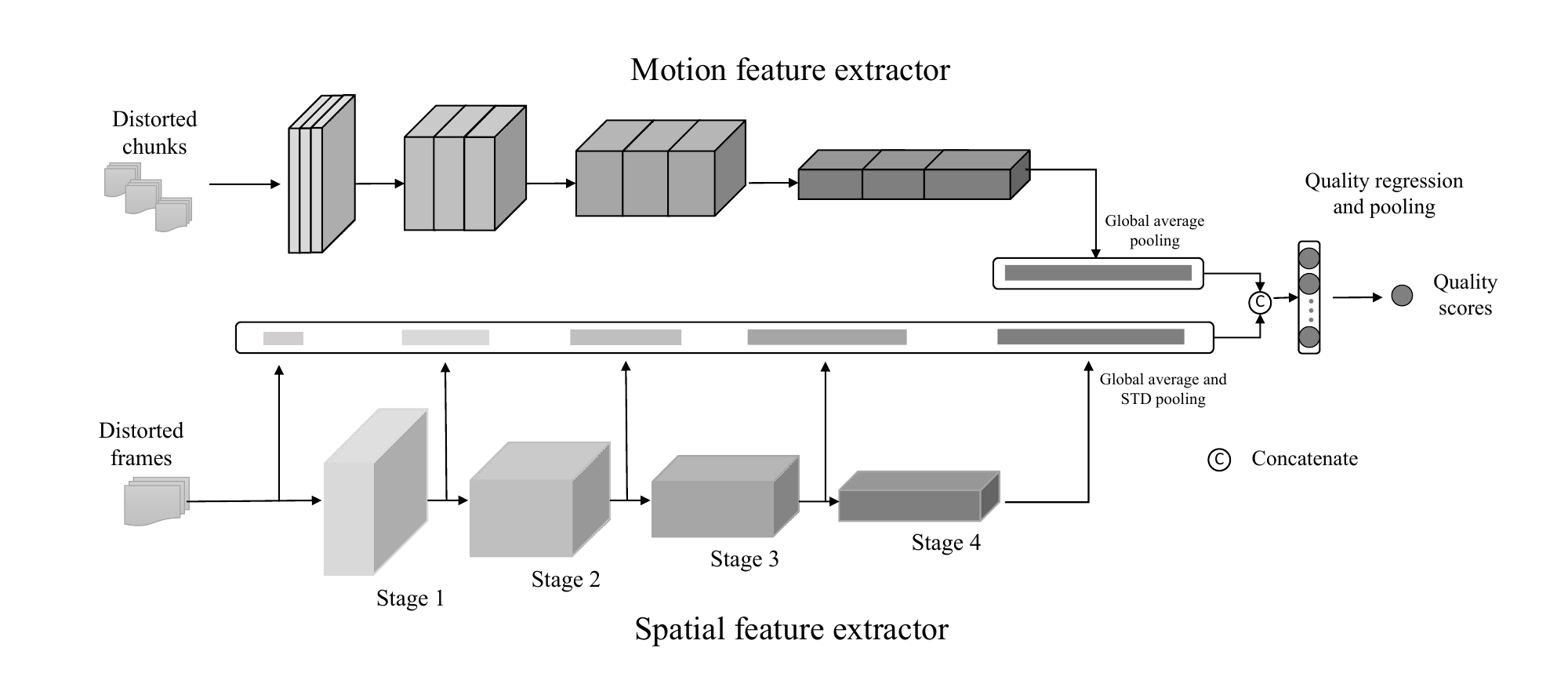}
    \caption{The framework of SimpleVQA~\cite{sun2022deep}, which trains an end-to-end spatial feature extraction module and utilizes a pre-trained motion feature extraction.}
    \label{SimpleVQA_framework}
\end{figure}

\textbf{(2) End-to-end training based methods:} The end-to-end training approach enables the BVQA model to directly learn the quality-aware feature representation from the raw pixels of a video. 
Liu \textit{et al.} \cite{liu2018end} proposed a multi-task BVQA model V-MEON by jointly optimizing the 3D-CNN model for quality assessment and compression distortion classification. 
You and Korhonen \cite{you2019deep} also employed a 3D-CNN model to extract features from a video clip and subsequently employed a LSTM network to regress the 3D-CNN features into the video quality scores. Note that the two network are trained independently. 
Yi \textit{et al.} \cite{yi2021attention} introduced an attention-based NR VQA model that tackles the problem of uneven spatial distortion by training the VGG network with a non-local operator in an end-to-end manner. 
Wen and Wang \cite{wen2021strong} developed an IQA-based VQA method, which uses a ResNet-18 to compute the frame-level quality scores and then averagely pools the frame-level quality scores into the video quality scores. They performed the L1 loss and the Rank loss to optimize the proposed VQA models. 

To better handle temporal-related distortions in videos, some NR VQA studies leverage the motion features or spatial-temporal modeling methods to improve the performance.  Sun \textit{et al.} \cite{sun2022deep} proposed SimpleVQA, a simple NR VQA framework consisting an end-to-end trained multi-scale spatial feature extraction module and a pre-trained motion extraction module. 
Sun \textit{et al.} \cite{sun2023analysis} further proposed a minimalistic VQA model, which includes four basic blocks:  a video preprocessor (for aggressive spatiotemporal downsampling), a spatial quality analyzer, an optional temporal quality analyzer, and a quality regressor, all with the simplest possible instantiations.
Shen \textit{et al.} \cite{shen2022end} presented an end-to-end NR VQA model that incorporates spatiotemporal feature fusion and hierarchical information integration. It includes a feature extraction model using 2D and 3D convolutional layers for gradual extraction of spatiotemporal features from raw video clips and a hierarchical branching network for fusing multiframe features. 
Xian \textit{et al.} \cite{xian2022spatiotemporal} proposed to generate a simulated video using a generative adversarial network (GAN)-based image restoration model as a pesudo reference video and then developed a pyramidal spatiotemporal feature hierarchy (PSFH) network to extract the multi-stage spatiotemporal features of the distorted videos and the differences between the distorted videos and the pesudo reference videos.
Guan \textit{et al.} \cite{guan2022end} developed a visual and memory attention-based NR VQA model. They proposed a visual attention module to derive spatial-temporal attention-guided representation for frame-level quality-aware features and a memory attention module to map the frame-level quality-aware features into the video-level quality scores.
Lu \textit{et al.} \cite{lu2022deep} proposed a grey-level co-occurrence matrix based text measure to select represent patches from high-resolution video content and subsequently employed a 2D-CNN backbone (\textit{i.e.} ResNet-18) to extract quality-aware features from these selected patches.

Recently, Vision Transformer have demonstrated outstanding performance in various computer vision tasks. Hence, an increasing number of NR VQA methods are adopting Transformer-based architectures.
Xing \textit{et al.} \cite{xing2022starvqa} introduced StarVQA, which constructs a Transformer model for NR VQA by combining divided space-time attention and then devises a vectorized regression loss that encodes the mean opinion score into a probability vector. They further developed StarVQA$+$ \cite{xing2023starvqa+} by co-training StarVQA on both images and videos across different kinds of datasets. 
Lin \textit{et al.} \cite{lin2022deep} took into account the visual saliency mechanism and employ holistically-nested edge detection \cite{xie2015holistically} to choose the saliency regions within the video. The selected video saliency clips are subsequently inputted into Inflated 3D ConvNet (I3D) \cite{carreira2017quo} to extract the features and a Transformer encoder was employed to regress the features into the video quality scores.
Wu \textit{et al.} \cite{wu2022fast} introduced FAST-VQA, a fragment attention network consisting of a video Swin Transformer and the gated relative position bias module, which is specifically designed to take mini-patches sampled from the video as the input.
Wu \textit{et al.} \cite{wu2022disentangling} further considered the video quality from two aspects: the technical and aesthetic perspectives, and proposed the disentangled objective video quality evaluator (DOVER) to independently learn two quality assessment models, each focusing on one of these perspectives. 
Kou \textit{et al.} \cite{kou2023stablevqa} introduced StableVQA, a NR VQA model designed to evaluate video stability. This model leverages Swin Transformer to capture video content, utilizes 3D ResNet to extract the motion information from the optical flow modality, and incorporates a blur encoder \cite{tsai2022stripformer} to measure the blur distortion. 
Yuan \textit{et al.} \cite{yuan2023capturing} developed the Visual Quality Transformer, which utilizes a multi-pathway temporal network consisting of multiple sparse temporal attention modules to sample keyframes and measure the degree of coexisting distortions of a video. 
Ke \textit{et al.} \cite{ke2023mret} presented a multi-resolution transformer for NR VQA, which first samples spatially aligned patches from the multi-resolution frames input to preserve high-resolution details and global content and then performs a factorized spatial-temporal transformer to derive the video quality scores.

\textbf{(3) Self-supervised learning based methods:} Both pre-trained model based methods or end-to-end learning based NR VQA models require a large-scale of VQA dataset to train a robust NR VQA model. However, obtaining high-quality labels for VQA datasets, typically acquired through subjective VQA experiments, is a time-consuming and expensive process. 
Therefore, some studies attempt to use self-supervised or unsupervised learning methods for NR VQA, which aim to learn quality-aware feature representation from large-scale unlabelled video data.
Liu \textit{et al.} \cite{wu2021no,liu2021spatiotemporal} proposed a weakly supervised learning method for NR VQA. They first constructed a large-scale VQA dataset via degrading the high-quality video clips by the video compression and transmission algorithms and calculating the quality scores of the degraded videos by multiple FR VQA methods. Subsequently, they introduced a NR VQA model with a heterogeneous knowledge ensemble to learn representation from the weakly labeled data.  
Chen \textit{et al.} \cite{chen2021unsupervised} proposed a curriculum-style unsupervised domain adaptation method to tackle the cross-domain VQA challenge. The approach consists of two main stages. First, they performed domain adaptation between the source and target domains to predict the rating distribution for target samples, which provides a more accurate understanding of the subjective aspects of VQA. Second, they treated the samples in the confident subset as the easier tasks in the curriculum, and conducted a fine-grained adaptation between these two subsets to refine the prediction model. 
Chen \textit{et al.} \cite{chen2021contrastive} presented a self-supervised pre-training method for video quality assessment using the contrastive learning approach. Specifically, they first generated a range of distorted video samples with diverse distortions and visual content through a carefully designed distortion augmentation strategy. Then, they applied contrastive learning to enhance feature representations by maximizing agreement between future frames and their corresponding predictions in the embedding space. Moreover, they introduced a distortion prediction task as an extra learning objective, encouraging the model to differentiate between various distortion categories in the input video. 

Madhusudana \textit{et al.} \cite{madhusudana2022conviqt} proposed to utilize distortion type identification and degradation level determination as the auxiliary tasks to train a NR VQA model consisting of a CNN for extracting spatial features and a GRU for extracting temporal information through the contrastive learning method. 
Mitra and Soundararajan \cite{mitra2022multiview} developed a self-supervised multi-view contrastive learning framework to learn quality-aware spatio-temporal representation by comparing features between frame differences and frames by treating them as a pair of views. The learned features were subsequently compared with a dataset of unaltered, high-quality natural video patches to derive the quality of the distorted video. 
Jiang \textit{et al.} \cite{jiang2022self} introduced a multi-task self-supervised representation learning framework for NR VQA. Three tasks including the distortion type classification, frame rate classification, and bitrate evaluation were used to train a Siamese network to capture spatiotemporal differences between the original video and the corresponding distorted ones. This model contains 3D-CNN and 2D-CNN to model short-term spatio-temporal dependencies and a Transformer to model the long-term spatio-temporal dependencies.

\section{Objective Video Quality Assessment: Specific-purpose Models}
\label{sec:objective_specific}
The following section provides an overview of emerging topics in the field of video quality assessment that have gained attention in recent years. These topics include compressed VQA, streaming VQA, stereoscopic VQA, VR VQA, framerate and frame interpolation VQA, audio-visual VQA, HDR or WCG VQA, screen or game VQA, and various other emerging topics. To ensure a clear organization, we have classified these surveyed algorithms based on their respective topics or applications.

\subsection{Compressed VQA}
In addition to the previously mentioned video quality assessment approaches, there exists a set of specialized methods tailored for evaluating compressed videos, which is the primary focus of this section review.
Compressed video assessment involves unique challenges and considerations due to the data reduction techniques applied during compression.
To address these aspects effectively, researchers and experts have developed various methodologies dedicated to this domain.

\subsubsection{FR and RR Methods}
Full reference and reduced reference methods compare the compressed video with its original, uncompressed version, allowing for a thorough and accurate analysis of the compression quality.
In \cite{xu2016free}, Xu \textit{et al.} proposed the FR free-energy principle inspired video quality metric (FePVQ), which is applied to optimize perceptual video coding. FePVQ separates videos into orderly and disorderly regions based on the free-energy principle, where fixation or visual attention is associated with objects exhibiting significant motion according to human visual speed perception, extending the principle into the spatio-temporal domain for VQA.
VMAF, developed by Netflix \cite{rassool2017vmaf}, is a full-reference, perceptual video quality metric designed to closely align with subjective Mean Opinion Score ratings. It employs machine learning techniques and a support vector machine to combine scores from multiple quality assessment algorithms, aiming to estimate the perceived quality of video content by considering degradation caused by compression and rescaling.
In \cite{sun2021deep, sun2022deep}, Sun proposed a FR deep learning-based VQA framework for evaluating the quality of compressed User-Generated Content videos. The proposed framework consists of three modules: a feature extraction module that fuses features from intermediate layers of a CNN to create quality-aware feature representation, a quality regression module that uses FC layers to regress the features into frame-level scores, and a subjectively-inspired temporal pooling strategy to aggregate frame-level scores into video-level scores.
In \cite{ma2012reduced}, Ma \textit{et al.} proposed a RR VQA method for compressed videos. In the model, the spatial aspect is measured using an energy variation descriptor that captures the energy change and texture masking property of the human visual system, while the temporal aspect is captured using the generalized Gaussian density function to model the interframe histogram distribution. The city-block distance is then used to calculate the histogram distance between the original video sequence and the encoded one.

\subsubsection{NR Methods}
No-reference video compression quality assessment methods are more generalized, as they do not require any reference information, such as the original uncompressed video, for evaluation.
In \cite{lee2012hybrid}, Lee \textit{et al.} proposed a NR video quality assessment method for scalable video coding that quantifies video quality using decoding parameters from compressed bitstreams, including both the base layer and enhancement layer. The proposed approach assesses the quality of the enhancement layers based on statistics of coding parameters and their relationship with the quality of the base layer, providing the assessment of the overall video quality.
In \cite{lin2012no}, Lin \textit{et al.} introduced a NR VQA algorithm, which operates in the compressed domain and considers three key factors: quantization parameter, motion, and bit allocation factor, extracted from the compressed bitstream. The algorithm also takes into account the characteristics of the human visual system for improved quality estimation.
In \cite{zhu2014no}, Zhu \textit{et al.} proposed a NR compressed video quality prediction model based on discrete cosine transform (DCT). The model has two stages: distortion measurement, where efficient frame-level features are extracted from DCT coefficients of decoded frames to quantify distortion, and nonlinear mapping, where a trained multilayer neural network takes video-level features obtained through temporal pooling as inputs and predicts the quality score of the video sequence.
In \cite{huang2017no}, Huang \textit{et al.} presented a NR VQA method for videos compressed using HEVC, without access to the bitstream. The proposed method estimates quantization levels based on transform coefficients extracted from the decoded video pixels, and models HEVC transform coefficients using a joint-Cauchy probability density function. These features are then used to predict Mean Opinion Scores for subjective video quality assessment using Elastic Net regression.
Liu \textit{et al.} \cite{liu2018end} proposed a NR VQA model named V-MEON. The proposed model uses a multi-task deep neural network framework to jointly estimate perceptual quality and codec type, leveraging complementary sets of labels obtained at low cost. The training process involves pre-training early convolutional layers with codec classification subtask, and jointly optimizing the entire network with the two subtasks together, while incorporating 3D convolutional layers for improved spatiotemporal feature extraction and performance enhancement.
In \cite{wang2021rich}, Wang \textit{et al.} introduced a NR VQA framework based on deep neural networks that comprehensively analyzes the significance of content, technical quality, and compression level in perceptual quality assessment.
In \cite{lin2023saliency}, Lin \textit{et al.} addressed the issue of Perceivable Encoding Artifacts (PEAs) in compressed videos, which significantly reduce video quality. The study investigates four spatial PEAs (blurring, blocking, bleeding, and ringing) and two temporal PEAs (flickering and floating) and proposes a compressed video quality index based on saliency-aware spatio-temporal artifact detection.

\begin{figure}[t]
    \centering
    \includegraphics[width=0.98\textwidth]{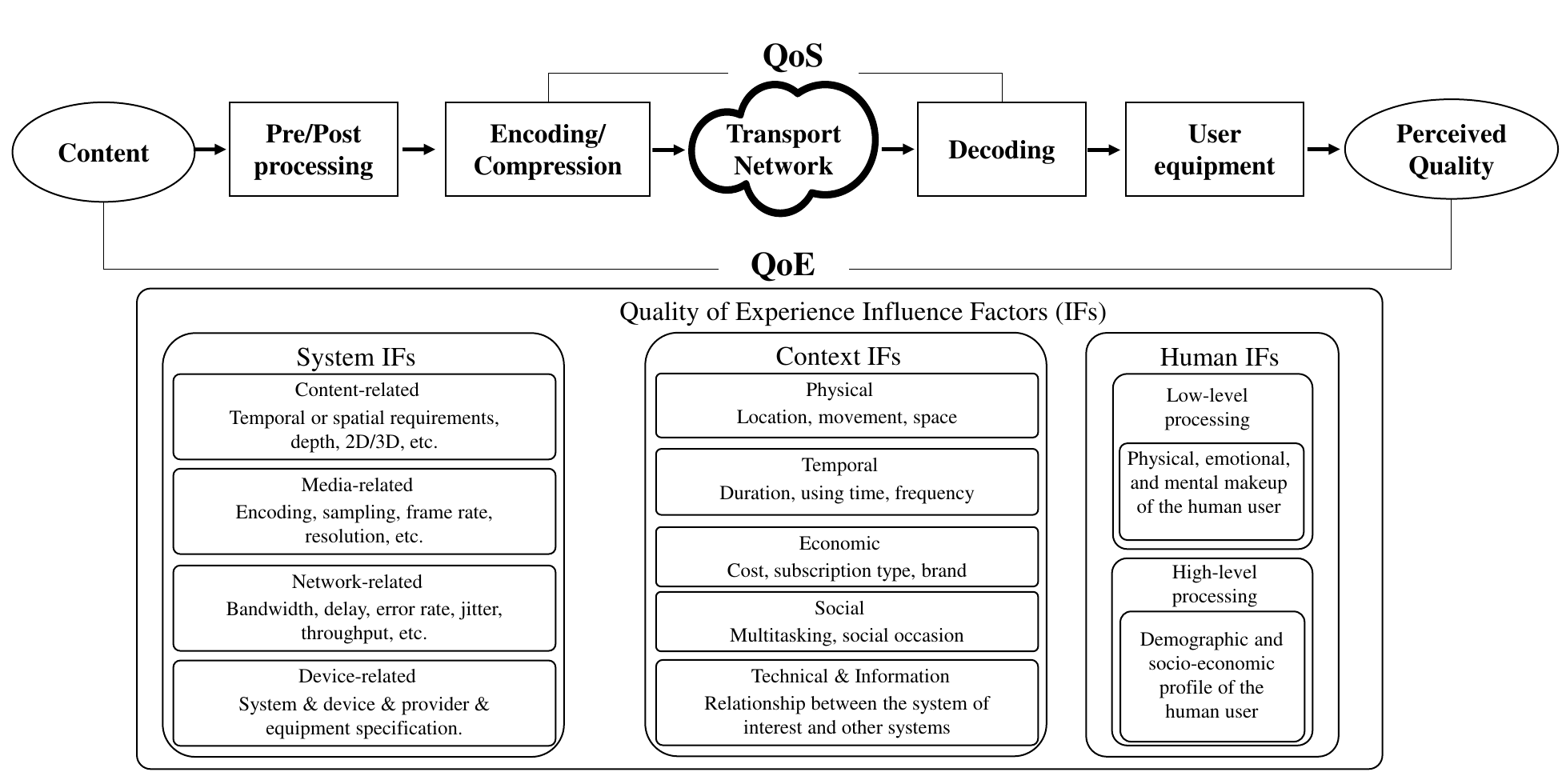}
    \caption{The scope of QoE and QoS}
    \label{qoe}
\end{figure}

\subsection{Streaming VQA}

The variability of streaming environments and the intricacies of human QoE responses have presented significant challenges for delivering optimal content distribution services.
In the past decade, there has been significant effort invested in the development of objective QoE models.

\subsubsection{QoS-driven User QoE Assessment}
The QoS driven user QoE assessment exploits the causal relationship between QoS and QoE problems.
Liu \textit{et al.} \cite{liu2012case} conducted an analysis of the effects of client-side, video coding, and CDN factors on QoE and proposed a video control plane capable of dynamically optimizing video delivery by considering a comprehensive perspective of the above mentioned client and network conditions.
Rodr{\'\i}guez \textit{et al.} \cite{rodriguez2014impact} addressed the impact of frequent video quality level (VQL) switching on QoE through subjective testing, objective modeling, and computer/network configurations. Contributions include identifying the strong impact of frequent VQL switching on users' attention, different impacts of spatial and temporal resolution switchings on QoE, identification of key factors in characterizing VQL switching impact, development of a switching degradation factor model to account for changes in QoE.
In \cite{nightingale2014impact}, Nightingale \textit{et al.} evaluated HEVC video streaming under network impairments, quantifying their impact on perceptual quality and offering insights into influencing factors, thus informing QoE-oriented HEVC streaming development.

\subsubsection{QoS and QA-driven User QoE Assessment}
Despite the diversity in the implementations of QoE models, recent studies have increasingly converged towards utilizing the QoS driven user QoE assessment and the visual quality measurement at the same time.
In \cite{bentaleb2016sdndash}, Bentaleb \textit{et al.} conducted an analysis of the effects of chunk quality, startup delay, number of stalls, average video quality, and video quality switches on QoE and introduced an architecture for dynamic resource allocation and management in DASH systems.
In \cite{duanmu2016quality}, Duanmu \textit{et al.} developed a unified QoE prediction model called Streaming QoE Index (SQI), which considers video presentation quality, initial buffering, and stalling events as combined factors in determining QoE. The SQI model takes into account the overall experience of video quality, stalling events, and their interaction for a more comprehensive QoE assessment.
In \cite{bampis2017learning}, Bampis \textit{et al.} introduced a machine learning framework called Video Assessment of Temporal Artifacts and Stalls (Video ATLAS) for accurately predicting user QoE. The framework combines multiple QoE-related features, including objective quality features, rebuffering-aware features, and memory-driven features, to make reliable QoE predictions.
In \cite{bampis2017continuous}, Bampis \textit{et al.} proposed a machine learning-based Nonlinear Autoregressive Network with Exogenous Inputs (NARX) model, which utilizes objective metrics, rebuffering-related information, and memory-related features for predicting QoE in video streaming.
In \cite{ghadiyaram2018learning}, Ghadiyaram \textit{et al.} developed a QoE evaluation tool, called the time-varying QoE Indexer, which considers interactions between stalling events, analyzes the spatial and temporal content of a video, predicts perceptual video quality, models the state of the client-side data buffer, and provides continuous-time quality scores that are in good agreement with human opinion scores.
In \cite{duanmu2018quality}, Duanmu \textit{et al.} proposed the ECT-QoE. The proposed framework is based on the expectation confirmation theory (ECT) to construct an ECT-based QoE measure (ECT-QoE) that considers spatial and temporal expectation confirmations separately.
The effects of adaptation intensity, adaptation type, intrinsic quality and content type on the end user QoE are considered in the method.
Eswara \textit{et al.} \cite{eswara2019streaming} introduced LSTM-QoE, a new dynamic model that utilizes LSTM networks for predicting continuous QoE. The model incorporates a network of LSTMs optimized for QoE prediction performance using advanced QoE features, and has the potential for real-time QoE computation.
Rao \textit{et al.} \cite{rao2022avqbits} proposed a bitstream-based video quality model that utilizes both metadata such as codec type, framerate, resoution and bitrate as well as the video pixel information.
Duanmu \textit{et al.} \cite{duanmu2023bayesian} proposed a Bayesian streaming quality index (BSQI) model that integrates prior knowledge on the human visual system and human annotated data in a principled manner to predict objective QoE. Through analysis of subjective characteristics in streaming videos from subjective studies, authors demonstrated that a family of QoE functions follows a convex set, and they optimized the BSQI model using a variant of projected gradient descent over a training video database.

\begin{figure}[t]
    \centering
    \includegraphics[width=0.98\textwidth]{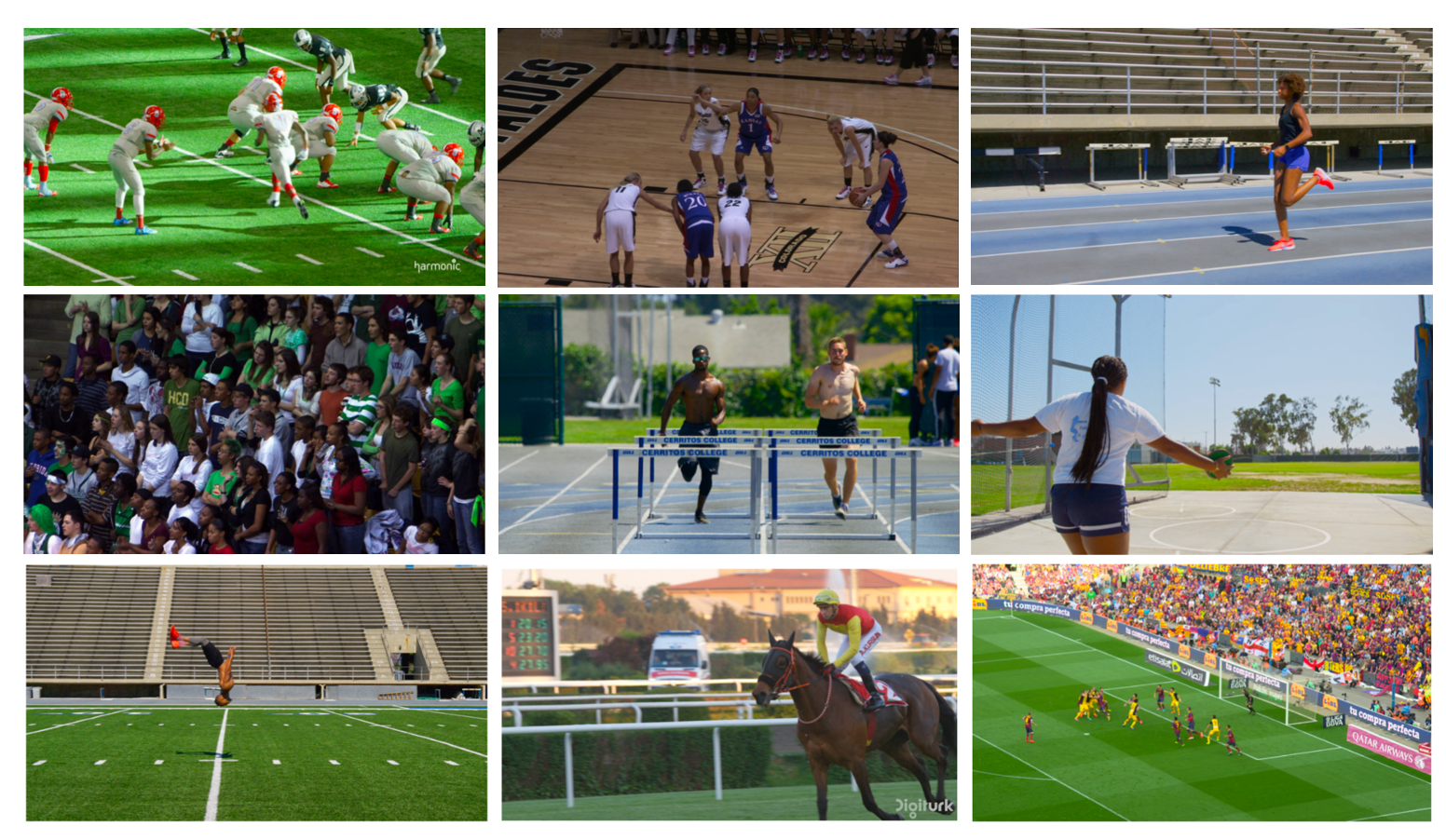}
    \caption{Sample frames of the video contents in the LIVE-APV Livestream Video Quality Assessment Database \cite{shang2021study}.}
    \label{25ERP}
\end{figure}

\subsubsection{Data-driven Approaches}
Another type of model utilizes data-driven approaches, employing machine learning models like random forest and neural network, which impose noninformative priors on the model parameters to achieve effective results.
In \cite{singh2012quality}, Singh \textit{et al.} developed a no-reference QoE monitoring module for HTTP/TCP video streaming using H.264/AVC video codec in the context of IPTV. The proposed approach utilizes pseudo-subjective quality assessment (PSQA) methodology based on random neural network (RNN), considering the quantization parameter (QP) used in video compression and playout interruptions as metrics impacting QoE, as these factors are directly related to perceived quality in adaptive HTTP streaming.
In \cite{li2022whole}, Li \textit{et al.} proposed a novel weakly-supervised domain adaptation approach for continuous-time QoE evaluation, utilizing a small amount of labeled data in the source domain and weakly-labeled data (retrospective QoE labels only) in the target domain. The approach involves learning effective spatiotemporal segment-level feature representations using a combination of 2D and 3D convolutional networks, and developing a multi-task prediction framework that simultaneously predicts continuous-time and retrospective QoE.

\subsection{Stereoscopic VQA}
The advancement of 3D movies and TV programs has popularized stereoscopic or 3D Video Quality Assessment. Research in this area holds both theoretical and practical significance, as the current state of 3D content, capture, and display devices still have considerable room for improvement in terms of delivering optimal visual experiences.

\subsubsection{2D Extension Methods}
The quality of stereoscopic 3D videos can be assessed by utilizing established algorithms for image quality assessment and video quality assessment that are traditionally used for 2D content. These models employ IQA and VQA algorithms on the distinct views, including the disparity view, of stereoscopic 3D videos. Typically, these IQA and VQA models are applied at the level of individual frames or views to estimate the perceptual quality of a stereoscopic 3D video.

In \cite{yasakethu2008quality}, Yasakethu \textit{et al.} explored the correlation between subjective quality measures and various objective quality measures, such as PSNR, SSIM, in the context of 3D video content.
In \cite{nur2011extended}, Nur \textit{et al.} utilized classical 2D algorithms by directly applying them to each frame of the stereoscopic video and obtaining an average predicted quality as the global quality score for the stereoscopic video.
In order to address the systematic deviation in quality prediction for asymmetric distortion in stereo videos using weighted average 2D evaluation methods, Wang \textit{et al.} \cite{wang2017asymmetrically} proposed a dynamic weight method based on binocular rivalry theory. The weighting strategy combines the local energy information of image patches and integrates the prediction quality of left and right videos, resulting in improved performance for existing FR quality evaluation methods.
In \cite{hong2017spatio}, Hong \textit{et al.} proposed the 3-D-PQI metric to quantify video compression distortion in stereoscopic videos. The proposed model incorporates the measurement of local video compression distortions in both the spatial and temporal domains for both the left and right views, taking into consideration the contrast and motion masking effects. To accumulate these local spatial and temporal distortions, a stereo saliency-based pooling strategy is employed. Finally, the 3-D-PQI is derived through a texture energy-based fusion of the distortion measurements obtained from the left and right views.

\subsubsection{Stereo Vision Perception Methods}
In addition to the 2D extension methods, researchers have explored stereo vision perception methods. Several full-reference models for 3D VQA have also been developed.
In \cite{galkandage2016stereoscopic}, Galkandage \textit{et al.} proposed a FR metric for stereoscopic video quality assessment. In the proposed model, binocular suppression and recurrent excitation are considered. A novel image quality metric based on the HVS is proposed. The metric is extended to the video domain by introducing an optimized temporal pooling strategy.
Appina \textit{et al.} \cite{appina2018full} proposed the DeMo3D model, where the Bivariate Generalized Gaussian Distribution is employed to measure the correlation between motion and disparity maps at three different scales and six directions. Subsequently, 2D evaluation methods are applied to extract spatial features to predict the quality of stereo videos.
In \cite{zhang2019sparse}, Zhang \textit{et al.} proposed a FR VQA method for synthesized 3D videos. The method involves decomposing the synthesized video into spatially neighboring temporal layers, using gradient features and strong edges of depth maps to detect flicker distortions, and applying dictionary learning and sparse representation to effectively represent temporal flicker distortion. A rank pooling method is then used to combine the temporal flicker distortion measurement with conventional spatial distortion measurement for overall quality assessment of synthesized 3D videos.
In \cite{galkandage2020full}, Galkandage \textit{et al.} proposed a FR VQA model. The model consider the motion sensitivity of HVS and extract both non-motion sensitive and motion sensitive energy terms to mimic the response of the HVS.

Researchers have also made progress in the development of reduced-reference models for 3D VQA. These models aim to efficiently assess the visual quality of 3D content while using only a limited set of reference information.
In \cite{hewage2011reduced}, Hewage \textit{et al.} proposed a RR quality metric for color plus depth 3D video transmission. The metric utilizes edge information from depth maps and corresponding color images in the areas near edges to assess video quality.
Yu \textit{et al.} \cite{yu2016binocular} proposed a RR VQA model. The proposed method uses motion intensity to extract RR frames for temporal characteristics in stereo video. Binocular fusion and rivalry portions are modeled based on the internal generative mechanism of human visual perception. RR frame quality indicators are computed for these portions, and then compared between the original and distorted frames. A temporal pooling strategy, with motion intensity influencing pooling parameters, is applied to obtain the final stereo video quality score.

No-reference models for 3D VQA enable the assessment of quality without relying on any reference information, making them particularly valuable for real-world scenarios.
In \cite{chen2017blind}, Chen \textit{et al.} proposed a NR VQA model. In the proposed method, auto-regressive prediction-based disparity entropy (ARDE) and energy weighted video content measurement features are introduced, inspired by the free-energy principle and binocular vision mechanism. Binocular summation and difference operations are combined with natural scene statistic measurement and ARDE measurement to assess the impact of texture and disparity in video quality evaluation.
Yang \textit{et al.} \cite{yang2018no} proposed a NR VQA model. In the model, the sum map is calculated that remains basic information of the 3D video. Then saliency map and sparse coefficients are calculated on the sum map to predict the video quality.
In \cite{yang2018stereoscopic}, a NR VQA model was proposed. The model is built on 3D CNN to extract local spatiotemporal information and global temporal information. The global temporal clues are considered in the quality fusion.
In \cite{yang2019no}, Yang \textit{et al.} proposed a NR VQA model. In the model, key frame sequences are extracted. The binocular summation and difference are calculated on extracted sequences, and then texture statistic measurement are conducted to predict the 3D video quality.

Statistical dependencies between motion and disparity information are employed in some methods.
In \cite{appina2019study}, Appina \textit{et al.} proposed a NR VQA model called MoDi3D. The BGGD parameters of the joint statistical dependencies between motion and disparity subband coefficients are estimated as the features, which are pooled to predict the 3D video quality.
In \cite{biswas2022jomodevi}, Biswas \textit{et al.} proposed a NR VQA model of stereoscopic 3D videos. In the model, the correlation between the motion and depth components is computed and represented as a correlation map. The correlation maps are then subjected to steerable pyramid decomposition at various scales and orientations. The resulting subband decompositions of the correlation map are modeled using the UGGD models. The parameters of the UGGD model are estimated to predict the quality of the video content.


\subsection{VR VQA}
The metrics for omnidirectional videos (ODV) need to consider the unique aspects of ODV, such as the spherical nature and viewing characteristics, and often address projection distortions through distortion weights or resampling techniques. Distortion weights in ODV metrics are determined based on the level of projection distortion at a specific location, while resampling techniques may involve extracting viewports with low projection distortions, converting ODV into a projection format with low distortions, or extracting uniformly distributed points on the sphere. 
To this end, many traditional technique-based and deep learning-based methods have been proposed for the VR VQA problem.

\subsubsection{Traditional Visual Computing Techniques}
Traditional visual computing techniques have been developed and applied for the purpose of assessing the quality of VR videos.
In \cite{sun2017weighted}, Sun \textit{et al.} proposed a FR VQA method. The method involves multiplying the error of each pixel on projection planes by a weight to ensure equivalent spherical area in observation space, thereby avoiding error propagation caused by conversion from resampling representation space to observation space and improving the accuracy and reliability of quality evaluation results.
In \cite{zakharchenko2016quality}, Zakharchenko \textit{et al.} introduced a new location invariant quality assurance metrics for spherical panoramic images/videos. Two methods are proposed: using PSNR in the Craster parabolic projection format or weighted PSNR calculation for ERP contents.
In \cite{yu2015framework}, Yu \textit{et al.} proposed a FR method named S-PSNR. The spherical PSNR metric (S-PSNR) estimates the PSNR for uniformly sampled points on the sphere, but the number of sampled points in the official implementation is too small compared to the resolution of ODVs, resulting in massive information loss. S-PSNR has two variants, S-PSNR-NN and S-PSNR-I, which use nearest neighbor or bicubic interpolation for pixel sampling, respectively.
In \cite{zhou2018weighted}, Zhou \textit{et al.} proposed a FR method called Weighted-to-Spherically-Uniform SSIM for evaluating the objective quality of panoramic video and images, where the structural similarity index is multiplied by different weights in different regions to ensure that spherical distortion corresponds linearly to plane distortion as observed by the user.
In \cite{chen2018spherical}, Chen \textit{et al.} presented a FR quality assessment method for omnidirectional video based on structural similarity in the spherical domain, taking into account the relationship between the structural similarity in the 2D plane and the sphere, which helps to handle the interference caused by projection in the assessment process.
In \cite{ozcinar2019visual}, Ozcinar \textit{et al.} proposed a FR quality metric based on PSNR that takes into consideration visual attention and projection distortions, with the objective of optimizing streaming of omnidirectional video.
In \cite{meng2021viewport}, Meng \textit{et al.} proposed a RR analytical model to connect the perceptual quality of compressed viewport videos with their spatial, temporal, and amplitude resolutions variables, using linearly weighted content features. Additionally, the model is extended to infer the overall video quality by weighing the saliency-aggregated qualities of salient viewports and the quality of non-salient areas.

In certain methods, human visual perceptual regularities and natural scene statistics are effectively utilized to enhance video quality assessment techniques.
In \cite{xu2018assessing}, Xu \textit{et al.} proposed the FR VQA methods for encoded omnidirectional video, taking into consideration human perception characteristics. One method weighs pixel distortion based on their distances to the center of front regions, accounting for human preference in panoramic viewing, while the other method predicts viewing directions from the video content and allocates weights to pixel distortion accordingly in our VQA method.
In \cite{gao2020quality}, Gao \textit{et al.} proposed a FR spatiotemporal modeling approach for evaluating the quality of omnidirectional videos. The approach involves constructing a spatiotemporal quality assessment unit that evaluates distortion at the eye fixation level, incorporating temporal variations to obtain smoothed distortion values. The paper also presents a solution for integrating existing spatial video quality metrics, as well as investigating cross-format omnidirectional video distortion measurement.
In \cite{azevedo2021multi}, Azevedo \textit{et al.} proposed a FR approach for assessing the quality of omnidirectional videos. This approach uses viewports regularly sampled from ODV frames with low projection distortions to better capture the user experience, supports different ODV projection formats, and applies different spatio-temporal metrics combined with a model of human visual system's temporal quality perception for computing the final quality score using a random forest regression trained on the VQA-ODV dataset.
In \cite{zhou2021no}, Zhou \textit{et al.} proposed a NR algorithm called MultiFrequency Information and Local-Global Naturalness (MFILGN). The approach decomposes the projected equirectangular projection maps into wavelet subbands using discrete Haar wavelet transform, and measures multifrequency information using entropy intensities of low-frequency and high-frequency subbands. The natural scene statistics features are extracted from each viewport image to measure local naturalness. The support vector regression is used to train the quality evaluation model.

\begin{figure}[t]
    \centering
    \includegraphics[width=0.98\textwidth]{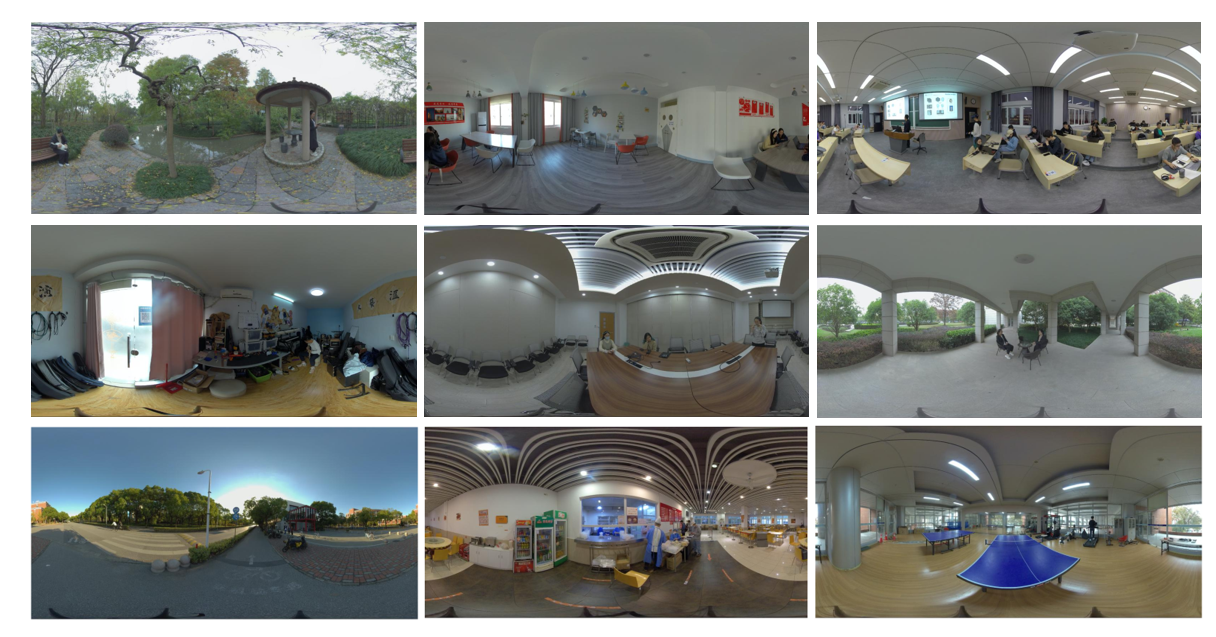}
    \caption{EPR format previews of sample ODVs in OAVQAD \cite{zhu2023perceptual}.}
    \label{25ERP}
\end{figure}

\subsubsection{Deep Learning-based Computing Techniques}
In recent years, in addition to traditional visual computing techniques, metrics based on deep learning have emerged and demonstrated state-of-the-art performance in video quality assessment. These deep learning-based methods encompass both FR and RR approaches.
In \cite{li2019viewport}, Li \textit{et al.} proposed a FR VQA approach that incorporates viewport proposal and saliency prediction as auxiliary tasks. The proposed approach consists of two stages - the first stage involves a viewport proposal network to generate potential viewports, and the second stage includes a Viewport Quality Network that rates the VQA score for each proposed viewport using predicted saliency maps.
In \cite{xu2020viewport}, Xu \textit{et al.} presented a FR approach using a viewport-based convolutional neural network (V-CNN) for VQA on $360^{\circ}$ videos. The V-CNN includes a multi-task architecture with a viewport proposal network for handling camera motion detection and viewport proposal, and a viewport quality network for handling viewport saliency prediction and the main VQA task.
In \cite{duan2023attentive}, Duan \textit{et al.} developed the FR metric that can predict the distortion caused by stitching in VR contents. The proposed method incorporates a subnetwork for spatial attention and introduces a spatial regularization component.
In the field of VR VQA, it is noticeable that utilizing saliency information is a widespread practice. Many emerging VR saliency methods have been introduced \cite{zhu2023toward, zhu2021viewing, zhu2020learning, zhu2019prediction, zhu2018prediction, li2022sound, yang2021salgfcn, ren2023children} to aid in predicting visual attention information. These methods play a crucial role in identifying the most relevant and visually significant regions within virtual environments, enhancing the overall VR experience and improving the accuracy of VQA tasks.

Besides, there are also video quality assessment metrics developed in the NR manner.
In \cite{yang20183d}, Yang \textit{et al.} proposed a NR approach for predicting VR video quality, using an end-to-end 3D convolutional neural network that extracts spatiotemporal features. The score fusion strategy is designed based on the characteristics of VR video projection, where local spatiotemporal features are captured from pre-processed VR video patches and combined to obtain the final quality score.
In \cite{fei2019qoe}, Fei \textit{et al.} proposed a NR two-step neural network model, leveraging features from physiological psychology and cognitive neurology, to capture the relationship between network parameters and perception in VR transmission for objective evaluation.
In \cite{yang2020panoramic}, Yang \textit{et al.} proposed a NR approach for predicting VR video quality. The proposed method combines spherical convolutional neural networks and non-local neural networks to extract spatiotemporal information from panoramic videos.
In \cite{xu2020blind}, Xu \textit{et al.} proposed a NR model. The proposed method includes a viewpoint detector to select viewports based on human visual system sensitivity, a viewport descriptor for feature extraction, and a spatial viewport graph to model mutual dependency among viewports. Graph convolutional networks are used for reasoning on the graph to obtain the global quality of the omnidirectional image, omitting the pseudo reconstruction step for simplicity and performance enhancement.
In \cite{guo2021no}, Guo \textit{et al.} proposed a NR omnidirectional video quality assessment approach based on generative adversarial networks, consisting of a reference video generator and a quality score predictor. To address the issue of varying reference image/video quality levels in existing GAN-based methods, a level loss is introduced, and the viewing direction of the omnidirectional video is incorporated in the quality and weight regression process.
In \cite{zhu2022eyeqoe}, Zhu \textit{et al.} proposed a NR approach. The proposed EyeQoE method is inspired by advanced techniques in deep neural networks and uses a graph-based approach to model eye-based cues for video quality assessment. The method organizes fixations and saccades into a graph, where edges represent temporal relations and additional edges are added for content-dependent features. A graph convolution network is used to learn useful feature representations from the graph, which are then used to compute the quality of the video clip.
In \cite{yang2022blind}, Yang \textit{et al.} proposed a NR approach called ProVQA for quality assessment of $360^{\circ}$ videos, taking into account the progressive paradigm of human perception. Three sub-nets are designed in ProVQA: the spherical perception aware quality prediction sub-net models spatial quality degradation based on human spherical perception mechanism, the motion perception aware quality prediction sub-net incorporates motion contextual information for quality assessment, and the multi-frame temporal non-local sub-net aggregates multi-frame quality degradation to yield the final quality score.
In \cite{an2022panoramic}, An \textit{et al.} proposed a method that uses both 2D-CNN and 3D-CNN to extract video features in both temporal and spatial domains. The input video is divided into patches and processed through convolutional, excitation, pooling, and fully connected layers to obtain a score for the video.

\subsection{Framerate \& Frame Interpolation VQA}

Altering the frame rate of a video can significantly influence its visual quality. Lower frame rates might introduce choppiness and reduced motion smoothness, while higher frame rates can enhance the viewing experience with improved clarity and realism. As a result, specific framerate VQA methods become essential to evaluate and ensure the perceptual quality of videos across different frame rates, helping content creators, streaming platforms, and viewers make informed decisions about frame rate selection to achieve the best visual experience.
In \cite{ou2010perceptual, ma2011modeling}, Ma \textit{et al.} proposed FR rate and quality model, which is analytically tractable and relys on content-dependent parameters, combines a spatial quality factor assessing decoded frames' quality and a temporal correction factor adjusting for the frame rate.
In \cite{ou2014q}, Ou \textit{et al.} explored the impact of spatial, temporal, and amplitude resolution on video's perceptual quality and related reductions in frame rate to perceptual quality through subjective and objective analyses.
Zhang \textit{et al.} proposed the FR method FRQM in \cite{zhang2017frame}. The method evaluates the relationship between frame rate variations and perceptual video quality. FRQM utilizes temporal wavelet decomposition, subband combination, and spatiotemporal pooling to estimate the relative quality of low frame rate videos compared to higher frame rate versions.
In \cite{madhusudana2021st}, Madhusudana \textit{et al.} proposed the objective VQA model, called Space-Time GeneRalized Entropic Difference (GREED), analyzes spatial and temporal band-pass video coefficient statistics using a generalized Gaussian distribution. GREED captures quality variations due to frame rate changes by calculating entropic differences across multiple temporal and spatial subbands
In \cite{madhusudana2021high}, Madhusudana \textit{et al.} focused on VQA for High Frame Rate videos with different frame rates and compression factors. They proposed a FR model that combines features from VMAF and GREED, offering improved efficiency in predicting frame rate dependent video quality.
Lee \textit{et al.} \cite{lee2022video} developed a FR video quality predictor sensitive to spatial, temporal, or space-time subsampling combined with compression. The predictor utilizes space-time natural video statistics models to capture regularities in motion trajectories and disturbances caused by space-time distortions.
In \cite{zheng2022no}, Zheng \textit{et al.} introduced FAVER, a NR VQA model tailored for high frame rate videos, utilizing the temporal natural video statistics of bandpass filtered videos to capture and represent aspects of temporal video quality.

Video frame interpolation results often show unique artifacts, which can lead to inconsistencies between existing quality metrics and human perception when assessing the interpolation outcomes.
In \cite{yang2008new}, Yang \textit{et al.} proposed a FR metric that quantifies interpolation artifacts, incorporates human visual factors, and provides a global quality measurement. The proposed metric takes into account blocking artifacts and potential areas of quality degradation to overcome the limitations of other commonly used metrics.
In \cite{danier2022flolpips}, Danier \textit{et al.} proposed FloLPIPS, a FR video quality metric for VFI, based on LPIPS, which incorporates temporal distortion through optical flow comparison to enhance performance.
In \cite{hou2022perceptual}, Hou \textit{et al.} proposed a dedicated and FR perceptual quality metric that learns features directly from videos and considers spatio-temporal information using Swin Transformer blocks.

\begin{figure}[t]
    \centering
    \includegraphics[width=0.98\textwidth]{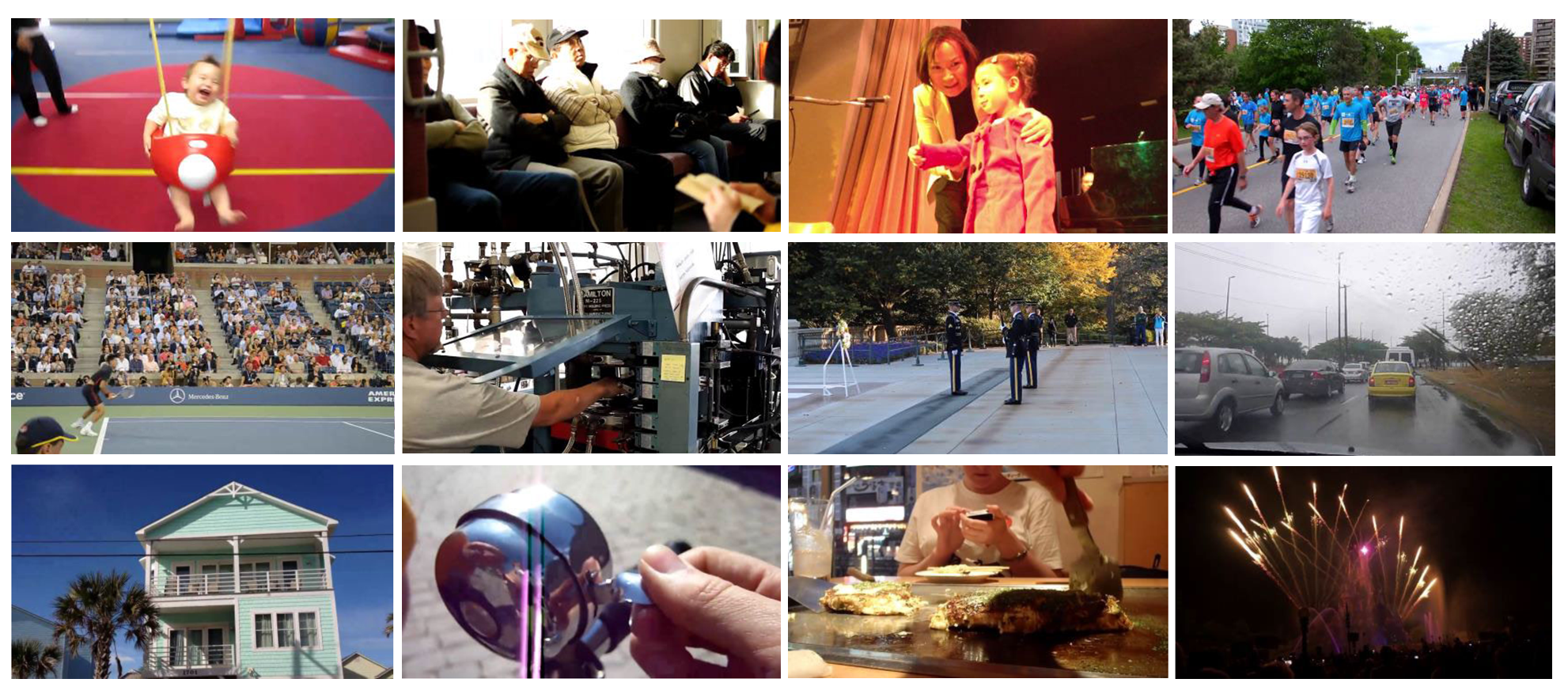}
    \caption{ Sample frames of the video contents in the SJTU-UAV database \cite{cao2023subjective}.}
    \label{25ERP}
\end{figure}

\subsection{Audio-Visual VQA}

With the increasing prevalence of mobile Internet, audio and video (A/V) are essential for everyday entertainment and social interactions. However, compression of A/V signals by service providers to reduce storage and transmission costs can result in distortions, negatively impacting end-users' QoE. Therefore, AVQA is a significant and attention-worthy area of research.

Most previous research has primarily focused on single-mode signals, overlooking the comprehensive impact of audio and video on consumers' QoE. Some studies have started to address the objective AVQA problems, recognizing the importance of jointly assessing the audio-visual aspects of multimedia content to enhance user satisfaction.
In \cite{beerends1999influence, hands2004basic}, researchers emphasized the significance of audiovisual quality and suggested that the overall audiovisual quality can be represented as a product of audio and video quality.
In \cite{winkler2006perceived}, Winkler \textit{et al.} conducted subjective experiments to assess audiovisual, audio-only, and video-only quality. The study analyzed the impact of video and audio coding parameters on quality, explored the optimal balance between audio and video bit allocation under global bitrate constraints, and investigated models for the interactions between audio and video in terms of perceived audiovisual quality.
In \cite{martinez2014full}, Martinez \textit{et al.} proposed a FR audio-visual quality metric. The FR audio-visual quality framework introduced three models based on the findings of psychophysical experiments: the linear model, the weighted Minkowski model, and the power model. These models offer different approaches to quantify the overall audio-visual quality based on audio and video components.
In the study by Martinez \textit{et al.} \cite{martinez2014no}, the three perceptual audio-visual models (linear, weighted Minkowski, and power models) were used to combine video and audio no-reference metrics. These combined metrics were then tested and evaluated in the research.
In \cite{martinez2018combining}, Martinez \textit{et al.} explored combination models to predict overall audio-visual quality by integrating audio and video quality estimates. It considers 7 video quality metrics (3 Full-Reference and 4 No-Reference) and 4 audio quality metrics (2 Full-Reference and 2 No-Reference), resulting in 18 Full-Reference and 24 No-Reference audio-visual combination metrics.
In \cite{min2020study}, four families of objective A/V quality prediction models were designed using a multimodal fusion strategy: product of video and audio quality predictors, fusion of video and audio quality predictors by SVR, A/V-QA models defined using 1D and 2D visual quality predictors, and deep neural families of A/V quality predictors.
In \cite{cao2023subjective}, a NR model was proposed. The model extracts audio features from the separable convolution network and visual features from the quality-aware ResNet-50, and learns temporal information through Bi-LSTM and fuses the features using FC layers.
In \cite{cao2023attention}, Cao \textit{et al.} proposed an objective model architecture based on attentional neural networks to consider both audio and video signals. The extended FR and NR models extract salient regions from video frames using an attention prediction model, utilize convolutional neural networks to extract short-time features, and employ gated recurrent unit (GRU) networks to model temporal relationships.

\begin{table}[t]
  \centering
  \renewcommand{\arraystretch}{1}
  \setlength{\tabcolsep}{0.4em}
  \caption{Overview of the objective video quality assessment for emerging topics}
  \resizebox{1\textwidth}{!}{
    \begin{tabular}{cccccc}
    \toprule
    \textbf{Applicable content} & \textbf{Type} & \textbf{Algorithm} & \textbf{Methodology} & \textbf{Extracted quality features} & \textbf{Quality fusion} \\
    \midrule
    \multirow{7}[6]{*}{Compressed VQA} & \multirow{2}[2]{*}{FR} & FePVQ \cite{xu2016free} & spatio-temporal similarity &  Motion, structure, and texture strength  & Weighted feature similarity \\
          &       & Sun \textit{et al.} \cite{sun2021deep} & spatio-temporal similarity & CNN features & CNN \\
\cmidrule{2-6}          & RR    &  Ma \textit{et al.} \cite{ma2012reduced} & spatio-temporal similarity & Energy and motion features &  Histogram distance \\
\cmidrule{2-6}          & \multirow{4}[2]{*}{NR} & Lin \textit{et al.} \cite{lin2012no} & spatio-temporal factors & QP, motion, bit allocation factors & Weighted average \\
          &       & Zhu \textit{et al.} \cite{zhu2014no} & spatio-temporal statistics & Frequency band features & CNN \\
          &       & V-MEON \cite{liu2018end} & DNN   &  CNN features & CNN \\
          &       & SSTAM \cite{lin2023saliency} & spatio-temporal features & Perceivable encoding artifacts & SVR \\
    \midrule
    \multirow{11}[4]{*}{Streaming VQA} & \multirow{9}[2]{*}{FR} & Liu \textit{et al.} \cite{liu2012case} & qiality of service factors & Client-side, video coding and CDN factors & Weighted average \\
          &       & Rodr{\'\i}guez \textit{et al.} \cite{rodriguez2014impact} & qiality of service factors & Video quality levels switching degradation factor & Weighted average \\
          &       & Bentaleb \textit{et al.} \cite{bentaleb2016sdndash} & content and qiality of service factors & Delay, stall, video quality and quality switch & Weighted average \\
          &       & SQI \cite{duanmu2016quality} & content and qiality of service factors & Video presentation quality, buffering, stalling & Weighted average \\
          &       & Video ATLAS \cite{bampis2017learning} & content and qiality of service factors & Video presentation quality, buffering, memory & Weighted average \\
          &       & Ghadiyaram \textit{et al.} \cite{ghadiyaram2018learning} & content and QoS factors, continuous time & Stalling, client buffering, video presentation & Wiener model and SVR \\
          &       &  ECT-QoE \cite{duanmu2018quality} &  Expectation confirmation theory & Video quality, adaptation type and intensity & Random forest regression \\
          &       &  LSTM-QoE \cite{eswara2019streaming} & content and QoS factors, continuous time & Video quality, playback indicator, rebuffering & CNN \\
          &       & BSQI \cite{duanmu2023bayesian} & content and qiality of service factors & Video presentation quality, buffering, adaptation & Piecewise linear \\
\cmidrule{2-6}          & \multirow{2}[2]{*}{NR} & Singh \textit{et al.} \cite{singh2012quality} & DNN   &  CNN features & CNN \\
          &       & Li \textit{et al.} \cite{li2022whole} & DNN, continuous time &  CNN features & CNN \\
    \midrule
    \multirow{11}[6]{*}{3D VQA} & \multirow{6}[2]{*}{FR} & Yasakethu \textit{et al.} \cite{yasakethu2008quality} & monocular spatial similarity & PSNR, SSIM, VQM & Weighted average \\
          &       & Wang \textit{et al.} \cite{wang2017asymmetrically} & monocular spatial similarity & 2D metrics, energy estimation & Binocular rivalry \\
          &       & Galkandage \textit{et al.} \cite{galkandage2016stereoscopic} & binocular spatial, frequency similarity & HVS features & Designed fusion function \\
          &       & DeMo3D \cite{appina2018full} & binocular spatio-temporal similarity & motion, depth, and spatial features & Designed fusion function \\
          &       & SR-3DVQA \cite{zhang2019sparse} & binocular spatio-temporal similarity &  gradient,  edges of depth maps & Weighted layer pooling \\
          &       & Galkandage \textit{et al.} \cite{galkandage2020full} & binocular spatio-temporal similarity & HVS features & Two-stage regression \\
\cmidrule{2-6}          & \multirow{2}[2]{*}{RR} & Hewage \textit{et al.} \cite{hewage2011reduced} & binocular spatial similarity & Edges of depth map & Weighted average \\
          &       & Yu \textit{et al.} \cite{yu2016binocular} & binocular spatio-temporal similarity & Statistical features & Designed fusion function \\
\cmidrule{2-6}          & \multirow{3}[2]{*}{NR} & Chen \textit{et al.} \cite{chen2017blind} & binocular spatio-temporal statistics  & Texture and disparity statistical features & SVR \\
          &       & Yang \textit{et al.} \cite{yang2019no} & binocular spatio-temporal statistics  &  Texture statistical features & SVR \\
          &       & Biswas \textit{et al.} \cite{biswas2022jomodevi} & binocular spatio-temporal statistics  & Statistical features & Designed fusion function \\
    \midrule
    \multirow{13}[6]{*}{VR VQA} & \multirow{7}[2]{*}{FR} & Sun \textit{et al.} \cite{sun2017weighted} & spatial similarity & PSNR weighted to sphere & Weighted average \\
          &       & Zakharchenko \textit{et al.} \cite{zakharchenko2016quality} & spatial similarity & Pixel errors weighted to sphere & Weighted average \\
          &       & Ozcinar \textit{et al.} \cite{ozcinar2019visual} & spatio-temporal similarity & PSNR, VMAF weighted to sphere and saliency & Weighted average \\
          &       & Xu \textit{et al.} \cite{xu2018assessing} & spatial similarity & PSNR, SSIM weighted to sphere and saliency & Weighted average \\
          &       & Gao \textit{et al.} \cite{gao2020quality} & spatio-temporal similarity & PSNR weighted to sphere and fixation & Weighted average \\
          &       & V-CNN \cite{xu2020viewport} & DNN   &  CNN features & CNN \\
          &       & Duan \textit{et al.} \cite{duan2023attentive} & DNN   &  CNN features & CNN \\
\cmidrule{2-6}          & RR    & Meng \textit{et al.} \cite{meng2021viewport} & spatio-temporal similarity & spatial, temporal and amplitude resolutions & Weighted average \\
\cmidrule{2-6}          & \multirow{5}[2]{*}{NR} & MFILGN \cite{zhou2021no} & spatial statistics & Statistical features & SVR \\
          &       & Fei \textit{et al.} \cite{fei2019qoe} & DNN   &  CNN features & CNN \\
          &       & Yang \textit{et al.} \cite{yang2020panoramic} & DNN   &  CNN features & CNN \\
          &       & Xu \textit{et al.} \cite{xu2020blind} & DNN   &  CNN features & CNN \\
          &       & Zhu \textit{et al.} \cite{zhu2022eyeqoe} & DNN   &  CNN features & CNN \\
    \midrule
    \multirow{5}[4]{*}{Framerate VQA} & \multirow{4}[2]{*}{FR} & Ou \textit{et al.} \cite{ou2014q} & spatio-temporal similarity & spatial, temporal and amplitude resolutions & Weighted average \\
          &       & FRQM \cite{zhang2017frame} & spatio-temporal similarity & Temporal wavelet decomposition & Designed fusion function \\
          &       & GREED \cite{madhusudana2021st} & spatio-temporal statistics similarity & Statistical features & SVR \\
          &       & Lee \textit{et al.} \cite{lee2022video} & spatio-temporal statistics similarity & Statistical features & SVR \\
\cmidrule{2-6}          & NR    & FAVER \cite{zheng2022no} & spatio-temporal statistics & Statistical features & SVR \\
    \midrule
    \multirow{5}[4]{*}{Audio-Visual VQA} & \multirow{3}[2]{*}{FR} & Martinez \textit{et al.} \cite{martinez2014full} & combination model & SESQA and VQM & Multiple fusion methods \\
          &       & Martinez \textit{et al.} \cite{martinez2018combining} & combination model & Audio and video quality estimates & Multiple fusion methods \\
          &       & Min \textit{et al.} \cite{min2020study} & combination model & Audio and video quality estimates & Multiple fusion methods \\
\cmidrule{2-6}          & \multirow{2}[2]{*}{NR} & Cao \textit{et al.} \cite{cao2023subjective} & DNN   &  CNN features & CNN \\
          &       & Cao \textit{et al.} \cite{cao2023attention} & DNN   &  CNN features & CNN \\
    \midrule
    \multirow{3}[4]{*}{HDR VQA} & \multirow{2}[2]{*}{FR} & HDR-VQM \cite{narwaria2015hdr} & spatio-temporal similarity & Subband errors & Weighted average \\
          &       & HDRMAX \cite{ebenezer2023making} & spatio-temporal similarity & Nonlinear features & SVR \\
\cmidrule{2-6}          & NR    & Hdr-chipqa \cite{ebenezer2023hdr0} & spatio-temporal statistics & Extended BRISQUE and ChipQA & SVR \\
    \midrule
    \multirow{5}[4]{*}{Screen and Game VQA} & \multirow{3}[2]{*}{FR} & MS-RSDS \cite{li2020subjective} & spatio-temporal similarity & Structural features & Designed fusion function \\
          &       & HSFM \cite{zeng2022screen} & spatio-temporal similarity & Screen and natural statistical features & Designed fusion function \\
          &       & SGFTM \cite{cheng2020screen} & spatio-temporal similarity &  Gabor features & Designed fusion function \\
\cmidrule{2-6}          & \multirow{2}[2]{*}{NR} & GAMIVAL \cite{saha2023study} & spatio-temporal statistics & Statistical and CNN features & SVR \\
          &       &  GAME-VQP \cite{yu2022perceptual} & spatio-temporal statistics & Statistical and CNN features & SVR \\
    \bottomrule
    \end{tabular}%
    }
  \label{tab:addlabel}%
\end{table}%

\subsection{HDR, WCG, iTMO and TMO VQA}
Due to rapid advancements in video acquisition, computational imaging, and display technologies, there is a growing interest in high dynamic range videos. HDR videos exhibit differences from SDR videos, which in turn pose new challenges for HDR VQA models.

Several studies have explored the factors that can influence the quality of HDR content.
In \cite{narwaria2015study}, Narwaria \textit{et al.} addressed key challenges in HDR video quality measurement, discussed practical aspects that make it challenging, and presented recent efforts in developing HDR video datasets subjectively annotated for visual quality.
In the study conducted by Shang \textit{et al.} \cite{shang2023subjective}, they explored the impact of live streaming challenges, such as resolution and frame rate crossover, intra-frame pulsing defects, and complex rate-control mode, on the quality of HDR content.
In \cite{athar2019perceptual}, Athar \textit{et al.} explored and analyzed the effects of compression on UHD-HDR-WCG videos. They aimed to understand how various compression techniques and settings influence the visual quality and overall user experience when viewing UHD-HDR-WCG videos.

HDR videos possess unique characteristics that differ from SDR videos, necessitating specialized techniques for HDR VQA models.
In \cite{narwaria2015hdr}, Narwaria \textit{et al.} proposed a FR HDR video quality measure approach that involves steps to convert input luminance to perceived luminance and then analyze the impact of distortions using frequency and orientation subbands, and error pooling through spatio-temporal processing of subband errors.
In \cite{ebenezer2023making}, Ebenezer \textit{et al.} proposed the HDRMAX feature set that enhances VQA algorithms designed for SDR videos, making them more sensitive to distortions in HDR videos and capture distortions in the brightest and darkest parts of videos. The nonlinear processing is designed to derive a set of nonlinear HDRMAX features for both FR and NR VQA models.
In \cite{ebenezer2023hdr0}, Ebenezer \textit{et al.} proposed a NR HDR VQA model. The approach involves a preprocessing step of local expansive nonlinearity that emphasizes distortions at the higher and lower ends of the luma range, allowing for the computation of additional quality-aware features and improves the prediction of HDR content quality using distortion-sensitive natural video statistics features.
In \cite{ebenezer2023hdr}, Ebenezer \textit{et al.} designed a HDR NR VQA algorithm. The proposed method utilizes features that are relevant to both SDR and HDR video quality, as well as features related to motion perception, which are NIQE features, PatchMAX features, HDRMAX features, and space-time features.

In the realm of HDR VQA, there are several works focused on the VQA for Tone Mapping Operators (TMOs), Inverse Tone Mapping Operators (ITMOs), and wide color gamut.
In \cite{melo2014evaluation}, a comparison was made between tone mapped HDR video shown on a tablet and an LCD display, compared to the same HDR video shown simultaneously on an HDR display.
In \cite{eilertsen2016evaluation}, Eilertsen \textit{et al.} provided an overview of various approaches for conducting evaluation of tone mapping operators for HDR video, including experimental setups, input data selection, tone mapping operator choices, and the significance of parameter adjustment for fair comparisons.
In \cite{yeganeh2016objective}, Yeganeh \textit{et al.} proposed a perceptual quality measure to compare different tone mapping operators. They presented a FR quality assessment model for tone-mapped videos that considers structural fidelity, statistical naturalness, and memory effect.
In the study by Mantiuk \textit{et al.} \cite{mantiuk2011hdr}, a model was developed to measure the visual color difference between test and reference HDR images. The model was designed to mimic the visual system's anatomy to improve accuracy in assessing HDR color differences.

\begin{figure}[t]
    \centering
    \includegraphics[width=0.98\textwidth]{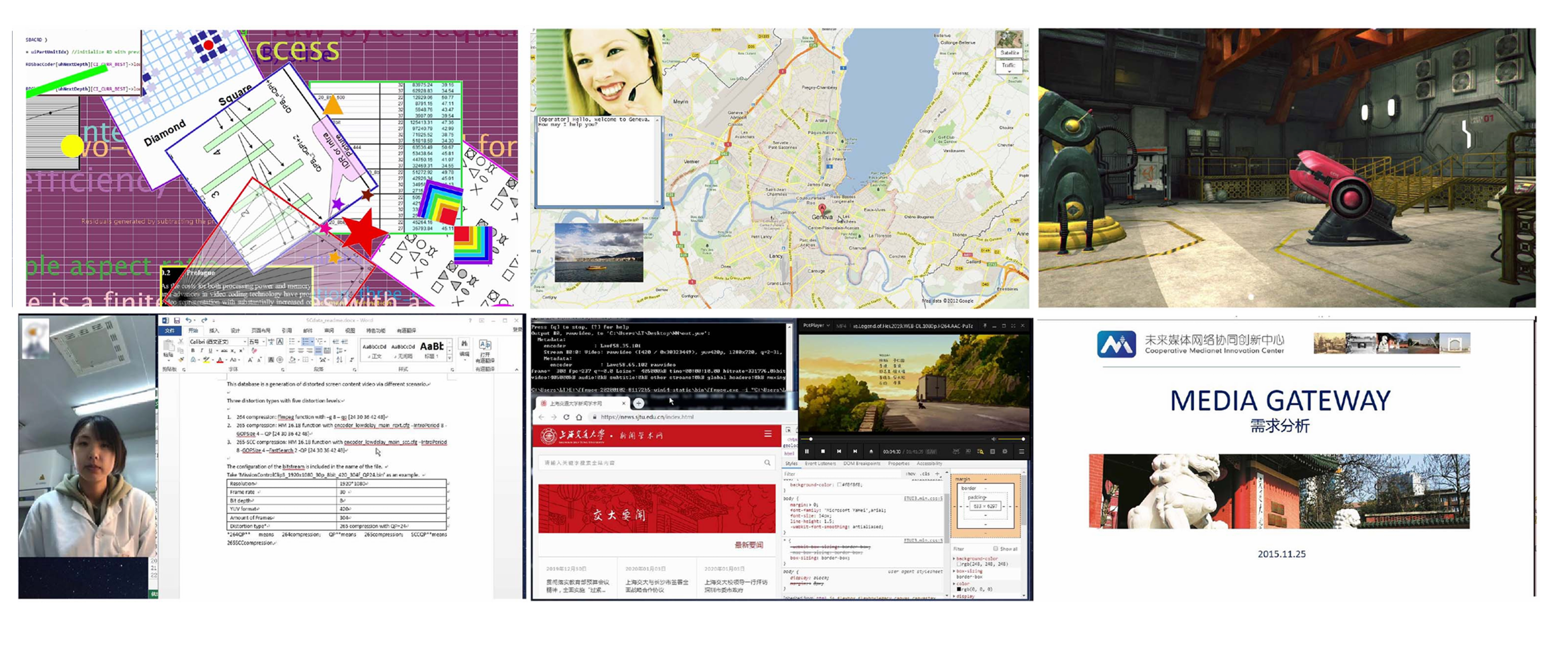}
    \caption{ Sample frames of the video contents in the CSCVQ database \cite{li2020subjective}.}
    \label{25ERP}
\end{figure}

\begin{figure}[t]
    \centering
    \includegraphics[width=0.6\textwidth]{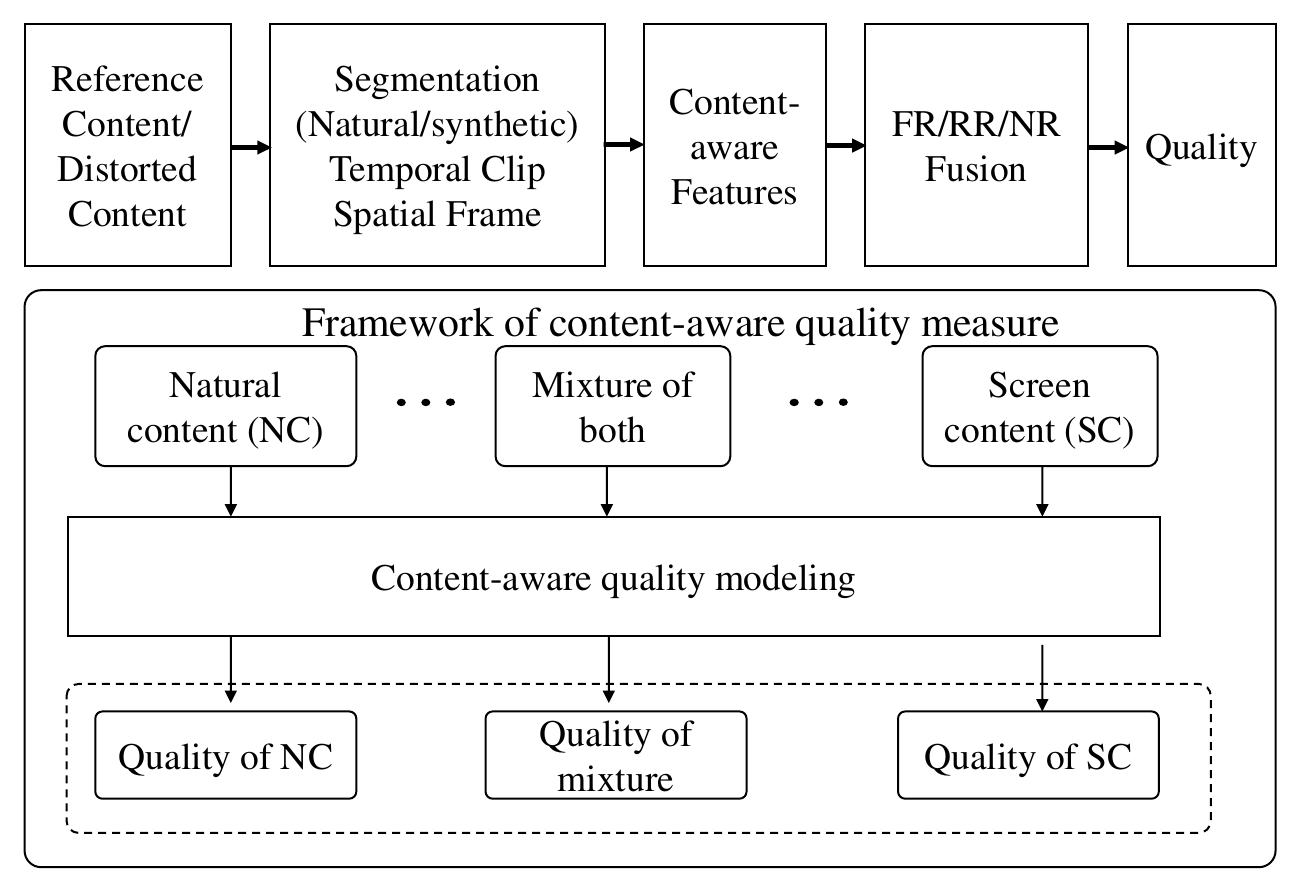}
    \caption{Typical frameworks of screen content video quality assessment}
    \label{screen2}
\end{figure}

\subsection{Screen and Game VQA}

The growing adoption of remote office and cloud collaboration scenarios has led to an increased interest in screen content videos (SCVs) and their processing. SCVs exhibit distinct characteristics from natural scene videos and have become a focus of attention among researchers.
In \cite{li2020subjective}, Li \textit{et al.} proposed a FR screen content VQA model. The proposed approach measures the relative standard deviation similarity between reference and distorted contents using frame differences to capture accurate spatiotemporal distortions and incorporates a multiscale strategy to enhance its performance.
In \cite{li2021no}, Li \textit{et al.} proposed a NR VQA model that utilizes a multi-scale approach to extract several intra-frame features and temporal features and employs support vector regressor for quality score prediction.
In \cite{zeng2022screen}, Zeng \textit{et al.} proposed a FR screen content VQA model. The model utilizes 3D-LOG and 3D-NSS filters to extract spatiotemporal features separately from reference and distorted SCV sequences, then computes similarities and generates quality scores for both screen and natural scenes. An adaptive fusion scheme combining screen and natural quality scores through local video activity is developed to arrive at the final VQA score for the distorted video.
In \cite{cheng2020screen}, Cheng \textit{et al.} proposed the FR VQA model for screen content videos based on the spatiotemporal gabor feature, which leverages 3D-Gabor filter to simulate the human visual system's perception of videos, particularly sensitive to edge and motion information.
In \cite{motamednia2023quality}, Motamednia \textit{et al.} devised the FR objective quality assessment metric for screen content videos. The proposed model utilizes horizontal and vertical subbands of the wavelet transform to characterize the structures present in the video.
In \cite{ying2022telepresence}, Ying \textit{et al.} proposed a NR VQA model for telepresence videos. The proposed model uses a multi-modal learning approach with separate pathways for visual and audio quality predictions. Features of frame-level, patch-level, clip-level, audio-level are extracted and fused to predict the quality of telepresence videos

In recent years, the video game industry has experienced significant growth, leading to a substantial increase in gaming videos on major platforms. Despite this surge, there has been limited research on automatically predicting the quality of gaming videos.
In \cite{xian2022content}, Xian \textit{et al.} proposed a NR VQA method for computer graphics animation videos. The proposed method extracts spatiotemporal features and visual perception information from the videos, which are then fed into an artificial neural network-based VQA model. Additionally, a convolutional neural network is applied to the VQA model to generate adaptive weight factors for the input features based on the different types of CG content in the videos.
In \cite{barman2018evaluation}, Barman \textit{et al.} investigated the performance of VQA metrics on gaming videos. The study considers eight widely used VQA metrics and evaluates their performance on a dataset of reference and compressed gaming videos.
In \cite{saha2023study}, Saha \textit{et al.} presented the outcomes and benchmark results of various FR and NR VQA methods on a large-scale subjective study on mobile cloud gaming.
In \cite{chen2023gamival}, Chen \textit{et al.} proposed a NR VQA model for gaming content. The proposed model combines spatial and temporal gaming distorted scene statistics models, a neural noise model, and deep semantic features.
In \cite{yu2023subjective, yu2022perceptual}, Yu \textit{et al.} proposed a NR VQA model for ugc gaming content. The model includes feature extraction, regression modeling, and score fusion modules. The feature extraction module computes low-level NSS features and high-level features from a pre-trained CNN model using training and test videos. Two separate SVR models are trained on the NSS and CNN features, respectively, as they represent different processing stages. The final video quality predictions are obtained by fusing the responses of these two models.

\section{Objective Video Quality Assessment Model Evaluation}
\label{sec:evaluation}

\subsection{Evaluation Criteria}
With the advancement of information technology research, many objective quality assessment models have been proposed in these years, thus it is important to consider how to evaluate the performance of an objective model.
On account of the reliability and accuracy of subjective quality assessment, its results are generally used as the verification criteria and optimization targets for objective quality evaluation methods.
As suggested by Video Quality Experts Group (VQEG) \cite{VQEG}, we can evaluate the performance of an objective model from the aspects of accuracy, monotonicity and consistency.
Using $o_i$ and $s_i$ to represent a subjective opinion score and an objective predicted score, respectively, where $i = 1, ..., N$ indicates video index, $N$ denotes the number of all videos, we first use a five-parameter logistic function to fit the quality scores:
\begin{equation}
    q(s)=\beta_{1}(\frac{1}{2}-\frac{1}{1+e^{\beta_{2}(s-\beta_{3})}})+\beta_{4}s+\beta_{5},
\end{equation}
where $s$ and $q(s)$ are the objective and best-fitting quality, $\beta_{i}(i=1,2,3,4,5)$ are the parameters to be fitted during the evaluation.
Then five traditional evaluation metrics are usually adopted to measure the consistency between the ground-truth (GT) subjective ratings and the fitted quality scores, including:
\begin{itemize}
    \item Spearman Rank-order Correlation Coefficient (SRCC)
            \begin{equation}
                \text{SRCC} = 1 - \frac{6\sum_{i=1}^{N}d_i^2}{N(N^2-1)}, 
            \end{equation}
          where $d_i$ indicates the difference value between the subjective and objective scores for the $i$-th video, $N$ denotes the number of all test videos.
    \item Kendall Rank-order Correlation Coefficient (KRCC)
            \begin{equation}
                \text{KRCC} = \frac{N_c-N_d}{\frac{1}{2}N(N-1)}, 
            \end{equation}
          where $N_c$ indicates the number of concordant pairs and $N_d$ denotes the number of discordant pairs.
    \item Pearson Linear Correlation Coefficient (PLCC)
            \begin{equation}
                \text{PLCC} = \frac{\sum_{i}^{N}(q_i-\bar{q}) \cdot (o_i-\bar{o})}{\sqrt{\sum_{i}^{N}(q_i-\bar{q})^2 \cdot (o_i-\bar{o})^2}},
            \end{equation}
          where $o_i$ and $q_i$ represent the subjective opinion score and the nonlinear-fitted objective score for the $i$-th video, $\bar{o}$ and $\bar{q}$ indicate the mean values of all $o_i$ and $q_i$ scores.
    \item Root Mean Square Error (RMSE)
            \begin{equation}
                \text{RMSE} = \sqrt{\frac{1}{N}\sum_{i}^{N}(q_i-o_i)^2}.
            \end{equation}
    \item Mean absolute error (MAE)
            \begin{equation}
                \text{MAE} = \frac{1}{N}\sum_{i}^{N}|q_i-o_i|.
            \end{equation}
    
\end{itemize}
Different statistical indexes demonstrate different aspects of the performance of the VQA model.
Among these traditional evaluation metrics, SRCC, KRCC, and PLCC calculate the correlation between the subjective quality ratings and the objective predicted scores, which demonstrate the prediction monotonicity, and RMSE and MAE compute the error between the subjective quality ratings and the objective predicted scores, which indicates the prediction accuracy.
The higher SRCC, KRCC, PLCC values (closer to 1) and the lower RMSE and MAE values (closer to 0) mean better performance.

\begin{table*}[!t]
    \centering
    \caption{Performance comparison of full-reference and reduced reference video quality assessment algorithms on LIVE VQA \cite{seshadrinathan2010study} database.}
    \label{tab:tab5.1}
    \setlength{\tabcolsep}{0.7em}
    \scalebox{0.7}{
    \renewcommand{\arraystretch}{1.5}
    \begin{tabular}{l l c c c c c c c c c c c}
    \toprule
    \multirow{2}{*}{Type} & \multirow{2}{*}{Metrics} & \multicolumn{5}{c}{SRCC} & & \multicolumn{5}{c}{PLCC} \\ 
    \cline{3-7}\cline{9-13}
     & & Wireless & IP & H.264 & MPEG-2 & All & & Wireless & IP & H.264 & MPEG-2 & All \\
     \hline
     \multirow{10}{*}{FR} & PSNR & 0.7381 & 0.6000 & 0.7143 & 0.6327 & 0.6958 & & 0.7274 & 0.6395 & 0.7359 & 0.6545 & 0.7499 \\
     & SSIM~\cite{wang2004image} & 0.7381 & 0.7751 & 0.6905 & 0.7846 & 0.7211 & & 0.7969 & 0.8269 & 0.7110 & 0.7849 & 0.7883 \\
     & VIF~\cite{sheikh2006image} & 0.7143 & 0.6000 & 0.5476 & 0.7319 & 0.6861 & & 0.7473 & 0.6925 & 0.6983 & 0.7504 & 0.7601 \\
     & STMAD~\cite{vu2011spatiotemporal} & 0.8257 & 0.7721 & 0.9323 & 0.8733 & 0.8301 & & 0.8887 & 0.8956 & 0.9209 & 0.8992 & 0.8774 \\
     & ViS3~\cite{vu2014vis} &  0.8257 & 0.7712 & 0.7657 & 0.7962 & 0.8156 & & 0.8597 & 0.8576 & 0.7809 & 0.7650 & 0.8251 \\
     & MOVIE~\cite{seshadrinathan2009motion} & 0.8113 & 0.7154 & 0.7644 & 0.7821 & 0.7895 & & 0.8392 & 0.7612 & 0.7902 & 0.7578 & 0.8112 \\
     & V-BLINDS~\cite{saad2014blind} & 0.8462 & 0.7829 & 0.8590 & 0.9371 & 0.8323 & & 0.9357 & 0.9291 & 0.9032 & 0.8757 & 0.8433\\
     & SACONVA~\cite{li2015no} & 0.8504 & 0.8018 & 0.9168 & 0.8614 & 0.8569 & & 0.8455 & 0.8280 & 0.9116 & 0.8778 & 0.8714 \\
     & DeepQA~\cite{kim2017deep} & 0.8290 & 0.7120 & 0.8600 & 0.8940 & 0.8678 & & 0.8070 & 0.8790 & 0.8820 & 0.8830 & 0.8692 \\
     & DeepVQA~\cite{kim2018deep} & 0.8674 & 0.8820 & 0.9200 & 0.9729 & 0.9152 & & 0.8979 & 0.8937 & 0.9421 & 0.9443 & 0.8952 \\
    \hline
    \multirow{3}{*}{RR} & TRRED~\cite{soundararajan2012video} & 0.7765 & 0.7513 & 0.8189 & 0.5879 & 0.7802 & & 0.7726 & 0.7619 & 0.8324 & 0.5998 & 0.7743 \\
     & SRRED~\cite{soundararajan2012video} & 0.7925 & 0.7624 & 0.7542 & 0.7249 & 0.7592 & & 0.8067 & 0.8033 & 0.7462 & 0.7281 & 0.7764 \\
     & STRRED~\cite{soundararajan2012video} & 0.7857 & 0.7722 & 0.8193 & 0.7193 & 0.8007 & & 0.8039 & 0.8020 & 0.8228 & 0.7467 & 0.8062 \\
    \bottomrule
    \end{tabular}
    }
\end{table*}

\begin{table*}[!t]
    \centering
    \caption{Performance comparison of no reference video quality assessment algorithms on KoNViD-1k~\cite{hosu2017konstanz}, LIVE-VQC~\cite{sinno2018large}, YouTube-UGC~\cite{wang2019youtube} databases. The top half: performances of IQA metrics; the bottom half: performances of VQA metrics.}
    \label{tab:tab5.2}
    \setlength{\tabcolsep}{0.5em}
    \scalebox{0.7}{
    \renewcommand{\arraystretch}{1.5}
    \begin{tabular}{l c c c c c c c c c c c c c c c c c c}
    \toprule
    \multirow{2}{*}{Metrics} & & \multicolumn{3}{c}{KoNViD-1k~\cite{hosu2017konstanz}} & & \multicolumn{3}{c}{LIVE-VQC~\cite{sinno2018large}} & & \multicolumn{3}{c}{YouTube-UGC~\cite{wang2019youtube}} & & \multicolumn{3}{c}{All-Combined} \\ 
    \cline{3-5}\cline{7-9}\cline{11-13}\cline{15-17}
    & & SRCC & PLCC & RMSE & & SRCC & PLCC & RMSE & & SRCC & PLCC & RMSE & & SRCC & PLCC & RMSE \\
    \hline
    NIQE \cite{mittal2012making} &
    & 0.5417 & 0.5530 & 0.5336 & & 0.5957 &  0.6286 & 13.110 & & 0.2379 & 0.2776 & 0.6174 & & 0.4622 & 0.4773 & 0.6112 \\
    BRISQUE \cite{mittal2012no} &
    & 0.6567 & 0.6576 & 0.4813 & & 0.5925 & 0.6380 & 13.100 &
    & 0.3820 & 0.3952 & 0.5919 & & 0.5695 & 0.5861 & 0.5617 \\
    GM-LOG \cite{xue2014blind} &
    & 0.6578 & 0.6636 & 0.4818 & & 0.5881 & 0.6212 & 13.223 &
    & 0.3678 & 0.3920 & 0.5896 & & 0.5650 & 0.5942 & 0.5588 \\
    HIGRADE \cite{kundu2017no} &
    & 0.7206 & 0.7269 & 0.4391 & & 0.6103 & 0.6332 & 13.027 & 
    & 0.7376 & 0.7216 & 0.4471 & & 0.7398 & 0.7368 & 0.4674 \\
    FRIQUEE \cite{ghadiyaram2017perceptual} &
    & 0.7472 & 0.7482 & 0.4252 & & 0.6579 & 0.7000 & 12.198 &
    & 0.7652 & 0.7571 & 0.4169 & & 0.7568 & 0.7550 & 0.4549 \\
    CORNIA \cite{ye2012unsupervised} &
    & 0.7169 & 0.7135 & 0.4486 & & 0.6719 & 0.7183 & 11.832 &
    & 0.5972 & 0.6057 & 0.5136 & & 0.6764 & 0.6974 & 0.4946 \\
    HOSA \cite{xu2016blind} &
    & 0.7654 & 0.7664 & 0.4142 & & 0.6873 & 0.7414 & 11.353 &
    & 0.6025 & 0.6047 & 0.5132 & & 0.6957 & 0.7082 & 0.4893 \\
    KonCept512 \cite{hosu2020koniq} &
    & 0.7349 & 0.7489 & 0.4260 & & 0.6645 & 0.7278 & 11.626 &
    & 0.5872 & 0.5940 & 0.5135 & & 0.6608 & 0.6763 & 0.5091 \\
    PaQ-2-PiQ \cite{ying2019patches} &
    & 0.6130 & 0.6014 & 0.5148 & & 0.6436 & 0.6683 & 12.619 &
    & 0.2658 & 0.2935 & 0.6153 & & 0.4727 & 0.4828 & 0.6081 \\
    \hline
    V-BLIINDS \cite{saad2014blind} &
    & 0.7101 & 0.7037 & 0.4595 & & 0.6939 & 0.7178 & 11.765 &
    & 0.5590 & 0.5551 & 0.5356 & & 0.6545 & 0.6599 & 0.5200 \\
    TLVQM \cite{korhonen2019two} &
    & 0.7729 & 0.7688 & 0.4102 & & 0.7988 & 0.8025 & 10.145 &
    & 0.6693 & {0.6590} & {0.4849} & & 0.7271 & 0.7342 & 0.4705  \\
    VMEON \cite{liu2018end} & 
    &  0.1118 & 0.1958 & 0.6322 &  
    &  0.4024 & 0.4088 & 15.524 &
    & 0.0634 & 0.1100 & 0.6304 &
    & 0.2578 & 0.2594 & 0.6657
    \\
    VSFA \cite{li2019quality} &
    & 0.7728 & 0.7754 & 0.4205 & & 0.6978 & 0.7426 & 11.649 &
    & - & - & - & & - & - & - \\
    MDVSFA \cite{li2020unified} &
    & 0.7812 & 0.7856 & - & & 0.7382 & 0.7728 & - &
    & - & - & - & & - & - & - \\
    VIDEVAL \cite{tu2020ugc} &
    & 0.7832 & 0.7803 & 0.4026 & & 0.7522 & 0.7514 & 11.100 &
    & 0.7787 &  0.7733 & 0.4049 & & 0.7960 & 0.7939 & 0.4268 \\
    RAPIQUE \cite{tu2021rapique} & 
    & 0.8031 & 0.8175 & 0.3623 & & 0.7548 & 0.7863 & 10.518 &  
    & 0.7591 & 0.7684 & 0.4060 & & 0.8070 & 0.8229 & 0.3968 \\
    PVQ~\cite{ying2021patch} & & 0.791 & 0.795 & - &&  0.770 & 0.807 & - && - & - & - && - & - & - \\
    Li \textit{el al.} \cite{li2022blindly} & & 0.836 & 0.834 & -  && - & - & - && 0.831 & 0.819 & - && - & - & - \\
    SimpleVQA \cite{sun2022deep} & & 0.856 & 0.860 & -  && - & - & - && 0.847 & 0.856 & - && - & - & - \\
    FastVQA \cite{wu2022fast} && 0.891 & 0.892 & -  && 0.849 & 0.865 & - && 0.855 & 0.852 & - && 0.865 & 0.869 & - \\
    \bottomrule
    \end{tabular}
    }
\end{table*}

\subsection{Performance Comparison}
We compare the performance of the surveyed VQA methods in this subsection.
Since not all reviewed algorithms are publicly available, for a fair comparison, we take the performance reported in the original papers.

Table \ref{tab:tab5.1} demonstrates the performance of FR and RR video quality assessment algorithms on the LIVE VQA \cite{seshadrinathan2010study} database.
It can be observed that general-purpose FR-IQA measures such as PSNR, SSIM \cite{wang2004image}, VIF \cite{sheikh2006image}, \textit{etc.,} perform worse than general-purpose FR-VQA metrics such as STMAD \cite{vu2011spatiotemporal}, ViS3 \cite{vu2014vis}, MOVIE \cite{seshadrinathan2009motion}, which demonstrates that hand-crafted temporal features are useful for the VQA task.
Moreover, deep learning-based VQA methods such as DeepVQA \cite{kim2018deep} achieve better performance compared to traditional models, which demonstrates the effectiveness of using DNN in the VQA task.

Table \ref{tab:tab5.2} demonstrates the performance of NR video quality assessment algorithms on three databases including KoNViD-1k~\cite{hosu2017konstanz}, LIVE-VQC~\cite{sinno2018large}, YouTube-UGC~\cite{wang2019youtube} databases.
It can be observed that traditional NR-IQA measures such as NIQE \cite{mittal2012making}, BRISQUE \cite{mittal2012no}, \textit{etc.}, performs worse than deep NR-IQA models such as KonCept512 \cite{hosu2020koniq} and PaQ-2-PiQ \cite{ying2019patches}.
Moreover, VQA models perform better than IQA models on the NR VQA task, which further demonstrates the importance and necessity of developing specific VQA algorithms.

\section{Future Research Directions}
\label{sec:future}

Though significant advancements have been achieved in field of video quality assessment in recent years, there remain unresolved challenges and promising research directions. In this section, we present an overview of promising research direction as follows.

\subsection{Human Perception Mechanism of Video Quality Assessment}

\noindent \textbf{Human perception} is a complex system and many scientific studies have conducted research on this problem \cite{tong2006neural,blake2002visual,duan2019dataset,duan2019visual,tu2022end,tu2022iwin,shi2021drawing}.
Due to the evolution of multimedia systems, many new capture, compression, transmission, and display techniques have been developed, which may have different influences on human visual perception \cite{shi2021semantic,duan2022confusing,duan2022saliency,duan2022develop,duan2022unified}.
It is necessary to study human perception in these new media systems, and conduct corresponding subjective quality assessment research.
Moreover, vision science-based models have been dominant methods in IQA and VQA for many years even with the revolution of deep learning \cite{wang2004image,li2016toward,saha2023study,mantiuk2021fovvideovdp}, and will continue to play a crucial role in the future video quality assessment realm since they can provide robust and reliable results through the comprehensive understanding of human visual perception.
Studying and integrating perception-based VQA models can enhance the interpretability and robustness of current VQA systems.



\subsection{Large Multi-modality Models for Video Quality Assessment}

Large multi-modality models (LMMs)~\cite{liu2023improved,ye2023mplug} have demonstrated excellent performance in various vision-language tasks, including image captioning, visual question answering, cross-modality grounding, as well as pure vision tasks such as image classification, object detection. Some studies~\cite{wu2023q,zhang2023q,wu2023q2,wu2023q3} have applied LMMs in the filed of image quality assessment. For instance, Wu \textit{et al.} created Q-Bench~\cite{wu2023q3}, a benchmark for evaluating the low-level visual perception ability of LMMs. They further introduced two LMMs, Q-Instruct~\cite{wu2023q2} and Q-Align~\cite{wu2023q}, which were respectively fine-tuned Q-Pathway, a low-level visual instruction dataset, and existing I/VQA datasets. While these models achieved remarkable performance on image quality assessment and image quality description, their performance in VQA still lags behind state-of-the-art methods. The primary reason is that current LLM-based quality assessment methods are tailored for images and overlook the distinctive characteristics of video content, such as various temporal distortions. Therefore, there is a necessity to design a LLM-based video quality assessment model.

\subsection{Quality Assessment of Emerging Video Media}

\noindent \textbf{Emerging media}, such as VR/AR/MR, HFR, HDR, gaming, \textit{etc.}, are becoming increasingly important in multimedia, which brings new challenges and opportunities for VQA research \cite{duan2017ivqad, sun2017cviqd, sun2018large, duan2018perceptual,duan2019perceptual,sun2019mc360iqa,duan2022confusing,duan2017assessment,yang2019predicting,duan2022saliency}.
Moreover, the advancement of communication systems, such as 5G/6G, semantic communication \cite{dong2022semantic}, also promotes the new video applications.
Thus, corresponding specific VQA systems are also required.
Specifically, multi-modal VQA is an important emerging topic, especially for immersive media.
For extended reality (XR), more multi-modal \cite{sun2023influence} quality assessment datasets and models are needed, which are not limited to the visual and auditory modalities, but also include other modalities, such as olfactory, gustatory, and tactile perception.
It is necessary to consider incorporating the immersive and interactive nature of XR contents in VQA \cite{zhu2023perceptual,zhu2023audio}, such as higher-order ambisonics.

\subsection{Quality Assessment of AIGC Videos}
\noindent \textbf{AI generated content} (AIGC) have achieved significant progress recently \cite{radford2018improving,rombach2022high,zheng2023judging}.
With the advancement of text-to-image \cite{ramesh2022hierarchical,rombach2022high} and text-to-video \cite{singer2022make,wu2023tune} techniques, AI-based image and video generation has been applied to various fields.
Some studies have investigated the unique distortions in AI generated images (AIGIs) and conducted IQA research \cite{wang2023aigciqa2023,zhang2023perceptual,xu2023imagereward}.
The further exploration of the quality assessment for AI generated videos (AIGVs) can help control and improve the quality of AIGVs, which is a new trend for future VQA research.

\subsection{Quality Assessment of Volumetric Videos}

\noindent \textbf{Volumetric video} has garnered increasing research interest since it can provide users immersive and realistic experiences by representing the complete volume of 3D content. Though there are numerous quality assessment studies for static 3D content~\cite{zhang2021no, zhang2022no, zhang2022treating, zhang2022mm, fan2022no, zhang2023eep, zhang2023gms, zhang2023advancing, zhou2023no, zhang2023simple, zhang2023evaluating}, there is a scarcity of studies~\cite{zerman2020textured, zhang2023ddh, zhang2023geometry, fan2023mv} on volumetric video, a format representing dynamic 3D content. Zerman \textit{et al.}~\cite{zerman2020textured} collected eight volumetric video sequences and investigated the subjective perception differences among different compression methods, including Darco, geometry-based point cloud compression (G-PCC)~\cite{schwarz2018emerging}, and video-based point cloud compression (V-PCC)~\cite{schwarz2018emerging}. Based on this databaset, Fan \textit{et al.}~\cite{fan2023mv} introduced a multi-view learning method for NR volumetric video quality assessment, utilizing 3D-CNN to extract features from multi-view projected video sequences.

\subsection{Green Learning for Video Quality Assessment}

\noindent \textbf{Green learning for VQA} is a promising study area because of its characteristics of low carbon footprints, lightweight model, low computational complexity, and logical transparency~\cite{kuo2023green}. The existing VQA models are typically based on DNNs, characterized by large model sizes and high computational complexity. So, they are hardly deployed in edge devices or real-time processing system. Mei \textit{et al.}~\cite{mei2023blind} have developed a lightweight NR VQA model called GreenBVQA, comprising four processing pipeline: video data cropping, unsuperised representation generation, superivsed feature selection, and MOS regression and ensembles. With the increasing demand for lightweight VQA models, green learning for VQA is becoming more and more important.

\section{Summary}
\label{sec:summary}
In this survey, we perform an extensive review of perceptual video quality assessment research.
Subjective video quality assessment methodologies and databases are first reviewed.
Then full-reference, reduced-reference and no-reference objective video quality assessment metrics are summarized and analyzed in sequence.
Emerging topics in the realm of objective video quality assessment for compressed videos, streaming videos, stereoscopic videos, VR/AR videos, HFR videos, audio-visual videos, HDR videos, screen and game videos, are also reviewed.
Finally, we evaluate and compare the performance of many objective video quality assessment models.
This survey provides a systematic overview of classical and recent progress in the VQA realm, which helps researchers in related areas quickly access the progress, and find solutions and trends in their study.
\Supplements{Appendix A.}


\bibliographystyle{scis}
\bibliography{ref}



\end{document}